\documentclass[12pt]{article}
\usepackage{amsmath}
\usepackage{graphicx}
\usepackage{enumerate}
\usepackage{natbib}
\usepackage{url} % not crucial - just used below for the URL 
\usepackage{multirow}
\usepackage{mathtools}
\usepackage{bm}
\usepackage{amsthm}
\usepackage{amssymb}
\usepackage{algorithm}
\usepackage{algpseudocode}
\usepackage{psfrag,epsf}
\usepackage{subcaption}
\usepackage{caption}
\usepackage{etoolbox}
\usepackage{xcolor}
\usepackage{makecell}
\usepackage{comment}
\usepackage{times}
\usepackage{setspace}

\makeatletter
\def\@seccntformat#1{\@ifundefined{#1@cntformat}%
   {\csname the#1\endcsname\quad}%    default
   {\csname #1@cntformat\endcsname}}% enable individual control
\apptocmd\appendix{%
    \newcommand\section@cntformat{\appendixname\ }
    \addtocontents{toc}{\bigskip\noindent\textbf{Appendix Material}\par}
    {}{}}
\makeatother

\usepackage{listings}
\lstnewenvironment{rc}[1][]{\lstset{language=R}}{}
\newenvironment{bessalgorithm}[2][htb] % Added a second parameter for custom numbering
  {%
   \floatname{algorithm}{BESS Algorithm} % Change algorithm name
   \def\thealgorithm{#2} % Set the algorithm number to the second parameter
   \begin{algorithm}[#1]% Start the algorithm with the specified floating option
  }%
  {\end{algorithm}\addtocounter{algorithm}{-1}} % End the algorithm and correct the global counter

\newenvironment{nminalgorithm}[2][htb] % Added a second parameter for custom numbering
  {%
   \floatname{algorithm}{Nmin Algorithm} % Change algorithm name
   \def\thealgorithm{#2} % Set the algorithm number to the second parameter
   \begin{algorithm}[#1]% Start the algorithm with the specified floating option
  }%
  {\end{algorithm}\addtocounter{algorithm}{-1}} % End the algorithm and correct the global counter

\DeclarePairedDelimiter\floor{\lfloor}{\rfloor}

\theoremstyle{plain}

\newtheorem{prop}{Proposition}
\newtheorem{theo}{Theorem}

\newtheorem{remark}{Remark}
\newtheorem{lemma}{Lemma}

\newcommand{\by}{\bm{y}}
\newcommand{\bth}{\bm{\theta}}

\newcommand{\cB}{\mathcal{B}}

\newcommand{\blind}{1}

\newcommand{\yy}{\color{red}}
\newcommand{\bb}{\color{blue}}
\newcommand{\jj}{\color{black}}

\newcommand{\tbph}{\tilde{\bm{\phi}}}

\begin{document}

\def\spacingset#1{\renewcommand{\baselinestretch}%
{#1}\small\normalsize} \spacingset{1}

\if1\blind
{
  \title{\LARGE\bf BESS: A Bayesian Estimator of Sample Size}
  \author{Dehua Bi \\
  %\thanks{
    %The authors gratefully acknowledge %\textit{please remember to list all relevant funding sources in the unblinded version}}\hspace{.2cm}\\
    Department of Biomedical Data Science, Stanford University, CA\\
    and \\
    Yuan Ji \\
    Department of Public Health Sciences, The University of Chicago, IL}
  \maketitle
} \fi

\if0\blind
{
  \bigskip
  \bigskip
  \bigskip
  \begin{center}
    {\LARGE\bf BESS: A Bayesian Estimator of Sample Size}
\end{center}
  \medskip
} \fi

\bigskip
\begin{abstract}
We consider a   Bayesian framework for estimating the %estimator of 
sample size of a clinical trial. %(BESS) and an application to  oncology    dose optimization clinical   trials.  
The new approach, called BESS,  is built upon three pillars:  {\bf S}ample size of the trial,  {\bf E}vidence from the observed data, and {\bf C}onfidence of the final decision in the posterior inference. 
It uses a simple logic of ``given the evidence from data, a specific sample size can  achieve a degree of confidence in trial success." %in the posterior inference." 
The key distinction between BESS and standard sample size estimation (SSE) is that SSE, typically based on Frequentist inference,  specifies the true parameters values in its calculation to achieve properties under repeated sampling while BESS assumes possible outcome from the observed data to achieve high posterior probabilities of trial success.  As a result, the calibration of the sample size is directly based on the probability of making a correct decision rather than \jj %not based on 
type I  or type II error rates. %, but on posterior probabilities. 
We demonstrate that BESS leads to a more interpretable statement for investigators, and can easily accommodates  prior information as well as sample size re-estimation.  We  explore its performance in comparison to the standard SSE and demonstrate its usage  through a case study of oncology optimization trial. 
An R tool is available at \url{https://ccte.uchicago.edu/BESS}.
\end{abstract}

\noindent%
\vfill

\newpage
\spacingset{1.9} % DON'T change the spacing!
%\section{Introduction}
%\label{sec:intro}

\section{Introduction}\label{s:intro}

\subsection{Motivation}
In  drug development,  
the clinical objective is to establish a treatment's  effectiveness and safety  
%Randomized clinical trials (RCTs) are the gold standard to achieve the objective 
based on the evidence manifested in the observed data. To achieve the objective,   
%As it is usually impractical to include all patients from a disease population, RCTs take 
a random sample of certain size  is taken from a target patient population to carry out the trial and statistical inference is performed to estimate the treatment effect.  % to address the clinical question through statistical inference. 
Due to the law of large numbers,  the larger the sample size,  the more precise is the estimated treatment effect  but also  more costly the trial.  Therefore, %in practice a 
sample size estimation (SSE) is typically %usually 
conducted prior to %before the
trial initiation %starts 
to balance the tradeoff between statistical precision %in statistical inference 
and study costs. %trial cost.  

The proposed research is motivated by the recent change in early-phase oncology drug development that recommends randomized comparison of multiple doses. In the past two decades,    oncology drug development has benefited from biological and genomics breakthroughs and  novel cancer drugs are no longer based on cytotoxic one-size-fits-all mechanism like chemicals or radiations. Instead, targeted, immune, and gene and cell  therapies eradicate  tumor cells by precisely altering oncogenic cellular or molecular pathways. Consequently, the traditional 
monotonic dose-response relationship is no longer valid for many novel cancer drugs.  Instead, efficacy often plateaus or even decreases after dose rises above certain level. 
To this end, the US FDA  launched  Project Optimus \citep{pjoptimus,shah2021drug,blumenthal2021optimizing} aiming to transform %the 
early-phase oncology drug development. Rather than defaulting to %The main objective of the project is to identify an optimal dose, 
%potentially lower than 
the maximum tolerated dose (MTD), the main objective is to identify an optimal dose that balances efficacy and tolerability. This new dose optimization paradigm recommends a %  but with at least comparable anti-tumor effect. The new dose optimization paradigm recommends  a 
randomized comparison of two or more doses following an initial dose escalation. While randomized controlled %clinical 
trials (RCTs) and SSE are routinely conducted in drug clinical trials, they are new in dose-optimization trials, usually restricted by limited resources. Controlling the type I/II error rates %for standard SSE 
is no longer %are \yy not \jj   
the main objective in %for 
early-phase  drug development. Instead, investigators are more concerned about the accuracy of %in the 
final decisions, such as selecting the right dose and determining whether to proceed with %including selection of the right dose or launching future 
confirmatory studies. %, both of which may be very costly.  
Existing  SSE for dose optimization trials is often based on ad-hoc choices, such as using a small sample size of 20 patients per dose. This leads to a question of how much is expected to learn from the 20 patients, or whatever the number may be.   
Moreover,  clinical trials that use human subjects for clinical research should always provide a reasonable justification for the number of subjects to be enrolled. We attempt to fill the gap by considering a Bayesian approach for sample size estimation, although the proposed approach is not restricted to dose optimization. 

\subsection{Review of SSE Methods}
In the literature,   standard SSE methods \citep{adcock1997sample,wittes2002sample,chow2008sample,desu2012sample} are based on Frequentist hypothesis testing aiming to control the type I  error rate  $\alpha$ and achieve a desirable  power $(1 - \beta)$, assuming the true population parameters are known. 
For example, a sample size statement for a two-arm RCT based on binary outcome is as follows:
\begin{itemize}
    \item[] {\bf Statement 1:} At type I  error rate of $\alpha$, with a clinically minimum effect size $\theta^*$, $n$ subjects are needed to achieve $(1-\beta)$ power when the response rates for the treatment and control arms are $\theta_1$ and $\theta_0$.
\end{itemize}
A %comprehensive 
review of SSE methods
can be found in \cite{wang2020sample}. In addition, sample size may be determined based on  
hybrid Frequentist-Bayesian inference  
\citep{ciarleglio2017sample,berry2010bayesian}. These methods   use Bayesian models for estimation but calibrate sample size based on type I error rate and power via simulations.  
In contrast, some methods use Bayesian properties for sample size estimation 
such as the length, coverage of posterior credible intervals, Bayes factors under sampling and fitting priors  \citep{chen2011bayesian,lin2022bayesian,chen2023biostatistical}, or loss functions related to false decision rates like false discovery %positive 
or false omission %negative 
rates \citep{muller2004optimal}. %like $k\cdot \text{False Positive Rate} + \text{False Negative Rate}$ . 
Novel Bayesian SSE approaches have also considered borrowing information from historical data \citep{chen2011bayesian,yuan2022bayesian}, mostly targeting device clinical trials. 

Several methods attempt to bridge the SSE philosophies between Frequentist and Bayesian methods. Notable contributions include \cite{kunzmann2021review}, \cite{inoue2005relationship}, and \cite{lee2000clinical}. In particular, \cite{lee2000clinical} propose to link posterior estimates with type I/II error rates, and  estimate sample size by providing posterior probabilities rather than the Frequentist error rates. The authors advocate defining ``statistical significance"  as the trial conclusion %outcome 
having a high %large 
posterior probability of being correct. %However, their sample size calculation still follows standard SSE approach.  

\subsection{Main Idea of BESS }
To this end,
we propose 
a new   Bayesian estimator of sample size  (BESS) based on a general hierarchical modeling framework and posterior inference. Motivated by but aiming to simplify the work in \cite{chen2011bayesian}, we consider a %\yy more \jj 
practical and simple  framework. Let $\by$  denote the observed data and  $\bth$ the unknown parameters as generic notation. We assume a Bayesian hierarchical model 
is to be used for the experimental design and data analysis. 
The  main idea is originated from noticing a simple trend between the sample size and observed %effect size from 
data. Consider a two-arm RCT with a binary outcome and sample size $n$ per arm.  We are interested in  testing if the effect size (difference in response rates), defined as $\theta  = \theta_1 - \theta_0$ between the treatment (1) and control (0) is greater than a clinically minimum effect size $\theta^*$. This can be formulated as testing the null hypothesis $H_0: \theta \le \theta^*$ versus the alternative $H_1: \theta > \theta^*.$ Suppose after the trial is completed, the observed response rates are  $\bar{y}_1$ and $\bar{y}_0$ for the treatment and control arms, respectively. A Frequentist $z$-test is used to test the hypotheses 
given by
$$
z^* = \frac{(\bar{y}_1 - \bar{y}_0) - \theta^*}{\sqrt{[\bar{y}_1(1-\bar{y}_1)+\bar{y}_0(1-\bar{y}_0)]/n}.}
$$
If we assume, for the sake of argument, that $\bar{y}_1 = 0.3$, we find in the table below  the relationship between the sample size $n$ and $(\bar{y}_1 - \bar{y}_0)$, for reaching the same $z^*$ value of 1.64, i.e., a one-sided $p$-value of 0.05. We call $(\bar{y}_1 - \bar{y}_0)$ the ``evidence." In order to reach the same Frequentist statistical significance, sample size needs to increase  %in order to achieve $z^*=1.64 \, (p \mbox{-value} =0.05)$ 
when evidence decreases.     

\begin{center}
    \begin{tabular}{c|lllllll}
    \hline
        $z^* \, (p \mbox{-value}) $ & \multicolumn{7}{c}{1.64 (0.05)}   \\ \hline
        Sample size ($n$) & 10 & 20 & 30 & 50 & 100 & 500 & 1000 \\
        Evidence ($\bar{y}_1 - \bar{y}_0$) & 0.29 & 0.24 & 0.21 & 0.18 & 0.15 & 0.10 & 0.08 \\ \hline
         
    \end{tabular}
\end{center}
   
\bigskip

Instead of using a Frequentist inference like the $z$-test,  we consider Bayesian hypothesis testing based on posterior probability of the alternative hypothesis, which, in the setting of clinical trials, refers to the probability of success (POS).  %of the trial.  
Decision of accepting the alternative hypothesis is by thresholding POS %the  posterior probability of $H_1$ 
at a relatively large value $c \in (0, 1)$.  
Assume a Bayesian hierarchical model is given by $f(\by \mid \bth)\pi(\bth \mid H)\pi(H)$, where $H=H_0$ or $H_1$ is a binary indicator of the null and alternative hypotheses. Using the proposed two-arm trial with binary outcomes as an example,  
the proposed BESS aims to find a balance between 1) ``Sample size" $n$ of the clinical trial, 2) ``Evidence"  defined as the observed  treatment effect $e= \bar{y}_1 - \bar{y}_0$, and 3) ``Confidence" %$c$ 
defined as the POS, i.e., $\text{Pr}(H = H_1 \mid \by).$ %posterior probability of the alternative hypothesis.

Considering testing whether %if 
the treatment effect $\theta=\theta_1 - \theta_0$ is greater than $\theta^*$,  a minimum effect size, BESS provides a sample size statement as follows:  
\begin{itemize}
    \item[] {\bf Statement 2:}  
    Assuming the evidence 
    is   at least   $e$, $n$ subjects are needed to declare with confidence $c$ that the treatment effect is at least $\theta^*$. 
\end{itemize}
%Note that ``evidence" is a function of the data, not parameters, and ``confidence"  refers to $\text{Pr}(H_1 | \by)$, the posterior probability of the alternative hypothesis, i.e., the treatment effect is at least $\theta^*.$   Comparing 
{\bf Statement 1} and {\bf Statement 2} %, one can see that the two statements 
are based on different statistical properties, Frequentist type I/II error rates for {\bf Statement 1} and Bayesian posterior probabilities for {\bf Statement 2}.  Also, while {\bf Statement 1} assumes the true values of the parameters are given, {\bf Statement 2} assumes what might be observed from the trial data. 
Alternatively, one may consider a predictive distribution of evidence by integrating the evidence over a prior of $\pi(\bm{\theta})$ as suggested in \cite{kunzmann2021review}. However, the type of approaches would inevitably invoke the probability inference based on the sampling or marginal distribution of $\by$, deviating from our focus using Bayesian inference based on posterior distributions.  
We will show that the  BESS and associated  {\bf Statement 2} is easier to interpret in practice since it directly addresses the uncertainty in the decision to be made for the trial at hand, measured by posterior probabilities. 

The remainder of the article is organized as follows: Section \ref{subsec:hminfprior} presents the proposed probability models and Section \ref{sec:sec_trio_alg} describes the  BESS method. Section \ref{sec:prop} illustrates some of the conditions and proprieties of BESS, namely, the ``$e$-Coherence" and the ``$n$-Coherence." %, including the relationships among sample size, evidence, and confidence, as well as \yy a property called  ``coherence". \jj % between BESS and Bayesian inference. 
%\note{Ed: I think we have renamed this part. Section 4 illustrates properties and conditions of BESS, which are the relationships among sample size, evidence, and confidence. The property is called e-coherence, and the condition is termed n-coherence.}
Section \ref{sec:oc} reports the operating characteristics of BESS with comparison to the standard SSE method. Section \ref{sec:na}  illustrates the applications of BESS  
to a hypothetical dose optimization trial with sample size re-estimation.  Finally, we conclude the article in Section \ref{sec:discuss}. Technical details are presented in the Appendix. 

\section{Probability Model } 
\label{subsec:hminfprior}

We consider BESS for both one-arm and two-arm trials. 
Denote   $y_{ij}$ the outcome of patient $i$ in arm $j$, where $i = 1, \ldots, n,$ index the patients, and $j=0$ and $ 1$ index the control and treatment arms, respectively.  When needed, we drop index $j$ and use $y_i$ for one-arm trials. Let  $\theta_0$ and $\theta_1$ denote the true response   parameters    for the control and treatment arms, respectively. %, and let 
%$\theta_0$ be the true response parameter for the control arm in a two-arm trial or the reference response parameter in a one-arm trial. 
Let $\theta = d(\theta_0, \theta_1) $, a function of $\theta_0$ and $\theta_1$, denote the treatment effect. For example, we consider $\theta =  \theta_1 - \theta_0  $ in this paper although it could be of other forms like $\theta = \theta_1 / \theta_0.$ %its form can be generalized.
Consider hypotheses    
\begin{equation}\label{eq:ht}
    H_0:   \theta \leq \theta^*    \text{ vs. } H_1: \theta >   \theta^*    ,
\end{equation}
where $\theta^*$ is a clinically minimum effect size. %minimum size %for treatment effect 
%deemed clinically meaningful.  
Let $H$ be the binary random variable taking $H_0$ or $H_1$ with probability $(1-q)$ and $q$, respectively. In the upcoming discussion, we will use three sets of notations for key trial parameters: %we use 
1) $\theta_0$ and $\theta_1$ to denote the true response parameters for the control and treatment arms, respectively,  2) $\theta = \theta_1 - \theta_0$ to denote the treatment effect, and 3) %later 
$\bm{\theta} = (\theta_0, \theta_1)$ to denote the parameter vector. %of \bb $\theta_j$'s for $j = 0,1$. \jj %\note{Ed: Originally, it was $\theta$'s. But this $\theta$ has been used in 2). So I think calling it $\theta_j$'s would be more clear?}

We propose a Bayesian hierarchical model for testing the hypotheses \eqref{eq:ht}. For $j=0$ or $1$, 
let 
\begin{eqnarray} \label{eq:bhm}
        y_{ij}\mid\theta_j &\sim& f(y_{ij} \mid \theta_j), \, i = 1, \ldots. n, \jj \nonumber \\ 
        %(\theta_0,\theta_1)
        \bm{\theta}\mid\tbph, H = H_j &\sim& \frac{1}{C_j} p( %\theta_0, \theta_1 
       \bm{\theta}\mid \tbph) I(\theta \in H_j), \\
         \text{Pr}(H = H_1) &=& q, \nonumber 
\end{eqnarray}
\noindent where $f(y_{ij} \mid \theta_j)$ represents a sampling model, $p(\bm{\theta}%\theta_0, \theta_1 
\mid \tbph)$ is a joint prior distribution for $(\theta_0, \theta_1)$, $\tbph$ are hyper-parameters, $C_j$ is the normalizing constant for the truncated priors under hypothesis $j$, i.e., $C_j = \iint_{\theta \in H_j}p(\bm{\theta}%\theta_0,\theta_1
\mid\tilde{\bm{\phi}})d\theta_0d\theta_1$, and    
$I(x \in A)$ is the indicator function which equals 1 if $x \in A$ and 0 otherwise.   For simplicity, we consider $p(\bm{\theta}%\theta_0, \theta_1 
\mid \tbph) = \pi(\theta_1 \mid \tbph) \pi(\theta_0 \mid \tbph) $ for some density $\pi$.  %, where $\pi_0 = \pi_1$.    

In this work, we consider   three specific types of outcome $y_{ij}$:  binary, continuous, and count. 
A summary of the parameter $\theta_j$, sampling model $f(\cdot)$, and prior distribution $\pi$ %\note{Ed: to be consistent with Table 1. Shall we use $\pi(\theta_j|\tilde{\bm{\phi}})$ or $p(\theta_0,\theta_1|\tilde{\bm{\phi}})$?}
%for $\theta_j$ 
is shown in Table \ref{tab:params}.  
We consider conjugate prior for simplicity and computational convenience although the proposed BESS framework can be extended to non-conjugate priors. %works for any general priors.  
While it is not the focus of this work to discuss choice of priors, we note the flexibility of BESS to incorporate various priors in real-life applications. For example, when little prior information is known vague priors like Jeffrey's prior may be considered; in contrast, it is also possible to use informative priors when prior information is available. 
For  instance, assume a previous trial with binary outcomes has been completed with a sample size of $n_0$  patients whose %of whom their 
outcome data are available, denoted as $\bm{y}^0 = \{y_{i}^0; i = 1, \ldots, n_0\}$. Then, one can %could 
consider an informative prior being $\text{Beta}(a,b)$ %\bb $\pi(\theta_1|\tilde{\bm{\phi}}) = \text{Beta}(a, b)$ \jj 
%\note{Ed: Here, we have not specify that the historical data of the previous trial is on the treatment. I think we may want to add this information to support $\pi(\theta_1) = \text{Beta}(a, b)$.} 
and set
$a = a^* + \sum_{i=1}^{n_0}y_{i}^0, \,\, b = b^* + n_0 - \sum_{i=1}^{n_0}y_{i}^0,$
where $a^*$ and $b^*$ are small (e.g., $a^* = b^* = 0.5$). 

\begin{table}[h!]
    \begin{center}
    \begin{tabular}{ll|lll}
    \hline 
       Model Name & Outcome type & Parameter $\theta_j$ & Sampling Model  $f(y\mid\cdot)$ & Prior $\pi(\theta_j %, \theta_1 
       \mid \tbph)$ \\
        \hline
       Binomial/Beta & Binary & Response rate & $\text{Bern}(\theta_j)$ & $\text{Beta}(a,b)$ \\ %\note{Ed: index $j$ for $a$ and $b$?} \\
      Normal/Normal &  Continuous & Mean response & $N(\theta_j, \sigma^2)$, $\sigma$ known &  $N(a,b)$  \\
       Poisson/Gamma  & Count-data & Event rate & $\text{Poisson}(\theta_j)$ &  $\text{Gamma}(a,b)$  \\
        \hline
    \end{tabular}
    \caption{Summary of model, parameter, sampling model %likelihood function, 
    and prior distribution for different outcome types. Hyperparameters $(a, b)$ are assumed to be specified by users. }\label{tab:params}
    \end{center}
\end{table}

\section{Proposed BESS Approach }%Confidence, Evidence, and Sample Size}
\label{sec:sec_trio_alg}

\subsection{Confidence}
We introduce three pillars of BESS, {\it confidence}, {\it evidence}, and {\it sample size}. We start with {\it confidence} %, the confidence 
in  posterior inference, expressed mathematically as %\bb POS, \jj %\note{Ed: we mentioned the term POS in introduction, I think we want to use it here for consistence.} %the posterior probability of the alternative hypothesis  
$\text{POS} = \text{Pr}(H=H_1 \mid \by_n),$ where $\by_n$ denotes the data with sample size $n$. 
The optimal decision rule under a variety of loss functions \citep{muller2004optimal} is to reject the null $H_0$ and accept the alternative $H_1$ if $\text{Pr}(H=H_1 \mid \by_n) \geq c$ for a high value of $c \in (0,1).$ 
The higher $c$ is, the more confident is the decision. 
To see this, one simply observes that $(1-c)$ is the upper bound of posterior probability of a wrong rejection since when $H_0$ is rejected, $\text{Pr}(H=H_0 \mid \by_n) < 1 - c$.

Denote $\by_n = \{y_{ij}, i=1,\ldots, n; j=0, 1\}$. %and $\bth = (\theta_0, \theta_1)$. 
Let $f(\by_n \mid \bth) = \prod_{i,j} f(y_{ij} \mid \theta_j)$ and 
$$
f(\bth \mid \by_n) = \frac{f(\by_n \mid \bth ) p(\bth)}{\int f(\by_n \mid \bth ) p(\bth) d \bth},
$$
where $p(\bth)$ is %= p(\theta_0, \theta_1)$ are
defined in model \eqref{eq:bhm}. 
According to model \eqref{eq:bhm}, it is straightforward to show that
\begin{equation}\label{eq:general_post_H1}
   \text{Pr}(H = H_1\mid\bm{y}_n) = \frac{q\cdot\frac{1}{C_1}\cdot \int_{\bm{\theta} \in H_1} f(\bm{\theta}\mid\bm{y}_n)d\bm{\theta}}{\frac{1 - q}{C_0} + \left[\frac{q}{C_1}-\frac{1-q}{C_0}\right]\cdot\int_{\bm{\theta} \in H_1} f(\bm{\theta}\mid\bm{y}_n)d\bm{\theta}}.
\end{equation}
%\yy where $C_0$ and $C_1$ are previously defined for model \eqref{eq:bhm}. 
If we assume \textit{a priori}, both   hypotheses    are equally likely, i.e., $q = 0.5$,  
then equation \eqref{eq:general_post_H1} may be further reduced to 
\begin{equation}\label{eq:post_H1_deri}
   \text{Pr}(H = H_1\mid\bm{y}_n) = \frac{C_0\cdot\int_{\bm{\theta} \in H_1} f(\bm{\theta}\mid\bm{y}_n)d\bm{\theta}}{C_1 + (C_0-C_1)\cdot\int_{\bm{\theta} \in H_1} f(\bm{\theta}\mid\bm{y}_n)d\bm{\theta}}.
\end{equation}
If conjugate priors 
in Table \ref{tab:params} are used,     $f(\bm{\theta} \mid \bm{y}_{n})$ have closed-form solutions.   Otherwise, numerical evaluation of \eqref{eq:general_post_H1} or \eqref{eq:post_H1_deri} is needed. 

\subsection{Evidence}
%Evidence is the main metric that differentiates BESS from a standard SSE.  In short, 
We define evidence  $e(\by_n)$ as a function of data that quantifies the treatment effect. This is similar to the ``population-level" summary in ICH E9R1 guidance \citep{ICH_E9R1_2019}.  In the settings listed in Table \ref{tab:params}, we define  evidence  as 
\begin{equation}\label{eq:evidence}
    e = \bar{y} - \theta_0 \text{ for one-arm trials, and } e = \bar{y}_1 - \bar{y}_0 \text{ for two-arm trials},
\end{equation}
where $\bar{y}$ is the sample mean in one-arm trials, and $\bar{y}_j$ for $j = 0, 1$ are the sample means for the control and treatment arms, respectively, in two-arm trials.
%In simple words, evidence $e$ is the observed effect size from the trial data before they are observed. This means that in 
In order to apply BESS, one needs to prespecify (and calibrate) the evidence, analogous to the calibration of %the potential observed effect size before the trial is conducted. This is analogous to the requirement of specifying 
the true parameter values $\bth$ in the standard  
SSE, except that BESS assumes what might be observed rather than what might be true. 
%We consider evidence $e$ as  a function of the sufficient \yy statistics \jj  in Table \ref{tab:params}. 
For the three outcomes in Table \ref{tab:params} in one-arm trials, $e + \theta_0$ is  the sufficient statistic. For two-arm trials, $e$ is the difference between the sufficient statistics $(\bar{y}_0, \bar{y}_1)$ of the two arms. Being a function of sufficient statistics allows for a search algorithm to find an appropriate sample size under BESS, which will be clear next. Another option is to define evidence as a test statistics, such as the $z$-statistics. However, this may reduce the interpretability of the evidence to nonstatisticians and may lead to more complicated calibration. We leave this option for future research.

\subsection{Sample Size of BESS} \label{subsec:bess_algs}

We first briefly review the standard SSE  
based on Frequentist inference. Considering a $z$-test for a two-arm trial with binary outcome,  the standard sample size approach   assumes  true values of  $\theta_1$ and $\theta_0$, and solves for $n$ based on desirable type I/II error rates 
$\alpha$/$\beta$ given by
\begin{equation}\label{eq:sse}
    n = \frac{(z_\alpha + z_{\beta})^2}{(\theta - \theta^*)^2}[\theta_1(1-\theta_1) + \theta_0(1-\theta_0)].
\end{equation}
In BESS,   we find the sample size through a similar argument but using posterior inference instead. Investigators specify the confidence $c$, so that   
\begin{equation}\label{eq:BESS_cond}
    \text{Pr}(H = H_1\mid\bm{y}_n) \geq c,
\end{equation}
  where $\text{Pr}(H = H_1\mid\bm{y}_n)$ is computed by equation \eqref{eq:general_post_H1}. In \cite{muller2004optimal} decision rule \eqref{eq:BESS_cond} is shown to be optimal for a variety of common loss functions, such as % the posterior expected loss of 
  $(k\cdot \text{FDR} + \text{FOR})$, where $\text{FDR}$ and $\text{FOR}$ are false discovery (positive) and omission (negative) rates, respectively.
  %\bb Additionally, analogous to the standard SSE property that, for fixed $\alpha$ and parameters $\theta_1$ and $\theta_0$, a larger sample size $n$ yields higher power $1 - \beta$, BESS possesses a property known as ``$n$-coherence". Specifically, under mild conditions, for fixed evidence $e$, increasing $n$ leads to greater confidence. This monotonic relationship enables the use of a grid-search-based algorithm to identify the appropriate sample size. The theoretical justification and detailed proofs for this property will be provided in the following section and in the Appendix.
  \jj
  Next,  
  we %\bb show that one may use a search based algorithm to find the sample size through  \jj  %
  provide three algorithms for sample size calculation based on BESS and models in Table \ref{tab:params}.

%\clearpage
%\newpage
%\yy
%\section{Prerequisite of BESS}
%\subsection{Coherence}
%BESS estimates a sample size $n$ under which when desirable evidence $e(\by_n)$ is demonstrated by the observed data, the POS $Pr(H=H_1 \mid \by_n)$ is at least as large as $c$, a high cutoff value. 
%\jj 

%\bigskip

\paragraph{One-arm trial} For one-arm trials, with the settings of likelihood and prior in Table \ref{tab:params}, assuming $\theta_0$ is known, we show that $\text{Pr}(H = H_1\mid\bm{y}_n) = \text{Pr}(H = H_1\mid e,n).$ 
See Appendix A.1  
for detail. Therefore, by fixing $e$ and $c$, we find the smallest sample size $n$ that satisfies \eqref{eq:BESS_cond}.   This leads to the proposed  BESS Algorithm \ref{alg:BESS1} and the corresponding  sample size statement {\bf BESS 1.}     

\bigskip

\fbox{\parbox{0.9\textwidth}{\bf 
BESS 1: Assuming the evidence is $e$, $n$ subjects are needed to declare with confidence $c$ that the treatment  effect is larger than $\theta^*$.}}

\bigskip
\noindent Here, ``declare with confidence $c$ that the treatment  effect is larger than $\theta^*$" means $\Pr(H=H_1 \mid \by_n) = \Pr(H = H_1\mid e,n) \ge c.$  

\paragraph{Two-arm trial; Continuous data, known variance} Second, for two-arm trials and continuous outcome with normal likelihood and known variance, we still have $\text{Pr}(H = H_1\mid\bm{y}_n) = \text{Pr}(H = H_1\mid e,n).$ See Appendix A.2 
for detail. Below is BESS statement 2.1 for sample size estimation in a two-arm trial with a normal sampling model and  known variance. %BESS Algorithm 1 applies. And we have the sample size statement BESS 2.1.     

\bigskip

\fbox{\parbox{0.9\textwidth}{\bf BESS 2.1: Assuming the evidence is $e$ and variance $\sigma^2$ is known for each arm, $n$ subjects per arm are needed to declare with confidence $c$ that the   treatment effect is larger than $\theta^*$.    }}

\bigskip

\noindent BESS Algorithm 1 lays out steps for  computing the sample size $n$ in \textbf{BESS 1} and \textbf{BESS 2.1}. 
The main idea is to keep increasing the sample size until the posterior probability $\Pr(H = H_1\mid e,n)$ is greater than the desirable level $c$, for fixed evidence $e$. This requires a condition called ``$n$-coherence" which will be established in the next section. In short, $n$-coherence ensures the monotone relationship between $n$ and $\Pr(H = H_1\mid e,n)$, for fixed $e$, thereby underwriting the validity of BESS Algorithm 1. However, the relationship only holds  when $n$ is greater than a small value $n_{\min}$, which can be identified by the Nmin Algorithm 1 in Appendix A.4. To summarize, BESS Algorithm 1 can be used to find a sample size $n$ for all three models in Table \ref{tab:params} (Binomial/Beta, Normal/Normal, and Poisson/Gamma) in one-arm trial and the Normal/Normal model in Table \ref{tab:params} in two-arm trial. 

%The use of Nmin Algorithm 1 in BESS Algorithm 1 is motivated by the theoretical condition underpinning $n$-coherence: for fixed $\theta^*$, hyperparameters, and evidence $e$, there exists a minimum sample size $n_{\min}$ beyond which the monotonic relationship between $n$ and $\Pr(H = H_1|e,n)$ is guaranteed to hold. Identifying this threshold is essential to ensure the validity of the grid-search approach for sample size determination. Accordingly, \textbf{Nmin Algorithm 1} (see Appendix A.12) is employed to determine $n_{\min}$ for any given configuration. %$n_{\min}$ such that, for any $n \geq n_{\min}$, the posterior probability $\Pr(H=H_1|e,n)$ is monotonically increasing in $n$ for fixed $e$. This property is essential for the validity of the search algorithm, and a detailed justification will be provided later. 
%Additionally, \bb in Step 5, we have $\Pr(H = H_1\mid \bm{y}_n) = \Pr(H = H_1\mid e,n)$ which corresponds to Equation (3). \jj

\bigskip
\begin{singlespace}
\begin{bessalgorithm}{1}
\caption{One-arm trials; two-arm trials with continuous data and known variance}\label{alg:BESS1}
\begin{algorithmic}[1]
    \State \textbf{input:} the %\note{Ed: This is not right. Algorithm 1 can be used for all three models in Table 1 in one-arm trial case. And it uses the normal/normal model in two-arm trial} \yy normal/normal \jj 
    hierarchical models in Table \ref{tab:params}. 
    \State \textbf{input:} clinically minimum %meaningful 
    effect size $\theta^*$, evidence $e$, confidence $c$, prior probability of $H_1$ $q$, reference %response rate 
    $\theta_0$ (for one-arm trials), maximum possible sample size $n_{\text{max}}$. 
    \State {\bf set} $n_{\text{max}}$ as the largest candidate sample size. 
    \State \textbf{find} $n_{\text{min}}$ using {Nmin Algorithm 1} (See Appendix A.4).   %For continuous outcome, find $n_{\text{min}}$ as the smallest $n$ such that the monotonic condition holds true, i.e.,
    %$$n_{\text{min}} = \min_n \{n;e > \theta^* + \frac{a}{b}\left(\frac{1}{b} + \frac{n}{\sigma^2}\right)^{-1},\; n = 1, 2, \ldots, n_{\text{max}}\}.$$\jj
    %\State {\bf Set} $n=n_{\text{min}}.$
    \While{$n \in \{n_{\text{min}},\ldots, n_{\text{max}}\}$} 
        \State compute $\text{Pr}(H=H_1 \mid e, n).$  %equation \eqref{eq:general_post_H1}.
        \If{$\text{Pr}(H=H_1 \mid e, n) \geq c $}
            \State stop and return the sample size $n$. 
        \Else
            \State $n=n+1$ 
        \EndIf
    \EndWhile 
    \If{$n > n_{\text{max}}$}
        \State return message: sample size is larger than $n_{\text{max}}.$
    \EndIf
\end{algorithmic}
\end{bessalgorithm}
\end{singlespace}
\medskip

\paragraph{Two-arm trial; Binary and Count data}

For two-arm trials with binary and count data outcomes, it can be shown that    
%in two-arm   trials,     
$\text{Pr}(H = H_1\mid\bm{y}_n) = \text{Pr}(H = H_1\mid\bar{y}_1,\bar{y}_0,n)$. See Appendix A.3 
for a quick proof. Since $e= \bar{y}_1 -\bar{y}_0$,  fixing evidence $e$ does not uniquely define $\text{Pr}(H = H_1\mid\bm{y}_n). $ 
However, for binary outcomes with $n$ samples per arm, $\bar{y}_1$ and $\bar{y}_0$ can only take on a finite number of values.
Hence, for binary outcomes, we   propose to specify evidence $e$ and find     
all pairs of $\bar{y}_1$ and $\bar{y}_0$ that satisfies $e= \bar{y}_1 -\bar{y}_0$.   
We then    find the minimum posterior probability of $H_1$ among these pairs of $\bar{y}_1$ and $\bar{y}_0$, i.e.,
\begin{equation}\label{eq:min_H1}
    \text{Pr}(H = H_1\mid e,n) =   \min_{\bar{y}_1, \bar{y}_0}    \{\text{Pr}(H = H_1\mid \bar{y}_1, \bar{y}_0,n ); \forall \bar{y}_1 - \bar{y}_0 = e\}.
\end{equation} 

\noindent Then the sample size statement for a two-arm trial is the same as \textbf{BESS 1}. And BESS Algorithm \ref{alg:BESS2} provides details of finding the sample size $n$ for a two-arm trial for binary outcomes. Note that, similar to BESS Algorithm \ref{alg:BESS1}, we use Nmin Algorithm 2 to determine $n_{\min}$ for BESS Algorithm \ref{alg:BESS2}, thereby ensuring the monotonic relationship between $n$ and $\Pr(H = H_1\mid e,n)$ for fixed $e$.  

\begin{bessalgorithm}{2}
\caption{Two-Arm Trials; Binary}\label{alg:BESS2}
\begin{algorithmic}[1]
    \State \textbf{input:} the Binomial/Beta hierarchical models in Table \ref{tab:params}.
    \State \textbf{input:} clinically minimum %meaningful 
    effect size $\theta^*$, evidence $e$, confidence $c$, prior probability of $H_1$ $q$, maximum possible sample size $n_{\text{max}}$. 
    \State \textbf{set} $n_{\text{max}}$ as the largest candidate sample size. 
    \State \textbf{find} $n_{\text{min}}$ using {Nmin Algorithm 2} (See Appendix A.4).
    %\State {\bf Set} $n = n_{\text{min}}$.
    \While{$n \in \{n_{\text{min}}, \ldots, n_{\text{max}}\}$}
    \State find all pairs of $(\bar{y}_1, \bar{y}_0)$ where $\bar{y}_1 - \bar{y}_0 = e$.
    \For{each pair of $(\bar{y}_1, \bar{y}_0)$}
    \State compute $\text{Pr}(H=H_1 \mid \bar{y}_1, \bar{y}_0, n ).$ 
    \EndFor
    \State compute equation \eqref{eq:min_H1}.
    \If{$\text{Pr}(H = H_1\mid e,n) \geq c$}
    \State stop and return the sample size $n$. 
    \Else
    \State $n = n+1$.
    \EndIf
    \EndWhile
    \If{$n > n_{\text{max}}$}
    \State return message: sample size is larger than $n_{\text{max}}$.
    \EndIf
\end{algorithmic}
\end{bessalgorithm}

%\bigskip

For count-data outcome, and as an alternative approach
for binary outcomes, one may   wish to specify   the values of 
$\bar{y}_1$ and $\bar{y}_0$ directly instead of their difference $e$. Since $(\bar{y}_1, \bar{y}_0)$ is sufficient, the sample size search algorithm becomes easier.    To this end, we propose a simple variation, BESS Algorithm \ref{alg:BESS2}' in Appendix A.5, which requires specification of the evidence as a vector $e = (\bar{y}_1, \bar{y}_0)$ rather than the scalar $e = \bar{y}_1 - \bar{y}_0. $
Statement \textbf{BESS 2.2} is used to interpret the sample size based on BESS algorithm 2'. %\note{Ed: We should only have BESS algorithm 2' here. BESS 1 is for BESS algorithm 2 (written explicitly in second-last line on page 12).}  % \yy below \jj assuming $(\bar{y}_1, \bar{y}_0)$ are given.  \bb In this case, Nmin Algorithm 1 is again used to identify the minimum sample size $n_{\min}$ that guarantees a monotinc relationship between $n$ and $\Pr(H = H_1|\bar{y}_1,\bar{y}_0,e)$ for fixed $\bar{y}_1$ and $\bar{y}_0$. \jj

\bigskip

\fbox{\parbox{0.9\textwidth}{\bf BESS 2.2 Assuming the response parameters in treatment and control arms are $\bar{y}_1$   and $\bar{y}_0$, respectively,     
$n$ subjects per arm are needed to declare with confidence $c$ that the treatment effect is larger than $\theta^*$.}}
\medskip

\bigskip

\paragraph{Discrete Data}

Lastly,   since  binary and count data are  
integer-valued, not all specified evidence value $e$ can be achieved  for a given sample size in  the proposed  BESS Algorithms \ref{alg:BESS1}, \ref{alg:BESS2} and 2'. For example, when the outcome is binary and  
sample size $n=10$, it is impossible to observe evidence $e=0.15$ since this requires having $n\cdot e = 1.5$ more responders in the treatment arm than the control. The number of responders cannot be a fraction.  
To this end, we propose to round down the specified  $e$  to the nearest possible value for a given $n$ by $\frac{1}{n}\floor{n \cdot e},$
where $\floor{.}$ is the floor function. This gives a conservative estimate of the sample size since even with smaller evidence, the sample size would still ensure the needed confidence. 

\section{Conditions and Properties of BESS}
\label{sec:prop}

It is important to explore the relationship among the three pillars of BESS, namely the sample size $n$, the evidence $e$, and the confidence $\text{Pr}(H=H_1 \mid e, n)$. Below we present two crucial results that empower the BESS algorithms.  % In order for BESS to work, the Bayesian models and defined evidence $e(\bm{y}_n)$ must satisfy some basic conditions. \jj
 
\subsection{Property: $e$-Coherence} %Correlation between sample size, evidence, and confidence}
\label{sec:relation}
%We explore the interrelationship among the three pillars of BESS: sample size, evidence, and confidence. Specifically, we fix one and investigate the correlation between the remaining two.
%using a two-arm trial with binary outcome. Similar results can be achieved   for other types of data or one-arm trials.  

%\paragraph{Positive correlation between confidence and evidence}%,   Figure \ref{fig:e_vs_s}(a)}

First, we demonstrate a monotone relationship %We first establish the positive correlation 
between confidence and evidence when sample size is fixed. %for a fixed sample size $n$, across all trial outcomes in both one-arm and two-arm trials. This correlation is formally stated in the following theorem. \jj

\begin{theo}\label{theo:pos_conf_evid}
    %\yy Consider \jj  the three hierarchical models in Table \ref{tab:params}. 
    Assume $e = \bar{y}_1 - \theta_0$ for one-arm trials and $e = \bar{y}_1 - \bar{y}_0$ for two-arm trials. Also assume a fixed $\bar{y}_0$  %Except 
    for the Poisson/Gamma model in Table \ref{tab:params} for two-arm trials. %where we assume, 
    The following property holds % is true 
    for all three models in Table \ref{tab:params} %the table 
    and for both one-arm and two-arm trials: %where $\bar{y}_0$ is replaced with $\theta_0$ for one-arm trials. % are proposed. 
    For a fixed clinically minimum %size for 
    %treatment 
    effect size $\theta^*$ %\bb a fixed or minimizer $\bar{y}_0$ (as of Equation \eqref{eq:min_H1}) in two-arm trials with binary or count-data outcome, \jj 
    and sample size $n$, the confidence $\text{Pr}(H = H_1\mid e,n)$ increases monotonically as % \yy $\bar{y}_1$ \jj increases, i.e., 
    $e$ increases. 
\end{theo}

\noindent The proof  is provided in Appendix A.6.  
%We compute confidence using equation \eqref{eq:general_post_H1} for various values of evidence $e$, while keeping the sample size $n$ constant \bb as an example to further demonstrate theorem \ref{theo:pos_conf_evid}. \jj   Figure \ref{fig:e_vs_s}(a)
%presents a line plot of the posterior probability of $H_1$ (the confidence) against various values of evidence, when sample size is set at $n = 10,$ $20,$ or $ 30$,   assuming     $\theta^* = 0.05$.
%This plot demonstrates that confidence increases monotonically from 0 to 1 as evidence shifts from -1 to 1.   When evidence is negative, the data does not support the alternative hypothesis and therefore the confidence (that the alternative is true) drops to zero. When evidence is positive, the confidence improves.  Interestingly, the order of confidence across different sample sizes flips when evidence is near 0, giving smaller sample size more confidence.  This is expected since when evidence is small, a larger sample size should imply less confidence of  the alternative hypothesis.    

%\begin{figure}
%    \centering
%    \includegraphics[width=\linewidth]{figs/updated/Updated_Sec4_d_1_ab.png}
%    \caption{The line plots of (left) confidence vs. evidence when sample size is fixed to be $n = 10, 20, \text{ and } 30$, and (right) confidence vs. sample size when $e = 0.1$. The result assumes binary outcome for a two-arm trial, with $\theta^* = 0.05$ \bb and $n_{\text{min}} = 1$. \jj}
%    \label{fig:e_vs_s}
%\end{figure}

 Theorem \ref{theo:pos_conf_evid} supports a property of BESS called  the \textbf{$e$-coherence}, which connects the sample size estimation with data analysis. %statements of BESS. 
 To see this,  %between \yy the design and \jj %BESS and Bayesian 
%inference. %stated below:
%\bigskip
%\fbox{\parbox{0.9\textwidth}{\yy 
suppose using BESS, one designs a trial of size $n$
by assuming evidence $e$ and desirable confidence $c$.
%\bb For two-arm trial with binary or count-data outcomes, when Algorithm 2' is used, then we further assume that $\bar{y}_0$ is fixed the same before and after the trial. \jj
Suppose the trial is conducted. 
% is used to design a trial by specifying a sample size $n$, assuming an evidence $e$ and confidence level $c$. After the trial is conducted and data observed, a Bayesian analysis of the observed data using posterior probability will be \textbf{coherent} with the BESS statement if the same prior is used for the sample size estimation and statistical inference. \jj
%\noindent More precisely, 
Let $\by^\star_n$ denote the observed data after the trial is completed, $e^\star$ the observed evidence, and $c^\star = \text{Pr}(H = H_1 \mid\by^\star_n)$ the POS %posterior probability of $H_1$ conditional 
on the observed data $\by^\star_n.$ %\bb Furthermore, if the trial is a two-arm trial with binary or count-data outcome, and BESS Algorithm 2' is utilized for trial planning, further assume that $\bar{y}_0$ stays the same before and after the trial. \jj

\begin{remark}\label{remark:coherence}
    \textbf{$e$-Coherence}: %The coherence between BESS and Bayesian inference implies that 
    If $e^\star \ge  e$, then $c^\star \ge  c.$
\end{remark}

\noindent The proof is immediate based on Theorem \ref{theo:pos_conf_evid}.
%\noindent To explain this, we first note that in the proposed BESS approach, the evidence $e(\by_n)$ is a function of the trial data $\by_n$. In addition, 
%the confidence, defined as  the posterior probability $\text{Pr}(H_1 | \by_n),$ is also a function of $\by_n$.  Let's take a look at the BESS sample size statement again. It can be generalized as 
%\bigskip
%\fbox{\parbox{0.9\textwidth}{For assumed evidence $e$, a sample size of $n$ will provide confidence $c$ that the alternative hypothesis is true. }}
%\bigskip
%\noindent Denote $\by^\star_n$ the observed data after the trial is completed, $e^\star$ the observed evidence, and $c^\star = \text{Pr}(H_1 |\by^\star_n)$ the posterior probability of $H_1$ conditional on the observed data $\by^\star_n.$  
% Then the coherence between BESS and Bayesian inference means:
%\begin{center}
%    If $e^\star \ge  e$, then $c^\star \ge  c.$
%\end{center}
The $e$-coherence property is important since it connects the sample size planning with data analysis in a coherent fashion. For example, the BESS sample size statement (e.g., \textbf{BESS 1}) promises a confidence value of $c$ assuming evidence $e$ and a sample size $n$. Therefore, when the actually observed evidence $e^\star$ is greater than or equal to the assumed value $e$, the actual posterior probability $c^\star$ is guaranteed to be greater than or equal to the promised level $c$.

%is important for  BESS to be adopted in practice 
%since investigators of  clinical trials \yy will be able to \jj %are usually not statisticians, and the coherence property of BESS allows them to 
%connect the design (i.e., sample size statement) of the trial with the statistical analysis of the observed data once the trial is carried out.  

The exception occurs when BESS %Lastly, it is important to note that $\bar{y}_0$ must be specified when using 
Algorithm 2' is used for the Poisson/Gamma model (Table \ref{tab:params}) in a two-arm trial and $\bar{y}_0$ is not fixed, which is illustrated by the following remark. %-- either for two-arm trials with count-data outcomes or as an alternative for two-arm trials with binary outcomes. As demonstrated in the following remark, the $e$-coherence property may fail to hold if $\bar{y}_0$ is not fixed, that is, when a different value of $\bar{y}_0$, denoted as $\bar{y}_0^\star$, is used. 
\begin{remark}\label{rm:2p_eco}
    % Counter example
    When applying BESS Algorithm 2' to estimate the sample size for two-arm trials with %binary or 
    %\note{I removed the binary data to be concise. We could assume for binary data, we always use BESS Algorithm 2.} 
    count-data %or binary  
    outcomes, since the evidence $e$ depends on both $\bar{y}_0$ and $\bar{y}_1$, when $e^\star > e$, $c^\star$ may be smaller than $c$. %if $\bar{y}_0$ is not fixed, then the confidence $\Pr(H = H_1|\bar{y}_0,e,n)$ may not increase monotonically as the evidence $e$ increases.
\end{remark}
\noindent We provide a counter example following Remark \ref{rm:2p_eco}. %Remark \ref{rm:2p_eco} can be demonstrated using counter-examples. 
Assume $q = 0.5$. %For binary outcomes, let $\theta^* = 0.05$, $a_0 = b_0 = a_1 = b_1 = 0.5$, and fix $n = 100$. Suppose $\bar{y}_0 = e = 0.1,$ we have $\Pr(H = H_1|\bar{y}_0,e,n) = 0.87$. If $e$ is increased to $e^\star = 0.11$, but $\bar{y}_0$ is changed to $\bar{y}_0^\star = 0.5$, then $\Pr(H = H_1|\bar{y}_0^\star,e^\star,n) = 0.84 < \Pr(H = H_1|\bar{y}_0,e,n)$. Similarly, f
For count data in a two-arm trial, let $\theta^* = 0.1$, $a_0 = a_1 = 1$, $b_0 = b_1 = 2$. Under BESS Algorithm 2', if we set $\bar{y}_0 = 1$ and $e = 0.5$ ($\bar{y}_1 = \bar{y}_0+e = 1.5$) we obtain a sample size  $n = 12$ that gives $\Pr(H = H_1\mid\bar{y}_1,\bar{y}_0,n) = 0.85$. However, if the observed data summaries are  $\bar{y}_0^\star = 5$ and $e^\star = 0.6$ ($\bar{y}_1^\star = \bar{y}_0^\star + e^\star = 5.6$), we have $\Pr(H = H_1\mid\bar{y}_1^\star, \bar{y}_0^\star,n) = 0.84 < 0.85$. %\note{Ed: I changed the notation as we have not used the notation of $\Pr(H = H_1\mid\bar{y}_0,e,n)$ before. The notation we used for Algorithm 2' is $\Pr(H = H_1\mid\bar{y}_1,\bar{y}_0,n)$}
%\yy We conduct a simulation study for a two-arm trial with binary outcomes using BESS Algorithm 2'. Specifically, we set $\theta^* = 0.05$, $n = 100$, and varied $e$ over a grid from $0.01$ to $0.15$. In each simulated trial, we generate $y_{i0} \overset{iid}{\sim} Bern(\theta_0=0.3)$ for $i = 1, \ldots, n$ and compute $\bar{y}_0$ as the average of the $y_{i0}$'s. We then set $\bar{y}_1 = \bar{y}_0 + e$ and calculate Equation \eqref{eq:general_post_H1} (since $(\bar{y}_0$, $\bar{y}_1)$ are sufficient). We obtain 100 posterior probabilities $\Pr(H=H_1 | y_n)$ for each value of $e$, and Figure \ref{fig:ecoh_demo} displays the corresponding boxplots. It can be seen overall the confidence increases with evidence, although when $e$ gets larger, sometimes a confidence value corresponding to a larger $e$ (say $e=0.13$) may be smaller than a confidence value corresponding to a smaller $e$ (say $e=0.12$). \jj 

%\note{Ed: I think this should be a new paragraph.} 
The counter example %usually 
requires the  data summary $\bar{y}_0^\star$ to deviate significantly from the assumed $\bar{y}_0$ in the BESS sample size estimation.  In real-life application, since $\bar{y}_0$ is estimated for the control arm that  has been extensively studied, it should be quite accurate and the observed $\bar{y}_0^\star$ is expected to be close to the estimated $\bar{y}_0$. And hence in reality, the $e$-coherence should still hold in general even for cases where BESS algorithm 2' is used.  

%the counter-example case rarely occurs. More commonly, $\bar{y}_0$ and $\bar{y}_0^\star$ are close to each other. In practice, $\bar{y}_0$ and $\bar{y}_0^\star$ may represent mean responses observed from different samples of size $n$, with each observation generated from some distribution with parameter $\theta_0$. Since the estimate of $\theta_0$ is typically accurate, large deviations between $\bar{y}_0$ and $\bar{y}_0^\star$ are uncommon. %before and after the trial while $e$ is also increasing, 

%\begin{figure}
%    \centering
%    \includegraphics[width=\linewidth]{figs/updated/Fig_ecoh.png}
%    \caption{Boxplot of Evidence versus Confidence when $\bar{y}_0$ is computed from simulated data and $\bar{y}_1 = \bar{y}_0 + e$. The red dashed line is the mean confidence for each $e$.}
%    \label{fig:ecoh_demo}
%\end{figure}

%It is evident that, even as $\bar{y}_0$ changes and $e$ increases from 0.01 to 0.15, the interquartile ranges (25th to 75th percentile) of the confidence for each $e$ do not overlap. This indicates that, when $\bar{y}_0$ values deviate only slightly from one another and are close to $\theta_0$, the $e$-coherence property generally holds in practice. 

\jj

\subsection{Condition: $n$-Coherence} 
%\paragraph{Positive correlation between sample size and confidence}%,   Figure \ref{fig:e_vs_s}(b)}

We now show that under mild conditions,  sample size and confidence are positively correlated when evidence is fixed and exceeds $\theta^*$.  Again assume one of the three probability models in Table \ref{tab:params} is used. It can be shown (Appendix A.1 - A.3) that %This property is demonstrated for binary, continuous, and count-data outcomes in one- and two-arm trials.
%In all cases, %if $e$ is fixed and $e > \theta^*$, then
\begin{eqnarray*}
    \eqref{eq:general_post_H1} &\equiv &\Pr(H=H_1 \mid \bm{y}_n)  \\ 
    &= &\frac{q}{C_1}\times\left(\frac{\frac{1-q}{C_0}}{\int_{\bm{\theta}\in H_1}f(\bm{\theta}\mid\bm{y}_n)d\bm{\theta}}+\left[\frac{q}{C_1}+\frac{1-q}{C_0}\right]\right)^{-1} = T_1\times\left(\frac{T_2}{\int_{\bm{\theta}\in H_1}f(\bm{\theta}\mid\bm{y}_n)d\bm{\theta}}+T_3\right)^{-1}, 
\end{eqnarray*}
where $T_1,T_2$, and $T_3$ are constants with respect to $n$. Therefore, the magnitude of \eqref{eq:general_post_H1}, i.e., the confidence value, depends on the sample size $n$ through 
\begin{equation}\label{eq:xi}
  \xi(n) \equiv  \int_{\bm{\theta}\in H_1}f(\bm{\theta}\mid\bm{y}_n)d\bm{\theta}. 
\end{equation}\ %As shown in Appendices A.1 to A.3. 
\noindent \textbf{Case 1}: Binary outcomes (Binomial/Beta)

\noindent
Referring to Table \ref{tab:params}, %we use Beta/Bernoulli model in this case. Specifically, we assume the likelihood function is 
we consider the sampling model and prior given by %$f(y|\theta_j) = f_{Bern}(y;\theta_j)$, and the prior distribution is $\pi(\tilde{\bm{\phi}}) = f_{Beta}(\theta_j;a_j,b_j)$, %are given by \jj
$y_{ij} \sim \text{Bern}(\theta_j)$ and 
$(\theta_0,\theta_1) \mid \tilde{\bm{\phi}}, H = H_j \sim \frac{1}{C_j}\prod_{j = 0,1}\text{Beta}(a_j,b_j)I(\theta \in H_j)$,
%$\theta_j \sim \text{Beta}(a_j,b_j)$, 
where %$f_{Bern}$ and $f_{Beta}$ denote the Bernoulli and Beta probability density functions (p.d.f.), respectively. The 
index $j = 0, 1$ is omitted for one-arm trials and retained for two-arm trials. 
%being the response rate, and $\pi_j(\tilde{\bm{\phi}}) = \text{Beta}(a,b)$. Using this model, one may then compute $\Pr(H = H_1|\bm{y}_n)$ (Equation \ref{eq:general_post_H1}), with $f(\bm{\theta}|\bm{y}_n) = \frac{\theta_1^{a+\sum \bm{y}_n}(1-\theta_1)^{b+n-\sum \bm{y}_n}}{B(a+\sum \bm{y}_n,b+n-\sum \bm{y}_n)}$ for one-arm trial and $f(\bm{\theta}|\bm{y}_n) = \frac{\theta_1^{a_1+\sum \bm{y}_n}(1-\theta_1)^{b_1+n-\sum \bm{y}_n}}{B(a_1+\sum \bm{y}_n,b_1+n-\sum \bm{y}_n)}\times \frac{\theta_0^{a_2+\sum \bm{y}_n}(1-\theta_0)^{b_2+n-\sum \bm{y}_n}}{B(a_2+\sum \bm{y}_n,b_2+n-\sum \bm{y}_n)}$ for two-arm trial. \jj
To show  $\Pr(H = H_1\mid\bm{y}_n)$ increases with $n$, we only need to show 
$\xi(n)$ % \equiv \int_{\bm{\theta}\in H_1} f(\bm{\theta}|\bm{y}_n) d\bm{\theta}$$
increases with $n$.

We begin by examining this property in a one-arm trial with binary outcomes. For a one-arm trial, we show $\xi(n) = \int_{\bm{\theta} \in H_1}  f(\bm{\theta}\mid e,n)d\bm{\theta}.$ See Appendix A.1 for a proof. 
%By \eqref{eq:evidence}, 
Let %$\theta_0$ be a known constant, set 
$e' = e + \theta_0$, and $\theta' = \theta^* + \theta_0$. Then %For a fixed $e$, consider Equation \eqref{eq:xi_def}, where %we have $\xi$ being a function of $n$:
$$\xi(n) = 1 - F_{\text{Beta}}(\theta';a+ne',b+n(1-e')),$$ 
where $F_{\text{Beta}}$ is the CDF of a beta distribution. The following proposition shows that under appropriate conditions, $\xi(n)$  monotonically increases in $n$. %, or equivalently, $1-\xi(n)$ is monotonically decreasing in $n$. 

\begin{prop}\label{prop:one_arm_binary_abv_area}
    Assume $\theta_0$ is known and $a$ and $b$ are fixed  hyperparameters. For a fixed $e$ and $\theta^* < e$, there exists a minimum sample size $n_{\text{min}}$ such that for all $n \geq n_{\text{min}}$, 
    $$\xi(n+1) - \xi(n) = F_{\text{Beta}}(\theta';a+ne',b+n(1-e')) - F_{\text{Beta}}(\theta';a+(n+1)e',b+(n+1)(1-e')) \geq 0,$$
    where $\theta' = \theta^* + \theta_0$ and $e' = e + \theta_0$.
\end{prop}
\noindent The proof of Proposition \ref{prop:one_arm_binary_abv_area} is provided in Appendix A.7. To find $n_{\text{min}}$, refer to  Appendix A.4 Nmin Algorithm 1.  %contains %BESS 
%Algorithm 3, which 
 % to %find such an $n_{\text{min}}$. \bb 
 For commonly used $\theta^*$, $\theta_0$, $e$ and hyperparameters $a$ and $b$, $n_{\min}$ is small, usually equal to 1 or 2. For instance, when $\theta^* = 0.3$, $\theta_0 = 0$, $e = 0.4$, and $a = b = 0.5$, %, and $e = 0.4$, 
 $n_{\min} = 1$.   %for one-arm trials with binary outcomes. 

%A similar result holds f
For a two-arm trial with binary outcomes, we show $\xi(n) \equiv \int_{\bm{\theta}\in H_1} f(\bm{\theta}\mid\bar{y}_0,e,n)d\bm{\theta}$. See Appendix A.3 for a proof.  %under the assumption that $\bar{y}_0$ is known. %and $\bar{y}_1 = e + \bar{y}_0$. 

\begin{prop}\label{prop:two_arm_binary_abv_area}
    Suppose $\bar{y}_0$ is prespecified %known 
    and the hyperparameters $a_0,\; b_0, \; a_1$ and $b_1$ are fixed.  
    For a fixed $e$ and a given $\theta^* < e$, there exists a minimum sample size $n_{\text{min}}$ such that for all $n \geq n_{\text{min}}$, 
    $$\xi(n+1) - \xi(n) %= \jj \int_{\bm{\theta} \in H_0} f(\bm{\theta}|\bar{y}_0,e,n)d\bm{\theta} - \int_{\bm{\theta} \in H_0} f(\bm{\theta}|\bar{y}_0,e,n+1)d\bm{\theta} 
    \geq 0.$$
    %where $H_0' = \{\theta_1 - \theta_0 \leq \eta\}$ and $\eta \geq \theta^*$.
\end{prop}
\noindent The proof of Proposition \ref{prop:two_arm_binary_abv_area} appears in Appendix A.8. Moreover, we show that the inequality holds even if % this result can be extended to the case in which 
$\bar{y}_0$ is not prespecified but is determined by the optimizer defined in Equation \eqref{eq:min_H1}. As a result, the following theorem  demonstrates that sample size and confidence remain positively correlated for fixed evidence in BESS Algorithm 2:

\begin{theo}\label{theo:two_arm_binary}
    For a fixed $e$, assume $\theta^* < e$ and let hyperparameters satisfy $a_0 = a_1 = \alpha$ and $b_0 = b_1 = \beta$. There exists a minimum sample size $n_{\text{min}}^*$ such that for $n \geq n_{\text{min}}^*$, $\text{Pr}(H = H_1\mid e,n)$, as defined in Equation \eqref{eq:min_H1}, increases monotonically with $n$.
\end{theo}
\noindent The proof of Theorem \ref{theo:two_arm_binary} is provided in Appendix A.9. %Similar to one-arm trial with binary outcome, 
Appendix A.4 provides Nmin Algorithm 2 to find $n_{\text{min}}$. Again, $n_{\min}$ is in general small for commonly used $e$ and hyperparameters $a_0,\; b_0, \; a_1$ and $b_1$. 

Propositions \ref{prop:one_arm_binary_abv_area} and \ref{prop:two_arm_binary_abv_area} and Theorem \ref{theo:two_arm_binary} establishes the $n$-coherence for the Binomial/Beta model since the confidence increases monotonically with $\xi(n)$. 
However, these results %of Proposition \ref{prop:two_arm_binary_abv_area} and Theorem \ref{theo:two_arm_binary} 
cannot be generalized to values of $\bar{y}_0$ that are not optimizers: 
\begin{remark}\label{rm:diff_ybar0_nco}
    Assume a fixed $e$, $\theta^* < e$, and let the hyperparameters satisfy $a_0 = a_1 = \alpha$ and $b_0 = b_1 = \beta$. Suppose there are two choices $\bar{y}_0^*$ and $\bar{y}_0$ (with $\bar{y}_1^\star = \bar{y}_0^\star + e$ and $\bar{y}_1 = \bar{y}_0+e$), which are used in BESS Algorithm 2' for computing sample sizes $n_1$ and $n_2$, respectively. % where neither $\bar{y}_0^\star$ nor $\bar{y}_0$ is the optimizer in Equation \eqref{eq:min_H1} for the sample sizes $n_1$ and $n_2$, respectively. Further, 
    Assume $n_2 > n_1 > n_{\min}$, where $n_{\min}$ is the minimum sample size such that, for all $n \geq n_{\min}$, both $\Pr(H = H_1\mid\bar{y}_1,\bar{y}_0,n)$ and $\Pr(H = H_1\mid\bar{y}_1^\star,\bar{y}_0^\star,n)$ are monotonically increasing in $n$. Then there exists  a tuple $(\bar{y}^\star_0, \bar{y}^\star_1, n_1, \bar{y}_0, \bar{y}_1, n_2)$ such that %cases where the following inequality holds:
    $$\Pr(H = H_1\mid\bar{y}_1^\star,\bar{y}_0^\star,n_1) > \Pr(H = H_1\mid\bar{y}_1,\bar{y}_0,n_2).$$
\end{remark}
Remark \ref{rm:diff_ybar0_nco} states that if $\bar{y}_0$ is not fixed in finding the sample size $n$ using BESS Algorithm 2', then increasing $n$ does not guarantee a larger posterior probability $\Pr(H=H_1 \mid\bar{y}_1, \bar{y}_0, n)$ where $\bar{y}_1 = \bar{y}_0 + e$ for fixed $e$.  We provide an \jj example. Let $\theta^* = 0.05$, $q = 0.5$, and fix $e = 0.1$, with $a_0 = a_1 = b_0 = b_1 = 0.5$. For $\bar{y}_0^\star = 0.1$ ($\bar{y}_1^\star = 0.2$) and $n_1 = 10$, we have $n_{\min} = 1$ and $\Pr(H = H_1\mid\bar{y}_1^\star,\bar{y}_0^\star,n_1) = 0.66$. If $n$ increases to $n_2 = 15$ but $\bar{y}_0 = 0.5$ ($\bar{y}_1 = 0.6$), we still have $n_{\min} = 1$, yet $\Pr(H = H_1\mid\bar{y}_1,\bar{y}_0,n_2) = 0.65 < \Pr(H = H_1\mid\bar{y}_1^\star,\bar{y}_0^\star,n_1) = 0.66$. %Consequently, in this case, increasing the sample size does not necessarily lead to increased confidence. This result implies that when Algorithm 2' is used, if $\bar{y}_0$ is not fixed across different sample sizes, there is no guarantee that larger sample size will yield greater confidence under fixed evidence.

\bigskip 

\noindent \textbf{Case 2:} Continuous Outcomes (Normal/Normal). 

Next, we establish the monotonic relationship for continuous outcomes in both one- and two-arm trials, using the Normal/Normal model. As shown in Table \ref{tab:params}, we assume %the likelihood is $f(y|\theta_j) = f_{N}(y;\theta_j,\sigma^2)$, where $\sigma^2$ is assumed known, and the prior distribution is %
$y_{ij} \sim N(\theta_j,\sigma^2).$ %and 
%$\theta_j \sim N(a_j, b_j).$ 
Here, $\sigma$ is assume to be known. For one-arm trials, index $j$ is omitted, i.e., $y_{i} \sim N(\theta, \sigma^2).$  For two-arm trials, we assume  
%$f_N$ denotes the normal p.d.f., and the 
%index $j=0,1$ is omitted for one-arm trials and retained for two-arm trials. \bb 
%Moreover, for two-arm trials, we specify both the likelihood and prior as normal distributions for the difference between the responses in the treatment and control groups. %with known variance, 
%assume 
a 1:1 randomization. We arbitrarily pair the outcomes from each arm and compute $\bm{y}_n^d = \{y_{i1} - y_{i0};i = 1, \ldots, n\}$. Then $y_{i}^d = y_{i1}-y_{i0}$ follows $N(\theta, 2\sigma^2),$  where $\theta=\theta_1 - \theta_0.$ In both cases, we assume $\theta\mid \bm{\tilde{\phi}},H = H_j \sim \frac{1}{C_j} N(a, b)I(\theta\in H_j).$ %This difference-in-response outcome again follows a Normal distribution, with mean parameter being $\theta = \theta_1 - \theta_0$. Assume a normal prior for $\theta$ with hyperparameters $a$ and $b$, the two-arm trial case reduces to the one-arm trial case. 
%See Appendix A.2 for further details. This leads to the follows. \jj 
%Next, we establish the monotonic relationship for continuous outcomes in both one- and two-arm trials. 
% For one-arm or two-arm trials with continuous outcomes, 
Appendix A.1 and A.2 provide details and show that $$\xi(n) = \int_{\bm{\theta} \in H_1} f(\bm{\theta}\mid\bm{y}_n)d\bm{\theta} = \int_{\bm{\theta} \in H_1} f(\bm{\theta}\mid e',n)d\bm{\theta}, \; $$ 
where $e'=e+\theta_0$ (one-arm) or $e' = e$ (two-arm). 
%\note{Ed: I think we need to add $e' = e$ for two arm trial. It is missing here.}
%See .
\jj  %in the next theorem:
\begin{theo} \label{theo:cont_one_two_arm}
    Assume the variance $\sigma^2$ is known. Let $\theta^*$ and hyperparameters $a$ and $b$ be given. For any fixed $e$ and $\theta^*$, %In a one-arm trial, define $e' = e + \theta_0$ and $\theta' = \theta^* + \theta_0$; in a two-arm trial, define $e' = e$ and $\theta' = \theta^*$. Suppose $e'$ and $\theta'$ are fixed. Then, 
    there exists a minimum sample size %n_{\text{min}}$ such that 
    $$n_{\min} = \max\left(\left\lfloor\frac{[a-(e-\theta^*)]\sigma^2}{(e-\theta^*)b}\right\rfloor,1\right).$$
    %$$e > \theta^* + \frac{a}{b}\left(\frac{1}{b}+\frac{n_{\text{min}}}{\sigma^2}\right)^{-1}.$$
    %Then, for 
    For all $n \geq n_{\text{min}}$, %Equation \eqref{eq:xi_def}, 
    $\xi(n)$ increases monotonically with $n$.
    Moreover, if the prior is centered at zero (i.e., $a = 0$), then $\xi(n)$ increases monotonically with all $n \geq 1$ if $e > \theta^*$.
\end{theo}
\noindent Theorem \ref{theo:cont_one_two_arm} establishes the $n$-coherence for the Normal/Normal model since the confidence monotonically increases with $\xi(n)$. Therefore when $n$ increases, according to Theorem \ref{theo:cont_one_two_arm} $\xi(n)$ increases, and therefore the confidence increases.  

The proof of Theorem \ref{theo:cont_one_two_arm} is provided in Appendix A.10. Note that $a = 0$ is sufficient, but not necessary, for $n_{\min} = 1$. In fact, when $a \leq e-\theta^*$, $n_{\min} = 1$. Furthermore, if $b$ is large such that $\frac{[a-(e-\theta^*)]\sigma^2}{(e-\theta^*)b} < 1$, $n_{\min} = 1$ as well. Consequently, for commonly used hyperparameters $a$ and $b$, the value of $n_{\min}$ is again in general small. 

\bigskip

\noindent \textbf{Case 3:} Count data (Poisson/Gamma). 

 Lastly, we establish the conditions for the monotonicity relationship between $n$ and confidence for count outcomes using the Poisson/Gamma model. Specifically, we assume %the likelihood function is 
$y_{ij} \sim \text{Poisson}(\theta_j)$ and $(\theta_0,\theta_1)\mid \tilde{\bm{\phi}}, H = H_j \sim \frac{1}{C_j}\prod_{j = 0,1}\text{Gamma}(a_j, b_j)I(\theta \in H_j)$, where index $j=0,1$ is omitted for one-arm trials and retained for two-arm trials. Here, the mean of $\text{Gamma}(a, b)$ is $a/b.$  

%Lastly, we establish \yy the \jj monotonicity relationship for count outcomes. 
We first consider conditions for a one-arm trial,  which is summarized in the following proposition. 

\begin{prop}\label{prop:count_one_arm}
    Assume $\theta_0$ is known and $a$ and $b$ are fixed  hyperparameters. Define $\theta' = \theta_0 + \theta^*$. For a fixed $e' = e + \theta_0$ and
    %if $a$ and $b$ satisfy
    %\begin{equation}\label{eq:count-two-arm}
    %    \frac{\Gamma(a+2e')\Gamma(a)}{\Gamma(a+e')^2}\left[\frac{(b+1)^2}{b(b+2)}\right]^a\left[\frac{b+1}{b+2}\right]^{2e'} \geq 1,
    %\end{equation}
    %then for 
    any $\theta^* < e$, there exists a minimum sample size $n_{\text{min}}$ such that for all $n \geq n_{\text{min}}$,
    $$\xi(n+1) - \xi(n) = F_{\text{Gamma}}(\theta';a+ne',b+n) - F_{\text{Gamma}}(\theta';a+(n+1)e',b+n+1) \geq 0.$$
\end{prop}

\noindent The proof of Proposition \ref{prop:count_one_arm} appears in Appendix A.11, and $n_{\min}$ can be found using Nmin Algorithm 1 (see Appendix A.4). %For count-data outcome in one-arm trial case, we found that 
%In general,  condition \eqref{eq:count-two-arm} %of $$\frac{\Gamma(a+2e')\Gamma(a)}{\Gamma(a+e')^2}\left[\frac{(b+1)^2}{b(b+2)}\right]^a\left[\frac{b+1}{b+2}\right]^{2e'} \geq 1$$ 
%can be easily satisfied in commonly used values of \bb $\theta^*$, $\theta_0$, $e$, \jj $a$ and $b$. 
%Additionally, f
For the commonly used values of $\theta^*$, $\theta_0$ and $e$, $n_{\min}$ is typically small (equal to or close to 1). 

For a two-arm trial, 
Appendix A.3 provide details and show that
$$\xi(n) = \int_{\bm{\theta}\in H_1}f(\bm{\theta}\mid\bm{y}_n)d\bm{\theta} = \int_{\bm{\theta}\in H_1}f(\bm{\theta}\mid\bar{y}_1,\bar{y}_0,n)d\bm{\theta}.$$
We %\note{Ed: I noticed that we do not have $\xi(n)$ defined in this case.}
%we 
require %with count-data outcome, 
the specification of $\bar{y}_0$, and set $\bar{y}_1 = e + \bar{y}_0$. %are required parameters for BESS to work. In this case, we assume that $\bar{y}_0$ is known and fixed, and $\bar{y}_1 = \bar{y}_0 + e$. 
Under a fixed $e$, we have the following theorem:%, with its proof is given in Appendix X.

\begin{theo}\label{theo:two_arm_count}
    Let $\bar{y}_0$ and $e$ be fixed. Suppose %$\bar{y}_0$ is known, 
    $\theta^* < e$, and the hyperparameters $a_0, b_0, a_1$ and $b_1$ are given. %Define $e_0 = \bar{y}_0+e$ and $e_1 = \bar{y}_0$. %If $a_j$ and $b_j$ satisfy 
    %$$\frac{\Gamma(a_j+2e_j)\Gamma(a_j)}{\Gamma(a_j+e_j)^2}\left[\frac{(b_j+1)^2}{b_j(b_j+2)}\right]^{a_j}\left[\frac{b_j+1}{b_j+2}\right]^{2e_j} \geq 1$$
    %for $j = 0, 1$, then
    Then, there exists a minimum sample size $n_{\text{min}}$ such that for all $n \geq n_{\text{min}}$, $\xi(n)$ increases monotonically with $n$. 
%    $$\int_{\bm{\theta} \in H_0} f(\bm{\theta}|e,n)d\bm{\theta} - \int_{\bm{\theta} \in H_0} f(\bm{\theta}|e,n+1)d\bm{\theta} \geq 0.$$
    %where $H_0' = \{\theta_1 - \theta_0 \leq {\theta^*}'\}$ and ${\theta^*}' \geq \theta^*$.
\end{theo}

\noindent The proof of Theorem \ref{theo:two_arm_count} is provided in Appendix A.12.  Moreover, %Similar to the one-arm trial with count-data outcome case, 
%the condition %is also 
%can be easily satisfied for commonly used $e$, \bb $\bar{y}_0$, \jj $a_0$ $b_0$, $a_1$, and $b_1$, and 
$n_{\min}$ is typically small for the commonly used values of $e$, $\bar{y}_0$, and the hyperparameters. Proposition \ref{prop:count_one_arm} and Theorem \ref{theo:two_arm_count} establish the condition for $n$-coherence for the one- and two-arm trials with count data and Poisson/Gamma model in Table \ref{tab:params}. 

Note that, similar to \textbf{Case 1} (Binary data with Binomial/Beta model), the monotonicity result in Theorem \ref{theo:two_arm_count} may not hold if $\bar{y}_0$ is not fixed, even if the %when the conditions on the hyperparameters are satisfied and all 
sample sizes are all greater than $n_{\min}$. 
This can again be demonstrated by a counter-example. Let $\theta^* = 0.1$, $q = 0.5$, $a_0 = a_1 = 1$, and $b_0 = b_1 = 2$. Assume $e = 0.5$. When $n_1 = 10$ and $\bar{y}_0 = 1$ ($\bar{y}_1 = 1.5$), we have $n_{\min} = 1$ and $\Pr(H = H_1\mid\bar{y}_1,\bar{y}_0,n_1) = 0.83$. If the sample size increases to $n_2 = 20$ but $\bar{y}_0^\star = 5$ ($\bar{y}_1^\star = 5.5$), we have $\Pr(H = H_1\mid \bar{y}_1^\star, \bar{y}_0^\star, n_2) = 0.78$, representing a clear decrease in confidence despite the increase in $n$ for fixed evidence $e$. Consequently, we again conclude that if $\bar{y}_0$ is not held fixed when using Algorithm 2', confidence may not increase as $e = \bar{y}_1- \bar{y}_0$ increases. %guarantee that a larger sample size leads to greater confidence does not generally hold.

%\bb For the BESS method to serve as a reliable sample size estimator, the relationship between confidence and sample size for a given evidence must be monotonic. We demonstrate this relationship for \textbf{Algorithm 1} (covering all three outcomes in one-arm trials and continuous outcomes in two-arm trials), \textbf{Algorithm 2} (binary outcomes in two-arms trial), and \textbf{Algorithm 2'} (count-data outcomes in two-arm trials, Appendix A.5) in Appendix A.4. \jj

%\bb As an example, \jj  
%we compute confidence using Equation \eqref{eq:general_post_H1}, varying the sample size $n$ while keeping the evidence $e$ fixed.   Figure \ref{fig:e_vs_s}(b) illustrates the positive relationship between sample size and confidence, when 
%$e = 0.1$ and $\theta^* = 0.05$.
%This behavior is expected when the evidence supports the alternative hypothesis, as a larger sample size increases the posterior probability of $H_1$. 

%\paragraph{Negative correlation between evidence and sample size}

\subsection{Negative Correlation Between Evidence and Sample Size}

Lastly, given $n$-coherence and $e$-coherence, %are true, 
it is immediate that sample size $n$ and evidence $e$ are negatively correlated for a fixed confidence. This is because with an increase in sample size, one must decrease the evidence to maintain the same confidence level.

\section{Comparison with Standard SSE} \label{sec:oc}

\subsection{  Simulation Setup    }\label{sec:sim_setup}

Through simulation, we compare BESS and the standard SSE 
for a two-arm trial with binary 
outcome. For fair comparison, we match the type I/II error rates for both methods through a numerical approach in %The matching is realized by 
three steps. %Step 1: \yy Given a null $H_0$ and alternative $H_1$ for fixed hypothesis in \eqref{eq:ht} and $e$ and $c$, \jj we obtain sample size estimated via the proposed BESS approach for a trial. Step 2: we repeatedly simulate trial data under the null and alternative hypotheses, perform the Bayesian inference \yy by rejecting $H_0$ if $\Pr(H = H_1|data) > c$ \jj using the same model as in BESS, and record the type I and type II error rates from the simulated trials. Step 3: using the two error rates \yy and for the same $e$, $H_0$, and $H_1$, \jj we apply the standard SSE  
%to arrive at a Frequentist sample size. %, and compare it with the BESS estimate. 
See Appendix A.13 
for an illustration. 

\paragraph{Step 1: Calculate BESS sample size}  %We 
%consider a two-arm trial with binary outcome. %and 
Let $\theta_1$ and $\theta_0$ be the response rates and $\theta^*$  the clinically minimum effect size. The trial aims to test   
$H_0: \theta_1 - \theta_0 \leq \theta^* \text{ vs. } H_1: \theta_1 - \theta_0 > \theta^*.$
For BESS,  
we  assume a binomial sampling model,  an improper prior $(\theta_1, \theta_0) \mid H_j \sim \text{Beta}(0, 0) \cdot \text{Beta}(0, 0) I((\theta_1, \theta_0) \in H_j)$, and $\text{Pr}(H = H_1) \equiv q = 0.5.$
We apply BESS Algorithm 2 to obtain a sample size for a pair of desirable evidence $e$ and confidence cutoff $c$. We try several different pairs of $e$ and $c$ as well, listed in  Table \ref{tab:vag_p_evc}. For these settings, we find  $n_{\min}$s to be equal to $1$ based on the Nmin Algorithm 2 in Appendix A.4. %\bb Note that the $n_{\text{min}}$s are all equal to 1. \jj

\begin{table}[h!]
    \caption{Simulation results compare BESS with Standard SSE with matched type I  error rate $\alpha$ and power $(1-\beta)$  in a two-arm trial with binary outcome. The results show the estimated sample sizes, false discovery  rate (FDR), and false omission  rate (FOR) of the two methods across various levels of evidence $e$ and confidence $c$. }
    \label{tab:vag_p_evc}
    \centering
    \begin{tabular}{cc|cc|ccc|ccc}
    \hline
        Evidence & Confidence & Type I & Power & \multicolumn{3}{c|}{BESS} & \multicolumn{3}{c}{Standard SSE} \\ 
        $e$ & $c$ & error rate $\alpha$ & $1 - \beta$ & $n$ &  FDR  &  FOR \jj  & $n$ &  FDR   &    FOR     \\
    \hline
        \multirow{3}{*}{0.10} & 0.7 & 0.36 & 0.73 & 40 & 0.33 & 0.29 & 42 & 0.33 & 0.29 \\
         & 0.8 & 0.23 & 0.82 & 120 & 0.22 & 0.19 & 117 & 0.22 & 0.19 \\
         & 0.9 & 0.12 & 0.92 & 290 & 0.12 & 0.08 & 284 & 0.12 & 0.08 \\
    \hline
        \multirow{3}{*}{0.15} & 0.7 & 0.35 & 0.56 & 14 & 0.38 & 0.39 & 14 & 0.38 & 0.39 \\
         & 0.8 & 0.23 & 0.56 & 34 & 0.29 & 0.36 & 36 & 0.29 & 0.36 \\
         & 0.9 & 0.12 & 0.56 & 74 & 0.18 & 0.34 & 74 & 0.18 & 0.34 \\
    \hline
        \multirow{3}{*}{0.20} & 0.7 & 0.45 & 0.59 & 5 & 0.43 & 0.43 & 6 & 0.43 & 0.43 \\
         & 0.8 & 0.28 & 0.51 & 15 & 0.35 & 0.41 & 16 & 0.35 & 0.41 \\
         & 0.9 & 0.19 & 0.48 & 35 & 0.28 & 0.39 & 39 & 0.28 & 0.39 \\
    \hline
    \end{tabular}
\end{table}

\paragraph{Step 2: Estimate type I/II error rates} Next,   for each BESS sample size in Step 1,    we repeatedly simulate multiple trials under the null $H_0$ and the alternative $H_1$ and  numerically    compute     the type I/II error rates.   Specifically, 
under the null  we let $\theta_1 = 0.3$ and $\theta_0 = 0.25$, and under the alternative, $\theta_1 = 0.4$ and $\theta_0 = 0.25$. %We set $\theta^*=0.05.$
%Based on each pair of $\theta_1$ and $\theta_0$ values, 
We 
then generate %the 
binary responses %of 
for $n$ %150 
patients in both the treatment and control arms %each %arm 
%for the treatment %arm 
%and control arms \bb 
under the null and alternative hypotheses, using the corresponding pairs of $\theta_1$ and $\theta_0$.    
We generate $y_{ij} \sim \text{Bern}(\theta_j)$, and denote the simulated outcomes $\bm{y}_{nj} = \{y_{ij}; i = 1, \ldots, n\}, \, j = 0, 1,$ and $\bm{y}_n = \{\bm{y}_{n1},   \bm{y}_{n0}\}$.
 %where $\text{Bern}(\theta)$ is a Bernoulli distribution with mean $\theta.$ 
The null is rejected   if    
$\text{Pr}(H = H_1\mid\bm{y}_n) \geq c.$ We repeat this procedure for each pair of $c$ and $n$ shown in Table \ref{tab:vag_p_evc}. For example, for a simulated trial with $n$ and $c$ corresponding to the first row of Table \ref{tab:vag_p_evc}, we generate $n = 40$ binary outcomes for each of the treatment and control arms based on true parameter values $\theta_1 = 0.3$ and $\theta_0 = 0.25$. We then computes $\text{Pr}(H = H_1\mid\bm{y}_{40})$ and reject the null hypothesis if this probability exceeds $c = 0.7$. This corresponds to a simulated trial under the null hypothesis.
%\note{Ed: This description was wrong before. The $n$ and $c$ are as shown in Table 2, not $n = 150$ and $c = 0.8$.}
%Here, $0.8$ is arbitrarily selected. %without specific intention. 
%A larger or smaller cutoff than $0.8$ arbitrarily affects the computed type I/II error rates, %which 
%\yy although it \jj does not affect  the objective of our simulation. 
We simulate 10,000 trials  each under the null and alternative hypotheses. 
A rejection of the null for a trial simulated under the null is recorded as an incidence of type I error, and a non-rejection of the null for a trial under the alternative is recorded as an incidence of type II error. The type I/II error rates $(\alpha / \beta)$ are then computed as the frequencies of the corresponding incidences over the 10,000 trials. Also, we compute the false discovery rate (FDR) and false omission rate (FOR) defined in Appendix A.14, both are quantification of the fraction of wrong decisions among those that are made for the simulated trials. 

\paragraph{Step 3: Estimate Frequentist sample size and compare with BESS}
  Based on the computed type I/II error rates $(\alpha / \beta)$ from Step 2,  denoted as $\alpha$ and $\beta$,  we  estimate a sample size using a standard SSE   approach. For example, we consider a superiority $z$-test for comparing $\theta_1$ and $\theta_0$, and a sample size can be estimated via equation \eqref{eq:sse}. In the estimation, we assume the true values for $\theta_1$ and $\theta_0$ under the alternative, which gives the standard approach an ``oracle" performance. In other words, the estimated sample size is guaranteed to achieve the target type I/II error rates $\alpha$ and $\beta$, since the assumed $\theta_1$ and $\theta_0$ in the sample size estimation match the true values. Even though this is typically not achievable in reality,   we decide to compare the oracle frequentist sample size with the BESS sample size. %nevertheless.   

\subsection{Simulation Result}

Table \ref{tab:vag_p_evc} presents the simulation results.  BESS sample sizes and standard SSE sample sizes are largely similar across different scenarios, which is %They are  
both surprising and reassuring. %that BESS and the standard SSE produce similar sample sizes across a variety of settings. 
It is surprising since the two approaches are based on different statistical metrics. As BESS aims %, it's aiming 
to balance between the anticipated evidence in the observed data and the confidence expressed as posterior probability,  the standard SSE approach trades %, it's trading 
off among the type I/II error rates given the presumed  true parameter values. On the other hand, the results are reassuring since despite using different metrics, when matching the type I/II error rates, both approaches produce highly similar sample size estimates. For other outcome and trial types, similar results can be found  in Appendix Table A.1 in Section A.13.

Several trends are worth noting in Table \ref{tab:vag_p_evc}. First, increasing confidence cutoff $c$ leads to lower type I error rate for fixed evidence $e.$ 
This is because a larger %increase in cutoff 
$c$  makes it harder to reject the null for BESS under equation \eqref{eq:BESS_cond}, and hence leading to fewer  rejected  simulated trials under $H_0$, %will be rejected, 
i.e., a lower %   decrease in the 
type I error rate.
Second,  note that the true effect size under $H_1$ equals $\theta \equiv \theta_1 - \theta_0 = 0.4 - 0.25 = 0.15.$ We observe that when $c$ increases 1) power increases when the evidence $e$ is less than the true effect size, i.e., $e=0.1,$ 2) power stays the same if evidence $e$ equals the true effect size, and 3) power decreases if $e$ is greater than the true effect size.  
%This complicated trend demonstrates an interaction  between \yy confidence level \jj $c$ and power conditional on whether $e$ is greater than, equal to, or less than the true effect size $\theta$. 
To understand the interaction between the confidence level $c$ and evidence $e$,  first recall that in Section \ref{sec:relation}, we used the notation $e^\star$ to denote the observed evidence after the trial is completed and the data are observed.  As the confidence level $c$ increases, the BESS-estimated sample size  also increases (see Table \ref{tab:vag_p_evc}). %Section \ref{sec:relation}). \jj 
%\note{Ed: This is not see Section 4.1, right? Should be See first three rows in table 2.}
Therefore, by the law of large numbers, the observed $e^\star$ will be closer to $\theta$. % due to large number theory. 
%Consequently, in simulations under the alternative hypothesis, more trials will have $e^\star$ close to $\theta$. 
If the assumed evidence $e$ is less than $\theta$, it is more likely that $e < e^\star$. This implies that the posterior probability $\text{Pr}(H = H_1 \mid \by_n)$ will be higher (due to the $e$-coherence property, see Section \ref{sec:relation}), %\note{Ed: I think here, we can say (see $e$-coherence in Section 4.1)} %(since there is stronger evidence supporting $H_1$), 
resulting in more rejections and thus higher power. Therefore, when $e < \theta$, a larger $c$ leads to higher power.  
The same logic applies when $e > \theta,$ in
which case a larger $c$ leads to lower power.  Finally, when $e=\theta$,  increasing the sample size (as a result of increasing $c$) causes $e^\star$ to approach $e$, and thus there is no substantial impact on power. \jj

Results in Table \ref{tab:vag_p_evc} imply that if one wishes to have  high power and a low type I error rate using BESS, one may want to specify an evidence that is smaller than the true effect size and a high confidence cutoff $c$. Since the true effect size is typically unknown,
one may either  construct an informative prior of $\theta_1$ and $\theta_0$ for BESS if prior information is available, or conduct interim analysis and sample size re-estimation  to better plan and conduct a trial. We will explore the latter option in the next section. 

Finally,  from a Bayesian perspective, BESS is concerned about the trial at hand, rather than hypothetical trials generated from the null or alternative (i.e., type I/II error rates). Therefore, while Table \ref{tab:vag_p_evc} illustrates a connection between BESS and standard SSE,
it does not imply that BESS needs to be calibrated based on type I/II error rates in practice. On the contrary, BESS focuses on the probability of making a right decision given the observed data, which can be measured by the FDR  and FOR.  In Table \ref{tab:vag_p_evc} the reported FDR  and FOR  for BESS and standard SSE are the same since 1) the prevalence of trials under the $H_0$ and $H_1$ is 50\% and 2) the standard SSE is oracle since it assumes the true $\theta_0$ and $\theta_1$ values in its computation. In Appendix Table A.2 in Section A.13,
we show that when the $\theta_0$ and $\theta_1$ are mis-specified in the standard SSE, the estimated sample sizes may be too large or too small, leading to over- or under-power, and deflated or inflated FDR/FOR 's.     
\paragraph{Sensitivity of Prior}
We demonstrate the sensitivity of incorporating  prior information through simulation, assuming there exists prior data of  $n_0$ patients per arm.  
The simulation details are presented in Appendix A.13. 

%For example, with %$n_0 = 10$,
To illustrate this approach, we consider a scenario with $e = 0.15$, and $c = 0.8$. Recall that a $\text{Beta}(0,0)$ prior was used for BESS in Table \ref{tab:vag_p_evc}, which yielded an average sample size of
%the average sample size from the %1,000 
%simulated trials was 
34 from the simulated trials. When we incorporate prior data from $n_0 = 10$ patients using an informative prior, the average sample size is %in Table \ref{tab:vag_p_evc}, and is 
reduced to 23.63 with a standard deviation of 9.22.

%when a prior data set with $n_0 = 10$ patients is used. Recall a Beta$(0,0)$ prior was used for BESS %sample size was 34 
%in Table \ref{tab:vag_p_evc}. 
The informative prior based on the data from the $n_0$ patients is given by Beta$(\sum_i y_{ij}^0,n_0-\sum_i y_{ij}^0)$, where $y_{ij}^0$ ($i = 1, \ldots, n_0; \, j = 0,1$) denotes the response for the $i$th patient in the $j$th arm. This reduction in required sample size demonstrates that BESS effectively incorporates prior information for sample size estimation. %\note{Ed: We mentioned recall the vague prior is used, but forget to state what is the informative prior} %for $e = 0.15$ and $c = 0.8$. 
%The results show that  
%is able to properly borrow prior information for sample size estimation. 

In practice, one may use different informative priors, e.g, power prior \citep{ibrahim2000power} or commensurate prior \citep{hobbs2011hierarchical}, for information borrowing. We leave these extensions for future research.   

\section{Demonstration of BESS with A Dose Optimization Trial} \label{sec:na}

\subsection{Fixed Sample Size}
Lastly, we consider a randomized comparison of two selected doses in an oncology phase I trial as part of FDA's Project Optimus initiative for dose optimization. % in oncology drug development. Suppose two doses are compared via a $1:1$ randomized design. 
%We apply BESS to estimate the sample size of the comparison. 
In dose optimization, the goal is to test if the lower dose is no worse than the higher dose in terms of efficacy, i.e., non-inferiority.    
Denoting $\theta_H$ and $\theta_L$ the response rates for the higher and lower doses, respectively, we want to test the following non-inferiority hypotheses 
\begin{equation}
    H_0: \theta_H - \theta_L \geq \theta^* \text{ vs. } H_1: \theta_H - \theta_L < \theta^*, \label{eq:doh}
\end{equation}
where $\theta^* \in (0, 1)$  is the non-inferiority margin.   
Let $\theta^* = 0.05$, i.e., the non-inferiority limit is 5\%. Assume the percent `success' in the higher and lower dose arms are both 30\%, with 10\% significant level and 70\% power, the Frequentist SSE estimates a sample size of $n = 548$ patients per arm \citep{blackwelder1982proving}, which is %a feasible sample size for a Phase III trial, but 
too large for a dose-optimization trial. 

Next, we apply BESS to estimate the sample size of the comparison.
To fit the setting in \eqref{eq:ht}, we rewrite the hypotheses  as 
$H_0: \theta_L - \theta_H \leq -\theta^* \text{ vs. } H_1: \theta_L - \theta_H > -\theta^*.$
We then use BESS for
%apply BESS algorithm 2 (Section \ref{subsec:bess_algs}) to estimate the sample size for the dose-optimization trial.  %Assuming $\theta^* = 0.05$, 
%We \bb find $n_{\text{min}} = 1$ \yy based on \bb $e = 0$ and $a_0 = a_1 = b_0 = b_1 = 0.5$ \jj %$n_{\min}$ 
%(See Nmin Algorithm 2 in Appendix A.12) \bb and \jj consider 
two related objectives: 
1) find sample size given evidence and confidence using BESS algorithm 2 (Section \ref{subsec:bess_algs}), or 2) find evidences and the corresponding confidences (i.e., Equation \eqref{eq:min_H1}) for a fixed sample size. 
%\note{Ed: the previous writing is problematic. We say apply BESS Algorithm 2 to estimate the sample size, and Nmin = 1. Then say the two objectives, which only the 1st one actually uses BESS Algorithm 2 for sample size. I changed the structure to better reflect what we did.}

\paragraph{Objective 1}

Assuming $e = 0$ and $a_0 = a_1 = b_0 = b_1 = 0.5$, we find $n_{\text{min}} = 1$ (See Nmin Algorithm 2 in Appendix A.4), and BESS provides the following sample size statement: 
{ %Assuming evidence $e = 0$, 
a sample size of 95 patients per arm is needed to declare with 70\% confidence that   the response rate of the higher dose is no higher than the lower dose by 0.05. } Larger or smaller sample size can be achieved by calibrating $e$ and $c$.

\paragraph{Objective 2}

Assume $n = 20$ patients per dose are randomized and    denote $\bar{y}_L$ and $\bar{y}_H$ the observed response rates for the lower and higher doses, respectively. Let $e^\star = \bar{y}_L - \bar{y}_H$ be the observed evidence. 
We compute the confidence, $\text{Pr}(H = H_1\mid e^\star,n)$ (see Equation \eqref{eq:min_H1}) for various values of $e^\star$ and $\theta^* = 0.05$, %$\text{Pr}(\theta_H - \theta_L < \theta^*|\bar{y}_L, \bar{y}_H, n)$, for various values of $(\bar{y}_L - \bar{y}_H)$ and $\theta^* = 0.05,$  
shown in Table \ref{tab:conf_tab}.   For example, if one observes the response rate of the lower dose is the same as the higher dose, i.e., $e^\star = 0$, %(\bar{y}_L - \bar{y}_H)=0$, 
with 20 patients per dose one gets about 57.46\% confidence to declare the lower dose response rate is within the non-inferiority margin  $0.05$ of the higher dose.   

\begin{table}[h!]
    \caption{List of various evidence and confidence for $\theta^* = 0.05$ with $n = 20$ patients per arm.}
    \label{tab:conf_tab}
    \centering
    \resizebox{\linewidth}{!}{
    \begin{tabular}{c|cccccccccc}
        \hline
         & \multicolumn{10}{c}{Noninferiority margin $\theta^* = 0.05$, Sample size $n = 20$} \\
         \hline
        Observed evidence $e^\star$ %$\bar{y}_L - \bar{y}_H$ 
        & $\leq -0.20$ & -0.15 & -0.10 & -0.05 & 0.00 & 0.05 & 0.10 & 0.15 & 0.20 & $\geq 0.25$ \\
        \hline
        %Confidence $c$ 
        $\Pr(H = H_1\mid e^\star,n)$
        %$\Pr(H_1|\bar{y}_L,\bar{y}_H,n)$ \jj
        & $< 0.05$ & 0.10 & 0.24 & 0.43 & 0.57 & 0.70 & 0.79 & 0.88 & 0.93 & $> 0.95$ \\
    \hline
    \end{tabular}}
\end{table}

\subsection{Sample Size Re-estimation (SSR)
%Adaptive Designs
} \label{subsec:bess_ssr}

We further consider adaptive  designs that allow interim analysis and early stopping for the randomized dose comparison in the previous section. We set the non-inferiority margin to $\theta^* = 0.07$, %(corresponding to $n_{\text{min}} = 1$) 
%so that the standard SSE \bb yields \jj a sample size of 100 patients per dose, \bb assuming type I and type II error rates of $\alpha = \beta = 0.3$, and response rates of $\theta_H = \theta_L = 0.335$ for the higher and lower doses under $H_1$. \jj  
%\note{Ed: We did not say where the 100 sample size is calculated based on. An alternative is to remove the sentence after $\theta^* = 0.07$. }
and consider four different designs based on either the standard SSE or BESS, assuming that an initial total sample size of $n$ patients per arm is planned for the entire trial.

\paragraph{Design 1. BESS SSR} The first design is %BESS with 
sample size re-estimation (SSR) using BESS. In this design, %BESS SSR estimates a sample size $n$ for the entire trial first using input of evidence $e$ and confidence $c$. 
%\bb Assume that a total sample size of $n$ patients per arm is planned for the entire trial.
%Then 
when $n/2$ patients are enrolled at each dose, an interim analysis is performed to allow trial stopping or SSR if the trial is not stopped. 
Denoting the interim data as $\bm{y}_{n/2}$, the trial is stopped early if 
$\text{Pr}(H = H_1\mid\bm{y}_{n/2}) \geq c$ or $\text{Pr}(H = H_1\mid\bm{y}_{n/2}) \leq c^*$, where $c$ close to 1 and $c^*$ close to 0 are probability thresholds for early stopping due to success or failure, respectively. If neither condition  is met, the trial proceeds with an SSR as follows. 

In the SSR, we %again 
use BESS to re-estimate the sample size based on updated evidence from the interim data, defined as $e_{\text{int}} = \text{E}[\theta_L\mid\bm{y}_{n/2}] - \text{E}[\theta_H\mid\bm{y}_{n/2}]$, which corresponds to the difference of posterior means. We use the same cutoff $c$ for the SSR. However, we use the posterior distribution  $\pi(\theta_L, \theta_H \mid \bm{y}_{n/2})$ as the prior in the SSR applying the BESS algorithm 2. Denote the additional sample size as $n^*$ based on SSR. At the end of the trial after $n^*$ more patients are randomized to each dose, BESS SSR rejects the null and accepts the alternative if $\text{Pr}(H = H_1\mid\bm{y}_{n^* + n/2}) \geq c$. Otherwise, it accepts the null and rejects the alternative.

\paragraph{Design 2. BESS SSR Cap} The design is the same as BESS SSR, except if $n^* > n/2$, we cap $n^*=n/2$. That is, we restrict the maximum sample size at $n$ for the entire trial.

\paragraph{Design 3. Standard SSE} %The standard SSE estimates a sample size $n$ based on a $z$-test for \eqref{eq:doh}, with a significance level $\alpha$, and a desired power $(1-\beta).$
This design uses the $z$-test for \eqref{eq:doh} with a significance level $\alpha$. Specifically,
the null hypothesis is rejected if $1 - \Phi(z) \leq \alpha$ using the trial data of $n$ patients.
%Formula \eqref{eq:sse} is used to compute $n.$  
 
 \paragraph{Design 4. Standard SSE with interim} This design %follows the previous standard SSE to estimate $n$, but at $n/2$ it 
 stops the trial  if at $n/2$,
$\text{Pr}(H = H_1\mid\bm{y}_{n/2}) \geq c$ or $\text{Pr}(H = H_1\mid\bm{y}_{n/2}) \leq c^*$.  
This is the same Bayesian interim analysis in design 1, BESS SSR. If the trial is not stopped at $n/2$,  the trial  adds  additional $n/2$ patients and use the $z$-test to make a final decision. We note that the overall type I error rate is not adjusted based on this design. One could but it's beyond the scope of this paper.

\noindent \paragraph{Scenarios}

The four designs are compared through simulated trials. For each trial,  we generate the alternative and null hypotheses indicators $H = 1$ or $0$ with probabilities $q$ or $(1-q)$, respectively. Then given $H$, we generate  true model parameters $\theta_L$ and $\theta_H$ based on two scenarios. Note here the letter ``H" in $\theta_H$ means ``high dose", not ``hypothesis". %\note{Ed: I think we can remove this sentence, it is clear from the context.}

In scenario 1,  $\theta_L$ and $\theta_H$ are random variables under the null or alternative, where $\theta_H \sim \text{Unif}(\theta^*=0.07, 0.6)$, \jj and 
    $$\theta_L\mid\theta_H,H \sim \left \{ \begin{matrix} \text{Unif}(0, \theta_H - \theta^*) & H = 0 \; (H_0) \\ \text{Unif}(\theta_H - \theta^*, 0.28)I(\theta_H - \theta^* \leq 0.28) + \delta_{\theta_H}I(\theta_H - \theta^* > 0.28) & H = 1  \; (H_1) \end{matrix} \right .$$ 
This setting ensures $\theta_L < \theta_H - \theta^*$ under $H_0$, and $E(\theta_L)=E(\theta_H)=0.335$ under $H_1$. 

In scenario 2, we assume $\theta_H = 0.335$, and $\theta_L = 0.265$ under $H_0$ or $\theta_L = 0.335$ under $H_1$, i.e., they are fixed. The two scenarios reflect two different practical considerations. Scenario 1 aims to assess the performance of the four designs assuming they are applied to different drug programs and trials in which the true responses of the drugs and doses are different. Scenario 2 is a classical frequentist setting to assess the type I/II error rates of a design assuming true response rates are fixed.  Finally, 
given $\theta_L$ and $\theta_H$, for each trial we simulate patients outcome data based on Binomial distributions and the corresponding designs. 

We assume $q = 0.5$. For Designs 1 \& 2, we let $e=-0.02,$ $c=0.7$ and $c^* = 0.3$. For designs 3 \& 4, we set $\alpha = 0.3$ and $\beta=0.3$, so that BESS and standard SSE produce the same sample size, %for the trial in the beginning, 
which is 100 patients. Next, we evaluate the designs performance for $n = 20,40,60,80,100$. %$n=100.$ 
%\note{Ed: I changed this part because this is really what we did for the simulation. The previous language sounds like we calibrated the $e$s, $\alpha$s, and $\beta$s to first find a sample size, which is not true, and is not consistent with what's shown in the figure. Reviewers will be confused by seeing that we have sample size $n$ from 20 to 100, while we sounds like there is only 1 interim and the entire trial is 100 patients per arm.}
Lastly, we use the prior $(\theta_L, \theta_H) \sim \text{Beta}(0.05, 0.05) \cdot \text{Beta}(0.05, 0.05) I (\theta \in H)$, where $\text{Beta}(0.05, 0.05)$ is chosen to reduce the prior effective sample size. 
A total of 2,000 simulated trials are generated, 1,000 each under $H_0$ or $H_1$, for each design in each scenario. 

\begin{figure}
    \centering
    \includegraphics[width = 0.9\linewidth]{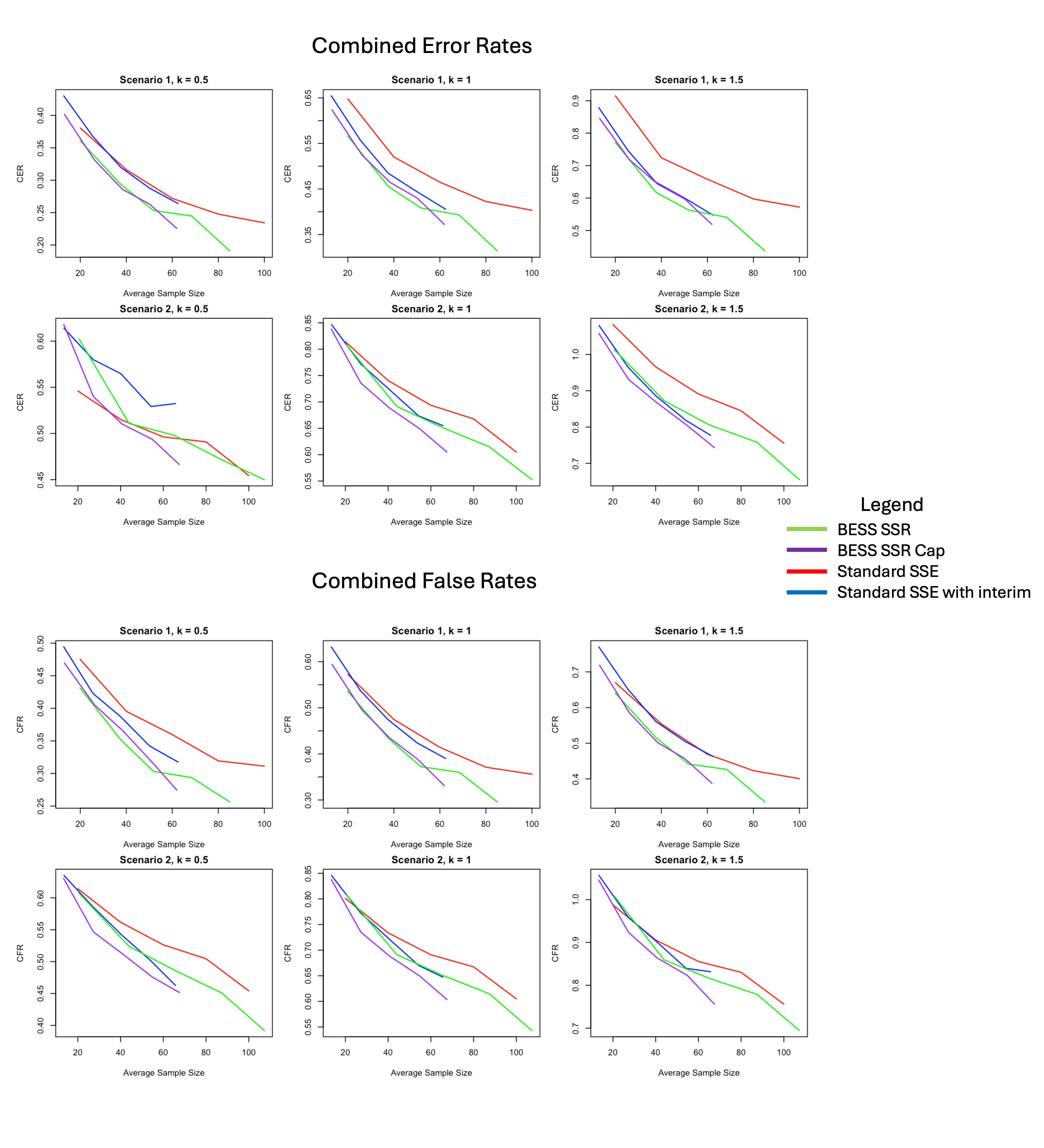}
    \caption{Combined Error Rates (CER) and Combined False Rates (CFR) across various sample sizes for the four designs under comparison. Different $k$ values are used to illustrate the importance of type I error rate over the type II error rate in CER or FDR  over FOR in CFR.}
    \label{fig:cer_cfr}
\end{figure}

\paragraph{Results}
We first present the operating characteristics of the four designs by evaluating their %type I and II 
error rates (see Appendix A.14 
for detail) across each scenario. To facilitate the comparison of designs, motivated by \cite{kim2021choosing} we consider a combined error rate defined as 
\begin{equation}\label{eq:cmbt12}
      \text{Combined Error Rate (CER) }  = \text{type I error rate} +  k \cdot \text{type II error rate}, \nonumber 
\end{equation}
where the weight $k \in [0,\infty)$ is a pre-determined factor that quantifies the relative weight between the type I and  type II error rates. We let $k \in \{0.5, 1, 1.5\}$. A lower CER is more desirable. 
The top two rows in Figure \ref{fig:cer_cfr} report the results 
for the four designs. 
Across all four designs, 
as the sample size increases, %the rate at which 
CER declines. %, \yy but at a slower rate \jj %indicating potential diminishing return 
%when sample size gets larger.  In other words, the gain in reduction of error rates may lessen when sample size continues to increase, \yy indicating diminishing return. \jj
%Considering the substantial costs associated with patient enrollment, these findings suggest finding a ``sweet spot" of sample size that achieves a desirable tradeoff between cost and statistical properties. 
Comparing across designs, 
we find that except for the Standard SSE design, the performance of the other designs are similar for the same average sample size.  Notably, the BESS SSR Cap design (design 2) seems to perform well %be in general the best 
across most cases as it shows slightly lower CERs for fixed sample sizes. In other words, under a same error rate, BESS SSR Cap design in general requires fewer patients on average than the other designs. In Scenario 2, when $k=0.5$, the standard SSE performs well and is comparable to both the BESS SSR and BESS SSR Cap designs. 
In contrast, it is the less desirable %losing 
design 
in the same scenario but with $k=1.5.$ This seems to suggest the standard SSE is a better design if the frequentist type I error rate is of important consideration in design evaluation. However, in early-phase dose optimization trials, type I error rate is not the primary concern. 
For example, one would be much more concerned if a ``GO" decision is made that recommends a wrong dose to further clinical development, or a ``No Go" decision that fails to recommend a promising dose.  These are measured by the FDR  and FOR   %(See Appendix A.8  for detail) 
in the bottom two rows of Figure \ref{fig:cer_cfr}, which we present next.

Following \cite{muller2004optimal}, we employ a loss function that integrates FDR  and FOR into a single metric given by
$$\text{Combined False Rate (CFR)} = k\cdot \text{FDR} + \text{FOR}.$$
Again, $k \in \{0.5, 1, 1.5\}$. 
The results are shown in Figure \ref{fig:cer_cfr}. Through careful examination, we find  
%again suggest 
the BESS SSR Cap design (design 2) performs favorable in most scenarios. %is the overall most desirable method, \bb i.e., in general, it has smaller sample size on average than the other designs for fixed false rate. \jj
%Therefore, to summarize, we recommend using the BESS with a sample size re-estimation capping the max sample size for the dose optimization trial. 

\section{Discussion}
\label{sec:discuss}

In this work we propose %BESS as 
a new and simple Bayesian  sample size estimator BESS that provides %. What \jj %BESS 
%leads to 
a straightforward interpretation of estimated sample size: if the observed data exhibits certain level of evidence $e$ that supports trial success, %the alternative hypothesis, 
with sample size $n$ one can conclude the alternative hypothesis with confidence $c$, measured by the POS, or the posterior probability of the alternative. The method exhibits $n$-coherent, where a larger sample size corresponds to higher POS under fixed evidence. It also demonstrates 
%The statement is \bb 
$e$-coherent when conducting %with a 
subsequent Bayesian analysis with the estimated sample size $n$. Specifically, 
%of the trial using $n$ as the sample size. 
if the observed trial data exhibits evidence $e^{\star} \geq e$, then the posterior probability of alternative will be at least %greater than or equal to 
$c$, thereby validating the sample size estimation. %corroborating with the sample size estimation.  
%For dose optimization trials, the BESS SSR Cap design shows a superior performance under both Frequentist and Bayesian properties. 
Furthermore, BESS %for continuous outcomes, this design 
enables sample size re-estimation at interim analysis, leveraging the observed sample variance for more precise sample size estimation.

Simulation results show that for matched type I/II error rates and using vague priors, BESS produces similar sample size estimates as the standard frequentist sample size estimation based on the oracle assumption, even though the two are based on different philosophies and metrics. In addition, the preliminary comparison of various designs based on BESS suggests a slight advantage of the Bayesian approach, requiring fewer patients to achieve comparable error and false discovery/remission rates. %positive/negative rates. 

In this work, we demonstrate  %have implemented 
BESS for three outcome types: binary, continuous (assuming known variance), and count-data, for superiority and non-inferiority tests and using simple Bayesian models. However, as a framework, BESS can be generalized %is flexible enough 
to accommodating different outcome types and hypotheses. For example, BESS can be extended to composite outcomes. Suppose $y_1$ and $y_2$ represent two outcome types (categorical and continuous) under a sampling model $f(y_1 \mid \theta_1) f(y_2 \mid \theta_2)$ with a prior $p(\theta_1, \theta_2)$. In this case, BESS can be extended based on the confidence level of $\text{Pr}((\theta_1, \theta_2) \in \Theta^* \mid data)$, where $\Theta^*$ is a promising zone that reflects the desirable treatment effects defined based on both $\theta$'s. Additionally, BESS can be adapted to other hypothesis tests, such as equivalence testing. For instance, when the parameter is on the boundary under the null, the hypothesis is $H_0:\theta = \theta^* \text{ vs. } H_1:\theta \neq \theta^*$. BESS remains applicable in this setting but requires different priors. For example, under $H_0$, a point null can be used for $\theta_1$, such that $\theta_0\mid a_0, b_0, H=H_0 \sim \frac{1}{C_0}Beta(\theta_0\mid a_0, b_0)I(\theta_0 < 1 - \theta^*)$, and $f(\theta_1\mid\theta_0,H = H_0) = I(\theta_1 = \theta_0 + \theta^*)$. This modification leads to different formula of Equation (3).

Many future directions may be considered to further develop BESS and corresponding adaptive designs, such as group sequential designs with sample size re-estimations, down-weight historical data in prior construction, and decision rules for futility stopping. %, etc. 
We leave these topics for future research.

\section*{Data Availability Statement}
The data that support the findings of this study are available from the corresponding author %, Dr. Yuan Ji, 
upon reasonable request.

\begin{singlespace}
\bibliographystyle{plainnat}
\bibliography{reference}
\end{singlespace}

\bigskip
\newpage

\appendix
\section{}

\renewcommand\thefigure{\thesection.\arabic{figure}} 
\setcounter{figure}{0}

\renewcommand\thetable{\thesection.\arabic{table}} 
\setcounter{table}{0}

\renewcommand\theequation{\thesection.\arabic{equation}} 
\setcounter{equation}{0}

\subsection{Posterior Probability of $H_1$ for One-arm Trials}
\label{subsec:pH1_onearm}

In this subsection, we show for Binomial, Normal with known variance, and Poisson sample models in one-arm trial, the POS (posterior probability of $H_1$) is a function of evidence $e$ (as defined in Equation (5)) %\eqref{eq:evidence} 
and sample size $n$, i.e., $\text{Pr}(H = H_1 \mid \bm{y}_n) = \text{Pr}(H = H_1 \mid e,n)$.

Assume $\theta_0$ is a known constant, the density  function $f(\bm{\theta} \mid \bm{y}_n)$ in Equation (3) reduces to the posterior $f(\theta_1 \mid \bm{y}_n)$.
From Table (1) %\eqref{tab:params} 
and Equation (5)%\eqref{eq:evidence}
, we have the posterior $f(\theta_1\mid \bm{y}_n)$ in Equation (3) %\eqref{eq:general_post_H1} 
for the three outcomes be as follows:
\begin{itemize}
    \item Binary: 
    $$f(\theta_1 \mid \bm{y}_n) = \frac{\theta_1^{a + n\bar{y}-1}(1-\theta_1)^{b+n(1-\bar{y})-1}}{B(a + n\bar{y},b+n(1-\bar{y}))} = \frac{\theta_1^{a + n[e+\theta_0]-1}(1-\theta_1)^{b+n(1-[e+\theta_0])-1}}{B(a + n[e+\theta_0],b+n(1-[e+\theta_0]))} = f(\theta_1\mid e,n),$$
    where $B(\cdot,\cdot)$ is the beta function;
    \item Continuous with known $\sigma^2$: 
    $$f(\theta_1\mid \bm{y}_n) = \frac{1}{\sqrt{\frac{2\pi}{\frac{1}{b}+\frac{n}{\sigma^2}}}}\exp{\left\{-\frac{1}{2}\left[\frac{1}{b}+\frac{n}{\sigma^2}\right]\left(\theta_1 - \frac{1}{\frac{1}{b}+\frac{n}{\sigma^2}}\left[\frac{a}{b}+\frac{n\bar{y}}{\sigma^2}\right]\right)^2\right\}}$$
    $$= \frac{1}{\sqrt{\frac{2\pi}{\frac{1}{b}+\frac{n}{\sigma^2}}}}\exp{\left\{-\frac{1}{2}\left[\frac{1}{b}+\frac{n}{\sigma^2}\right]\left(\theta_1 - \frac{1}{\frac{1}{b}+\frac{n}{\sigma^2}}\left[\frac{a}{b}+\frac{n[e+\theta_0]}{\sigma^2}\right]\right)^2\right\}} = f(\theta_1\mid e,n),$$
    \item Count-data:
    $$f(\theta_1\mid \bm{y}_n) = \frac{(b+n)^{a+n\bar{y}}}{\Gamma(a+n\bar{y})}\theta_1^{a+n\bar{y}-1}\exp{\{-(b+n)\theta_1\}} $$
    $$= \frac{(b+n)^{a+n[e+\theta_0]}}{\Gamma(a+n[e+\theta_0])}\theta_1^{a+n[e+\theta_0]-1}\exp{\{-(b+n)\theta_1\}} = f(\theta_1\mid e,n),$$
    where $\Gamma{(\cdot)}$ is the gamma function.
\end{itemize}
Hence, for all three outcomes in one-arm trial, we have $f(\theta_1\mid \bm{y}_n) = f(\theta_1\mid e,n)$. As a result, when both $\theta_0$ and $\theta^*$ are given as fixed constants, we see that the integration term in Equation (3) %\eqref{eq:general_post_H1} 
becomes a function of $e$ and $n$, i.e., 
$$\int_{\bm{\theta} \in H_1} f(\bm{\theta}\mid \bm{y}_n)d\bm{\theta} = \int_{\theta_1 - \theta_0 > \theta^*} f(\theta_1\mid e,n) d\theta_1.$$
Consequently, Equation (3) %\eqref{eq:general_post_H1} 
is a function of $e$ and $n$: 
$$\text{Pr}(H = H_1\mid \bm{y}_n) = \text{Pr}(H = H_1\mid e,n).$$
Note that $e$ is the sufficient statistic for $\bm{y}_n$ for the three outcome types specified in this work.

\bigskip
\newpage

\subsection{Posterior Probability of $H_1$ for Two-arm Trials with Continuous outcome and known variance}
\label{subsec:pH1_twoarm_cont}

%For two-arm trial with continuous outcome and known variance, equation \eqref{eq:general_post_H1} is again a function of $e$ and $n$. 
Assuming a 1:1 randomized clinical trial, denote $y_{ij}$ the continuous outcomes of patient $i$ in arm $j$. 
We arbitrarily pair the outcomes from each arm and
compute $\bm{y}_n^d = \{y_i^d; i = 1, \ldots, n\}$ where $y_i^d = y_{i1} - y_{i0}$. Since $y_{ij}\mid \theta_j \sim N(\theta_j,\sigma^2), \, j = 0, 1$ in Model (2) %\eqref{eq:bhm} 
for continuous outcome with known variance $\sigma^2$, %which is common between $j = 0$ and $j = 1$, 
the likelihood of $y_i^d$ follows a normal distribution index by the parameters $\theta = \theta_1 - \theta_0$ and $2\sigma^2$: 
$y_i^d\mid \theta \sim N(\theta, 2\sigma^2), \,  i = 1, \ldots, n$.
Note that here we slightly abuse the notation since $y_{i1}$ and $y_{i0}$ may refer to two different subjects $i$ in the treatment or control groups. So technically, we should use different notations, such as $i1$ and $i0$ to denote the two subjects. However, since the difference of the two $y$'s still follow the same distribution as $y_i^d$, we do not differentiate their indices.  

Let $\pi = N(a,b)$ be the prior for $\theta$, we have: 
$$\theta\mid a,b, H = H_j \sim \frac{1}{C_j}N(a,b)I(\theta \in H_j), \,\, j = 0, 1,$$
and we have $\text{Pr}(H = H_1) = q$ and $\text{Pr}(H = H_0) = 1 - q$ as in Model (2).%\eqref{eq:bhm}. 

Therefore, the Bayesian model above leads to
%The posterior probability of $H_1$ is again given by equation (3)%\eqref{eq:general_post_H1}
%, which we have 
%$$f(\bm{\theta}|\bm{y}_n) = 
$$f(\theta\mid \bm{y}_n^d) \propto f(\bm{y}_n^d\mid \theta)\pi(\theta).$$ 
Since the variance $\sigma^2$ is known, and both $f(.)$ and $\pi(.)$ follows normal distribution, we have $f(\theta\mid \bm{y}_n^d)$ follows a normal distribution and is similar to the continuous with known variance outcome in one-arm trial. Since $\bar{y}^d = \frac{1}{n}\sum_{i=1}^n y_i^d = \bar{y}_1 - \bar{y}_0 = e$, by Equation (5), we have
$$f(\theta\mid \bm{y}_n^d) = \frac{1}{\sqrt{\frac{2\pi}{\frac{1}{b}+\frac{n}{2\sigma^2}}}}\exp{\left\{-\frac{1}{2}\left[\frac{1}{b}+\frac{n}{2\sigma^2}\right]\left(\theta - \frac{1}{\frac{1}{b}+\frac{n}{2\sigma^2}}\left[\frac{a}{b}+\frac{n\bar{y}^d}{2\sigma^2}\right]\right)^2\right\}}$$
$$= \frac{1}{\sqrt{\frac{2\pi}{\frac{1}{b}+\frac{n}{2\sigma^2}}}}\exp{\left\{-\frac{1}{2}\left[\frac{1}{b}+\frac{n}{2\sigma^2}\right]\left(\theta - \frac{1}{\frac{1}{b}+\frac{n}{2\sigma^2}}\left[\frac{a}{b}+\frac{ne}{2\sigma^2}\right]\right)^2\right\}} = f(\theta\mid e,n).$$
%where by equation (5)%\eqref{eq:evidence}
%, $\bar{y}^d = e$. 
With an argument similar to section \ref{subsec:pH1_onearm}, we have 
$$\text{Pr}(H = H_1\mid \bm{y}_n) = \text{Pr}(H = H_1\mid e,n)$$
for the continuous with known variance outcome for two-arm trial.

\bigskip
\newpage

\subsection{Posterior Probability of $H_1$ for Two-arm Trial with Binary and Count-data Outcomes}
\label{subsec:pH1_twoarm}

From section \ref{subsec:pH1_onearm}, we see that for $\theta_j$, $j = 0, 1$, we have $f(\theta_j \mid \bm{y}_n) = f(\theta_j\mid \bar{y}_j,n)$ for both the binary and count-data outcomes. Hence, we have 
$$f(\bm{\theta}\mid \bm{y}_n) = \frac{f(\bm{y}_n\mid \bm{\theta})p(\bm{\theta})}{\int f(\bm{y}_n\mid \bm{\theta})p(\bm{\theta})d\bm{\theta}} = \frac{f(\theta_0\mid \bm{y}_{0n})f(\theta_1\mid \bm{y}_{1n})\pi(\theta_0)\pi(\theta_1)}{\iint f(\theta_0\mid \bm{y}_{0n})f(\theta_1\mid \bm{y}_{1n})\pi(\theta_0)\pi(\theta_1) d\theta_0d\theta_1} = $$
$$f(\theta_0\mid \bm{y}_{0n})f(\theta_1\mid \bm{y}_{1n}) = f(\theta_0\mid \bar{y}_0,n)f(\theta_1\mid \bar{y}_1,n) = f(\bm{\theta}\mid \bar{y}_1,\bar{y}_0,n),$$
where $\bm{y}_{jn} = \{y_{ij};i = 1, \ldots, n\}$. Therefore, according to Equation (3) in the manuscript, it follows
that  $\text{Pr}(H = H_1|\bm{y}_n) = \text{Pr}(H =H_1| \bar{y}_1, \bar{y}_0,n)$.  
\bigskip
\newpage

\subsection{Nmin Algorithms}\label{subsec:nmin_alg}

Let $\xi(n) \equiv \int_{\bm{\theta} \in H_1} f(\bm{\theta} \mid \bm{y}_n) d\bm{\theta}$.

\begin{nminalgorithm}{1}
    \caption{Determine $n_{\text{min}}$ for Algorithms 1 and 2'.}\label{alg:find_nmin_v}
\begin{algorithmic}[1]
    \State \textbf{Input:} $\theta^*$, $e$, hyperparemters, %maximum sample size 
    $n_{\text{max}}$, and $\bar{y}_0$ if for Algorithm 2'.
    \If{$\xi(n_{\max}+1) - \xi(n_{\max}) < 0$}
    \State stop, and return message asking user to increase $n_{\max}$.
    \Else
    \For{$n \in \{1, \ldots, n_{\text{max}}\}$}
    \State compute
    $\xi_d = \xi(n+1) - \xi(n)$. %= \int_{\bm{\theta} \in H_0}f(\bm{\theta}|e,n)d\bm{\theta} - \int_{\bm{\theta} \in H_0}f(\bm{\theta}|e,n+1)d\bm{\theta}.$$
    \If{$\xi_d \geq 0$}
    \State stop, and return $n_{\text{min}} = n$.
    \Else
    \State proceed to next $n$.
    \EndIf
    \EndFor
    \EndIf
    %\State \textbf{return} $n_{\min}$.
\end{algorithmic}
\end{nminalgorithm}

\begin{nminalgorithm}{2}
    \caption{Determine $n_{\text{min}}$ for Algorithm 2.}\label{alg:find_nmin_v}
\begin{algorithmic}[1]
    \State \textbf{Input:} $\theta^*$, $e$, hyperparemters, %maximum sample size 
    $n_{\text{max}}$, and grid size $\epsilon_s$.
    \For{$n \in \{1, \ldots, n_{\max}\}$}
        \For{$\bar{y}_0 \in [0,\epsilon_s,2\epsilon_s,\ldots,1-e]$}
            \State \textbf{set} $\bar{y}_1 = e+\bar{y}_0$ 
            \State \textbf{compute} and save $\Pr(H = H_1\mid \bar{y}_1,\bar{y}_0,n)$.
        \EndFor
        \State \textbf{find} $\bar{y}_0$ that results in minimum value of $\Pr(H = H_1\mid \bar{y}_1,\bar{y}_0,n)$.
        \State \textbf{compute} $n_{\min}$ using Nmin Algorithm 1.
        \If{$n_{\min} \leq n$}
        \State save the $n_{\min}$.
        \Else
        \State proceed to the next $n$.
        \EndIf
    \EndFor
    \State \textbf{return} $n_{\min}^* = \max\{n_{\min}\}$.
\end{algorithmic}
\end{nminalgorithm}

\bigskip
\clearpage

\subsection{BESS Algorithm 2'} \label{subsec:alg2_p}

Below is BESS Algorithm 2', where $\bar{y}_1$ and $\bar{y}_0$ are provided instead of evidence $e$.

\begin{bessalgorithm}{2'}
\caption{Two-Arm Trials; Binary or Count Data}\label{alg:BESS2p}
\begin{algorithmic}[1]
    \State \textbf{input:} the hierarchical models in Table 1.
    \State \textbf{input:} clinically minimum effect size $\theta^*$, response parameters for the treatment and control   $\bar{y}_1$ and $\bar{y}_0$,     confidence $c$, prior probability of $H_1$ $q$, maximum possible sample size $n_{\max}$. 
    \State {\bf set} $n_{\text{max}}$ as the largest candidate sample size. 
    \State \textbf{find} $n_{\text{min}}$ using \textbf{Nmin Algorithm 1}. %, which is the smallest sample size.\jj
    %\State {\bf Set} $n = n_{\text{min}}$.
    \While{$n \leq n_{\text{max}}$}
    \State compute Equation (3).
    \If{$\text{Pr}(H = H_1 \mid e,n) \geq c$}
    \State stop and return the sample size $n$. 
    \Else
    \State n = n+1
    \EndIf
    \EndWhile
    \If{$n > n_{\text{max}}$}
    \State return message: sample size is larger than $n_{\text{max}}$.
    \EndIf
\end{algorithmic}
\end{bessalgorithm}

\bigskip
\newpage

\subsection{Proof of Theorem 1}
\label{subsec:coherence}

For fixed $\theta^*$ and sample size $n$, denote $c(e) \equiv \Pr(H = H_1\mid e,n)$, and %define:
$$c(e) = %\text{Pr}(H = H_1|e,n) = 
\frac{\frac{q}{C_1}}{\frac{\frac{1-q}{C_0}}{\int_{\bm{\theta}\in H_1}f(\bm{\theta}|e,n)d\bm{\theta}}+\left[\frac{q}{C_1}+\frac{1-q}{C_0}\right]} = T_1\cdot\left(\frac{T_2}{\int_{\bm{\theta}\in H_1}f(\bm{\theta}\mid e,n)d\bm{\theta}}+T_3\right)^{-1},$$
where $T_1 = \frac{q}{C_1},T_2 = \frac{1-q}{C_0},$ and $T_3 = \left[\frac{q}{C_1}+\frac{1-q}{C_0}\right]$. To show that $c(e)$ is monotonically increasing in $e$, it suffices to prove $\int_{\bm{\theta}\in H_1}f(\bm{\theta}\mid e,n)d\bm{\theta}$ is monotonically increasing in $e$ since $T_1$, $T_2$, and $T_3$ are constants. 

We demonstrate this for the three models in Table 1 %binary, continuous, and count-data outcomes 
separately for one- and two-arm trials.

\paragraph{Binary Outcome (Binomial/Beta)} \hfill

\noindent \textbf{One-arm trial:}

%\bb
%$f(y_i|\theta_1) = f_{Bern}(y_i;\theta_1), \, p(\theta_1|\tilde{\bm{\phi}}) = f_{Beta}(\theta_1;a,b),$ where $f_{Bern}(\cdot|\theta_1)$ is the Bernoulli p.d.f parameterized by $\theta_1$ and $f_{Beta}$
$y_{i}\mid \theta_1 \sim \text{Bern}(\theta_j), \, \theta_1 \sim \text{Beta}(a,b),$ 
$\theta_0$ fixed.
%We first demonstrate this relationship for binary outcomes in one- and two-arms trial.
%aim to show that the coherence property between BESS and Bayesian inference holds for Binary outcomes. Specifically, for
%$$c = \text{Pr}(H = H_1|e,n) = T_1\cdot\left(\frac{T_2}{\int_{\bm{\theta}\in H_1}f(\bm{\theta}|e,n)d\bm{\theta}}+T_3\right)^{-1},$$
%where $T_1,T_2,$ and $T_3$ are constants, if an evidence of $e^* \geq e$ is observed, then $c^* \geq c$ where $c^* = \text{Pr}(H = H_1|e^*,n)$. In other words, the goal is to demonstrate that $\text{Pr}(H = H_1|e,n)$ is monotonically increasing with $e$ for fixed $n$ and $\theta^*$.
%In one-arm trial, assume a fixed $\theta_0$. 
Let $e' = e + \theta_0$ and $\theta' = \theta^* + \theta_0$. Then,
$$\int_{\theta_1\in H_1}f(\theta_1\mid e,n)d\theta_1 = 1 - F_{\text{Beta}}(\theta';a+ne',b+n(1-e')).$$
%Since $T_1,T_2,$ and $T_3$ are constants, it is suffice to proof that $F_{\text{Beta}}(\theta^*;a+ne,b+n(1-e))$ is monotonically decreasing with $e$. 
For evidence $e^* \geq e$, denote $\epsilon = e^* +\theta_0 \geq e'$. Thus,
$$\int_{\bm{\theta}\in H_1}f(\bm{\theta}\mid e^*,n)d\bm{\theta} = 1 - F_{\text{Beta}}(\theta';a+n\epsilon,b+n(1-\epsilon)).$$
%For fixed $\theta^*$ and $n$ (as well as the hyperparameters $a$ and $b$), it is straightforward to see that $\int_{\bm{\theta}\in H_1}f(\bm{\theta}|e,n)d\bm{\theta}$ is monotonically increasing with $e$ if and only if
%$$F_{\text{Beta}}(\theta^*;a+n{e^*}',b+n(1-{e^*}')) \leq F_{\text{Beta}}(\theta^*;a+ne',b+n(1-e')).$$
Define parameters: $\alpha = a+ne'$, $\alpha^* = a+n\epsilon$, $\beta = b+n(1-e')$, and $\beta^* = b+n(1-\epsilon)$. 
Let
$$X \sim \text{Beta}(\alpha,\beta), \text{ and } Y \sim \text{Beta}(\alpha^*,\beta^*).$$
Since $\alpha \leq \alpha^*$ and $\beta \geq \beta^*$, by Theorem 1 of \cite{arab2021convex}, it follows that $X \leq_{\text{st}} Y$%, where $\leq_{\text{st}}$ means that $X$ is 
(stochastically dominated). %by $Y$. 
Thus, by the definition of stochastic dominance \citep{arab2021convex}, %we have
$$F_{\text{Beta}}(\theta';\alpha,\beta) \geq F_{\text{Beta}}(\theta';\alpha^*,\beta^*)).$$
%for all $\theta^*$. 
Hence, %for any fixed $\theta^*$, we have proved that 
$\int_{\bm{\theta}\in H_1}f(\bm{\theta}\mid e,n)d\bm{\theta}$ is monotonically increasing in $e$. %for fixed $n$. 

\bigskip

\noindent \textbf{Two-arm trial:}

$y_{ij}\mid \theta_j \sim \text{Bern}(\theta_j), \, \theta_j \sim \text{Beta}(a_j,b_j)$, $j = 0, 1$. 
%In two-arms trial, we first 
First, let's assume $\bar{y}_0$ is fixed for now. 
%Assuming fixed $\bar{y}_0$, %is given and $\bar{y}_1 = e + \bar{y}_0$ (i.e., as required by Algorithm 2'). In this case, 
We have %$\int_{\bm{\theta}\in H_1}f(\bm{\theta}|e,n)d\bm{\theta}$ as a double integral:
$$\int_{\bm{\theta}\in H_1}f(\bm{\theta}\mid \bm{y}_n)d\bm{\theta} = \int_{\bm{\theta}\in H_1}f(\bm{\theta}\mid \bar{y}_0,e,n)d\bm{\theta} = 1 - $$
\begin{equation}\label{eq:2arm_iint}
    \iint_{\theta_1 - \theta_0 \leq \theta^*} f(\theta_1;a_1+n(\bar{y}_0+e),b_1+n(1-\bar{y}_0-e))f(\theta_0;a_0+n\bar{y}_0,b_0+n(1-\bar{y}_0))d\theta_1d\theta_0.
\end{equation}
%increasing monotonically with respect to $e$ for fixed $n, \bar{y}_0, \text{ and } \theta^*$. 
%We can 
Rewrite as an expectation: %equation \eqref{eq:2arm_iint} as a 1-dimensional integration over $\theta_1$:
$$\int_{\bm{\theta}\in H_1}f(\bm{\theta}\mid \bar{y}_0,e,n)d\bm{\theta} = 1 - $$
$$\int_0^1[1-F_{\text{Beta}}(\max(0,\theta_1-\theta^*),a_0+n\bar{y}_0,b_0+n(1-\bar{y}_0))]f(\theta_1;a_1+n(\bar{y}_0+e),b_1+n(1-\bar{y}_0-e))d\theta_1$$
$$= \int_0^1 [F_{\text{Beta}}(\max(0,x-\theta^*),a_0+n\bar{y}_0,b_0+n(1-\bar{y}_0))] f(x;a_1+n(\bar{y}_0+e),b_1+n(1-\bar{y}_0-e))dx$$
$$=E_Z[F_{\text{Beta}}(\max(0,x-\theta^*),a_0+n\bar{y}_0,b_0+n(1-\bar{y}_0))]$$
where %$E$ is the expectation with respect to the random variable 
$Z \sim \text{Beta}(a_1+n(\bar{y}_0+e),b_1+n(1-\bar{y}_0-e))$.

For %fixed $\bar{y}_0$, $\theta^*$, $n$, and the hyperparameter values (i.e., $a_1, a_2, b_1,$ and $b_2$), let 
$e^* > e$, denote %and 
$$Z^* \sim \text{Beta}(a_1+n(\bar{y}_0+e^*),b_1+n(1-\bar{y}_0-e^*)).$$
Again, by stochastic dominance, we have %$a_1+n(\bar{y}_0+e) \leq a_1+n(\bar{y}_0+e^*)$ and $b_1+n(1-\bar{y}_0-e) \geq b_1+n(1-\bar{y}_0-e^*)$, again by Theorem 1 of \cite{arab2021convex}, we can conclude that 
$Z \leq_{\text{st}} Z^*$. %, i.e., 
%$$F_{\text{Beta}}(x;a_1+n(\bar{y}_0+e^*),b_1+n(1-\bar{y}_0-e^*)) = F_{Z^*}(x) \leq $$
%$$F_{Z}(x) = F_{\text{Beta}}(x;a_1+n(\bar{y}_0+e),b_1+n(1-\bar{y}_0-e)).$$
%Next, by definition 1 of \cite{seth2014stochastic}, 
As $u(x) = F_{\text{Beta}}(\max(0,x-\theta^*),a_0+n\bar{y}_0,b_0+n(1-\bar{y}_0))$
%, since $Z \leq_{\text{st}} Z^*$ and $u(x)$ 
is strictly increasing and continuous in $x$, it follows that: %function in $x$, we have
$E_{Z^*}[u(x)] \geq E_{Z}[u(x)]$, establishing monotonicity. %. In other words, we have shown that for fixed $\theta^*,n$ and $\bar{y}_0$,
%\begin{equation}\label{eq:ineq_2arm_fixy}
%    \int_{\bm{\theta}\in H_1}f(\bm{\theta}|\bar{y}_0,e^*,n) \geq \int_{\bm{\theta}\in H_1}f(\bm{\theta}|\bar{y}_0,e,n),
%\end{equation}
%i.e., $\int_{\bm{\theta}\in H_1}f(\bm{\theta}|e,n)$ is monotonically increasing with $e$.

%Lastly, we 
Relaxing the assumption of a fixed $\bar{y}_0$,  
we utilize the minimizer derived from Algorithm 2.  
Let $\bar{y}_0^*$ denote the minimizer of Equation (8) with evidence $e^*$, and $\bar{y}_0'$ denote the minimizer with evidence $e$. To demonstrate monotonicity, we must show that for fixed $\theta^*$ and $n$, and for $e^* > e$,
$$\int_{\bm{\theta}\in H_1}f(\bm{\theta}\mid \bar{y}_0^*,e^*,n)d\bm{\theta} \geq \int_{\bm{\theta}\in H_1}f(\bm{\theta}\mid \bar{y}_0',e,n) d\bm{\theta}.$$ 

Since $\bar{y}_0'$ minimizes the integral when evidence is $e$, it follows that if $\bar{y}_0^* \neq \bar{y}_0'$, 
$$\int_{\bm{\theta}\in H_1}f(\bm{\theta}\mid \bar{y}_0^*,e,n)d\bm{\theta} \geq \int_{\bm{\theta}\in H_1}f(\bm{\theta}\mid \bar{y}_0',e,n)d\bm{\theta}.$$
Furthermore, previously we established that for fixed $\bar{y}_0^*$ and $e^* > e$,
$$\int_{\bm{\theta}\in H_1}f(\bm{\theta}\mid \bar{y}_0^*,e^*,n)d\bm{\theta} \geq \int_{\bm{\theta}\in H_1}f(\bm{\theta}\mid \bar{y}_0^*,e,n)d\bm{\theta}.$$
Combining these inequalities, we conclude:
$$\int_{\bm{\theta}\in H_1}f(\bm{\theta}\mid \bar{y}_0^*,e^*,n)d\bm{\theta} \geq \int_{\bm{\theta}\in H_1}f(\bm{\theta}\mid \bar{y}_0',e,n)d\bm{\theta},$$
thus confirming monotonicity for the minimizer.

%\bb
%However, note that if $\bar{y}_0$ is not the minimizer derived from Algorithm 2, then Theorem 1 may not hold. We demonstrate this with a counter example. Let $\theta^* = 0.05$, $q = 0.5$, $n$ be fixed at $100$, and $a_1 = b_1 = a_0 = b_0 = 0.5$. Suppose $\bar{y}_0 = e = 0.1$, $\Pr(H = H_1|\bar{y}_0,e,n) = 0.87.$ Now, let $e^\star = 0.11 > e$ but $\bar{y}_0^{\star} = 0.5$, we have $\Pr(H = H_1|\bar{y}_0^{\star},e^\star,n) = 0.84 < \Pr(H = H_1|\bar{y}_0,e,n) = 0.87,$ despite an increase in $e$ by 0.01. 
%\jj

\bigskip

\paragraph{Continuous Outcome (Normal/Normal)} \hfill

Assume $\sigma$ is known, for one-arm trial,
$y_{i}|\theta_1 \sim N(\theta_1,\sigma^2), \, \theta_1 \sim N(a,b), \, \theta_0$ is fixed. For two-arm trial, $y_i^d|\theta \sim N(\theta,2\sigma^2), \, \theta \sim N(a,b)$ (see Appendix A.2).
%As shown in binary outcome case, it is suffice to show that $\int_{\bm{\theta} \in H_1} f(\bm{\theta}|e,n)d\bm{\theta}$ is monotonically increasing with $e$. 
%For a continuous outcome, 
We define $e' = e + \theta_0$, $\theta' = \theta^* + \theta_0$, and ${\sigma'}^2 = \sigma^2$ in the one-arm trial, and $e' = e$, $\theta' = \theta^*$, and ${\sigma'}^2 = 2\sigma^2$ in the two-arm trial. Then, we have
$$\int_{\bm{\theta}\in H_1}f(\bm{\theta}\mid e,n)d\bm{\theta} = 1 - \Phi\left(\theta';\frac{1}{\frac{1}{b}+\frac{n}{{\sigma'}^2}}\left[\frac{a}{b}+\frac{ne'}{{\sigma'}^2}\right], \left[\frac{1}{b}+\frac{n}{{\sigma'}^2}\right]^{-1}\right),$$
where $\Phi(\cdot|\mu,\Sigma)$ denotes the CDF of a normal distribution with mean $\mu$ and variance $\Sigma$. Therefore, showing monotonicity of $c(e)$ with respect to $e$ is equivalent to demonstrating that
$$\Phi\left(\theta';\frac{1}{\frac{1}{b}+\frac{n}{{\sigma'}^2}}\left[\frac{a}{b}+\frac{ne'}{{\sigma'}^2}\right], \left[\frac{1}{b}+\frac{n}{{\sigma'}^2}\right]^{-1}\right)$$
decreases monotonically with increasing $e'$ for fixed $n$, $\theta^*$, $\sigma$, and hyperparameters $a$ and $b$.

We analyze the monotonicity by examining the z-score:
$$z(e) = \frac{\theta' - \frac{ne' + \frac{a{\sigma'}^2}{b}}{\frac{{\sigma'}^2}{b}+n}}{\frac{1}{\frac{1}{\sigma'}\sqrt{\frac{{\sigma'}^2}{b}+n}}} = \left(\frac{1}{\sigma'}\sqrt{\frac{{\sigma'}^2}{b}+n}\right)\left(\theta' - \frac{ne'}{\frac{{\sigma'}^2}{b}+n} - \frac{\frac{a{\sigma'}^2}{b}}{\frac{{\sigma'}^2}{b}+n}\right).$$
This expression simplifies to
$$z(e) = C_A - \frac{ne'}{\sigma'\sqrt{\frac{{\sigma'}^2}{b}+n}} - C_B,$$
where $C_A \text{ and } C_B$ are constants independent of $e$. Clearly, %The monotonic decreasing relationship we seek to proof then corresponds to demonstrating that $z(e)$ is monotonically decreasing with an increase in $e$. Since $C_A \text{ and } C_B$ are constants with respect to $e$, it is straightforward to see that an increase in 
as $e$ increases, the term involving $e'$ increases, causing $z(e)$ to decrease. Thus, we conclude that $z(e)$ is monotonically decreasing with increasing $e$, confirming the montonic increase of %. Consequently, we have shown that an increase in $e$ corresponds to an increase in 
$$\int_{\bm{\theta}\in H_1}f(\bm{\theta}\mid e,n)d\bm{\theta}.$$

\bigskip

\paragraph{Count-data Outcome (Poisson/Gamma)}
\hfill

\noindent \textbf{One-arm trial:}

$y_i|\theta_1 \sim \text{Poi}(\theta_1), \, \theta_1 \sim \text{Gamma}(a,b),\, \theta_0$ is fixed.
%In one-arm trial, let 
Define $e' = e + \theta_0$ and $\theta' = \theta^* + \theta_0$. Then, %we have 
$$\int_{\bm{\theta}\in H_1} f(\bm{\theta}\mid e,n)d\bm{\theta} = 1 -F_{\text{Gamma}}(\theta';a+ne',b+n).$$ 

To demonstrate monotonicity, it suffices to show that %$\int_{\bm{\theta}\in H_1} f(\bm{\theta}|e,n)d\bm{\theta}$ is monotonically increasing with $e$, or equivalently, 
$F_{\text{Gamma}}(\theta';a+ne',b+n)$
decreases monotonically with increasing $e$ for fixed $n$ and $\theta^*$.
%To show this property, l
Consider $e^* \geq e$ and define $e_d = e^* - e$. Let $X \sim \text{Gamma}(a+ne',b+n)$ and $Y \sim \text{Gamma}(ne_d,b+n)$. By the summation property of Gamma distributions 
$$Z = X+Y \sim \text{Gamma}(a+n(e^*+\theta_0),b+n).$$
%where
%$$X \sim \text{Gamma}(a+ne_1,b+n), \text{ and } Y \sim \text{Gamma}(ne^*,b+n).$$

\noindent Thus, %for a fixed point $\theta^*$, we have
$$F_{\text{Gamma}}(\theta';a+n(e^* + \theta_0),b+n) = \text{Pr}(Z \leq \theta') = \text{Pr}(X+Y \leq \theta').$$
Since $Y \geq 0$, it follows directly that %we have 
$$\{X+Y \leq \theta'\} \subseteq \{X \leq \theta'\},$$
%as all events of $X+Y \leq \theta^*$ are events of $X \leq \theta^*$, but the events $X \leq \theta^*$ does not necessarily imply the events of $X + Y \leq \theta^*$. 
%This subset 
leading to %the following relationship: 
$$F_{\text{Gamma}}(\theta';a+n(e^*+\theta_0),b+n) = \text{Pr}(X+Y \leq \theta') \leq \text{Pr}(X \leq \theta') = F_{\text{Gamma}}(\theta';a+ne',b+n).$$
Therefore, %for fixed $n$ and $\theta^*$, 
$F_{\text{Gamma}}(\theta';a+ne',b+n)$ monotonically decreases with increasing $e$.

\bigskip

\noindent \textbf{Two-arm trial:}

$y_{ij}\mid \theta_j \sim \text{Poi}(\theta_j), \, \theta_j \sim \text{Gamma}(a_j,b_j)$ for $j = 0, 1$.
%In two-arms trial, 
%Assume $\bar{y}_0$ is fixed. 
Note $\bar{y}_0$ is %, $\bar{y}_1$ are 
assumed fixed.
Then,
$$\int_{\bm{\theta}\in H_1} f(\bm{\theta}\mid e,n)d\bm{\theta} = \int_{\bm{\theta}\in H_1} f(\bm{\theta}\mid \bar{y}_0,e,n)d\bm{\theta} = 1 - $$
$$\iint_{\theta_1-\theta_0 \leq \theta^* }f_{\text{Gamma}}(\theta_1;a_1+n(\bar{y}_0+e),b_1+n)f_{\text{Gamma}}(\theta_0;a_0+n\bar{y}_0,b_0+n)d\theta_1d\theta_0$$
$$= 1 - \text{Pr}(\Theta_1(e) \leq \Theta_0 + \theta^*),$$
where $\Theta_1(e) \sim \text{Gamma}(a_1+n(\bar{y}_0+e),b_1+n)$ and $\Theta_0 \sim \text{Gamma}(a_0+n\bar{y}_0,b_0+n)$.

Let $e^* \geq e$. From the previous one-arm trial analysis, %(with $\theta_0 = 0$), 
we know %proved in the one-arm trial case for count-data outcome (let $\theta_0 = 0$), we have
$$F_{\text{Gamma}}(x;a_1+n(e^*+\bar{y}_0),b_1+n) \leq F_{\text{Gamma}}(x;a_1+n(e+\bar{y}_0),b_1+n)$$ 
for fixed $x$ and $n$. Defining $f_{\Theta_0}(x) = f_{\text{Gamma}}(x;a_0+n\bar{y}_0,b_0+n)$, %, and $\Theta_1(e) \sim Gamma(a_1+n(\bar{y}_0+e),b_1+n)$ %being the random variable $\Theta_1$ that follows a Gamma distribution with its shape parameter being $a_1+n(\bar{y}_0+e')$ and rate parameter being $b_1+n$. 
we have
$$\text{Pr}(\Theta_1(e^*) \leq \Theta_0 + \theta^*) = \int_0^{\infty}\text{Pr}(\Theta_1(e^*)\leq x+\theta^*)f_{\Theta_0}(x)dx = $$
$$\int_0^\infty [F_{\text{Gamma}}(x+\theta^*;a_1+n(\bar{y}_0+e^*),b_1+n)]f_{\Theta_0}(x)dx \leq \int_0^\infty [F_{\text{Gamma}}(x+\theta^*;a_1+n(\bar{y}_0+e),b_1+n)]f_{\Theta_0}(x)dx$$
$$= \int_0^{\infty}\text{Pr}(\Theta_1(e)\leq x+\theta^*)f_{\Theta_0}(x)dx = \text{Pr}(\Theta_1(e)\leq \Theta_0+\theta^*).$$
Thus, %we have shown that $1 - \int_{\bm{\theta}\in H_1} f(\bm{\theta}|e,n)d\bm{\theta}$ is monotonically decreasing with an increase in $e$ for fixed $n$ and $\bar{y}_0$, i.e., 
$\int_{\bm{\theta}\in H_1} f(\bm{\theta}\mid e,n)d\bm{\theta}$ is monotonically increasing with $e$, confirming the desired result.
%\bb However, Theorem 1 may not hold if $\bar{y}_0$ is not fixed. Using the same argument as in Binary Outcome two-arm trial case, we can demonstrate that again, an increase in $e$ may not results in an increase in $\Pr(H = H_1|\bar{y}_0,e,n)$ if $\bar{y}_0$ is also varied from its initial value. \jj

$\hfill \square$

\bigskip
\newpage

\subsection{Proof of Proposition 1}\label{subsec:proof_prop1}

Using the definition of the beta CDF, we can write:
$$F_{\text{Beta}}(\theta';a+ne',b+n(1-e')) - F_{\text{Beta}}(\theta';a+(n+1)e',b+(n+1)(1-e'))$$
\begin{equation}\label{eq:A_and_B_b1}
= \int_{0}^{\theta'} \underbrace{\frac{x^{a+ne'-1}(1-x)^{b+n(1-e')-1}}{B(a+(n+1)e',b+(n+1)(1-e'))}}_{\cal{A}}\left[\underbrace{\frac{B(a+(n+1)e',b+(n+1)(1-e'))}{B(a+ne',b+n(1-e'))}-x^{e'}(1-x)^{1-e'}}_{\cal{B}}\right]dx.
\end{equation}

The term $\cal{A}$ is always positive for any $x \in (0,1)$, with $a,b > 0$, $n \geq 0$, and $e + \theta_0 = e' \in (0,1)$. The term $\cal{B}$ is the difference between a ratio of two Beta functions and the function $x^{e'}(1-x)^{1-e'}$. Clearly, the integrand is non-negative if ${\cal B} \geq 0$; that is, 
$$\int_0^{\theta'} {\cal A} \cdot {\cal B} \; dx \geq \int_0^{\theta'} {\cal A} \cdot 0 \; dx = 0.$$

We first analyze ${\cal B}$ as a function of $x$ (for fixed $e'$ and $n$). Taking derivatives with respect to $x$, we obtain: 
$$\frac{d \cal{B}}{dx} = x^{e'-1}(1-x)^{-e'}(x-e'), \;\;
\frac{d^2 \cal{B}}{dx^2} = (1-e')e'x^{e'-2}(1-x)^{-1-e'} > 0.$$
Thus, ${\cal B}$ is decreasing for $x < e'$, increasing for $x > e'$, and convex on $(0,1)$ since %. And for any values of $x$, 
$\frac{d^2{\cal B}}{dx^2} \geq 0$. It follows that ${\cal B}$ has a global minimum at $x = e'$.

Next, we analyze the asymptotic behavior of $\cal{B}$ as $n \rightarrow \infty$. First, we have
$$\lim_{n \rightarrow \infty} \frac{B(a+(n+1)e',b+(n+1)(1-e')}{B(a+ne',b+n(1-e'))} = \lim_{n \rightarrow \infty} \frac{\Gamma(a+ne'+e')}{\Gamma(a+ne')}\frac{\Gamma(b+n(1-e')+(1-e'))}{\Gamma(b+n(1-e'))(a+b+n)}.$$
Using the asymptotic expansion $\Gamma(x+\delta)/\Gamma(x) \rightarrow x^{\delta}$ as $x \rightarrow \infty$, above becomes:
$$\lim_{n \rightarrow \infty} \frac{(a+ne')^{e'}(b+n(1-e'))^{1-e'}}{a+b+n} = \lim_{n \rightarrow \infty} \frac{\left(\frac{a}{n}+e'\right)^{e'}\left(\frac{b}{n}+(1-e')\right)^{1-e'}}{\frac{a+b}{n}+1} = {e'}^{e'}(1-e')^{1-e'}.$$
%$$\frac{(ne')^{e'}(n(1-e'))^{1-e'}}{n} = {e'}^{e'}(1-e')^{1-e'}.$$
Therefore, $\lim_{n \rightarrow \infty}{\cal B}(n, x = e') = 0$ and ${\cal B}(n,x)$ attains its global minimum at $x = e'$ for a fixed $n$. %\yy the term $\cal{B}$ \jj attains
%its minimum value at $x = e'$ and has a single root at $x = e'$,
%\bb i.e., \yy $\lim_{n \to \infty}{\cal B}(n, x = e') = 0$. \jj

Now, we analyze $\cal B$ as a function of $n$ for fixed $x$. Since $x^{e'}(1-x)^{1-e'}$ is constant with respect to $n$, define 
$$R(n) = \frac{B(a+(n+1)e,b+(n+1)(1-e))}{B(a+ne,b+n(1-e))},$$
% \yy is non-decreasing in $n$. \jj To show this, 
we aim to show that $R(n)$ is non-decreasing in $n$. It suffices to prove:
\begin{equation}\label{eq:rn}
    \frac{R(n+1)}{R(n)} = \frac{B(a+(n+2)e',b+(n+2)(1-e'))B(a+ne',b+n(1-e'))}{[B(a+(n+1)e',b+(n+1)(1-e'))]^2} \geq 1.
\end{equation}
%\noindent Letting $t_1 = a+n$
%Rearranging terms, the inequality \yy becomes \jj %we need to prove is
%$$B(a+(n+2)e',b+(n+2)(1-e'))B(a+ne',b+n(1-e')) \geq [B(a+(n+1)e',b+(n+1)(1-e'))]^2.$$
Letting $t_1 = a+ne'$ and $t_2 = b+n(1-e')$, we must show: 
%\begin{equation}\label{eq:nondec}
    $$B(t_1+2e',t_2+2(1-e'))B(t_1,t_2) \geq [B(t_1+e',t_2+(1-e'))]^2, $$
%\end{equation}
or equivalently, 
\begin{equation}\label{eq:nondec}
%$$
\log\left(B(t_1+2e',t_2+2(1-e'))\right) + \log\left(B(t_1,t_2)\right) \geq 2\log\left(B\left(t_1+e',t_2+(1-e'\right)\right).
%$$
\end{equation}
We prove this using the following Lemma. %to prove this statement.

\bigskip

\begin{lemma}\label{lemma:conv_log_beta}
    The Beta function $B(a,b)$ is log-convex with respect to both parameters $a$ and $b$.
\end{lemma}

% LEMMA 2: beta function  is log-convex with respect to its parameters.
\paragraph{Proof of Lemma \ref{lemma:conv_log_beta}:}
%The beta function \yy $B(a, b)$ is log-convex with respect to its parameters $a$ and $b$ if for parameters $(a_1,b_1)$
%\begin{equation}
%    \log B(\lambda a_1+(1-\lambda)a_2,\lambda b_1+(1-\lambda)b_2) \leq \lambda\log(B(a_1,b_1)) + (1-\lambda)\log(B(a_2,b_2)) \label{eq:logc}
%\end{equation}
   %This can be seen as follows. 
%for $\lambda \in[0, 1].$ \jj 
By definition, 
$$B(a,b) = \int_0^1 x^{a-1}(1-x)^{b-1}dx.$$ 
For any $0 \leq \lambda \leq 1$, define %Express the beta function as a convex combination of parameters $(a_1,b_1)$ and $(a_2,b_2)$ with $\lambda \in [0,1]$, we have
$$B(\lambda a_1+(1-\lambda)a_2,\lambda b_1+(1-\lambda)b_2) = \int_0^1 x^{\lambda a_1 - \lambda +(1-\lambda)a_2-1+\lambda}(1-x)^{\lambda b_1-\lambda+(1-\lambda)b_2-1+\lambda}dx$$
%This is equal to:
%$$= \int_0^1 x^{\lambda(a_1-1)+(1-\lambda)(a_2-1))}(1-x)^{\lambda(b1-1)+(1-\lambda)(b_2-1)}dx = \int_0^1\left[x^{a_1-1}(1-x)^{b_1-1}\right]^{\lambda}\left[x^{a_2-1}(1-x)^{b_2-1}\right]^{1-\lambda}dx.$$
$$=\int_0^1\left[x^{a_1-1}(1-x)^{b_1-1}\right]^{\lambda}\left[x^{a_2-1}(1-x)^{b_2-1}\right]^{1-\lambda}dx.$$
Applying H\"older's inequality (with exponents $p = \frac{1}{\lambda}$, $q = \frac{1}{1-\lambda}$) yields
$$\int_0^1\left[x^{a_1-1}(1-x)^{b_1-1}\right]^{\lambda}\left[x^{a_2-1}(1-x)^{b_2-1}\right]^{1-\lambda}dx \leq $$
$$\left(\int_0^1 x^{a_1-1}(1-x)^{b_1-1} dx\right)^{\lambda}\left(\int_0^1 x^{a_2-1}(1-x)^{b_2-1}dx \right)^{1-\lambda},$$
or 
$$B(\lambda a_1+(1-\lambda)a_2,\lambda b_1+(1-\lambda)b_2) \leq [B(a_1,b_1)]^{\lambda}[B(a_2,b_2)]^{1-\lambda}.$$ 
Taking logarithms:
$$\log B(\lambda a_1+(1-\lambda)a_2,\lambda b_1+(1-\lambda)b_2) \leq \lambda\log(B(a_1,b_1)) + (1-\lambda)\log(B(a_2,b_2)),$$
which confirms that $B(a,b)$ is log-convex. $\hfill \square$ %\yy and therefore \eqref{eq:logc} is true. \jj 

\bigskip

%Let $\lambda = \frac{1}{2}$, $a_1 = t_1 + 2e'$, $a_2 = t_1$, $b_1 = t_2 + 2(1-e')$, and $b_2 = t_2$, we have
%$$t_1 + e' = \frac{(t_1 + 2e')+t_1}{2} \text{ and } t_2 + (1 - e') = \frac{(t_2 + 2(1-e'))+t_2}{2}.$$
%Consequently, 
%$$\frac{1}{2}\log\left(B(t_1+2e',t_2+2(1-e'))\right) + \frac{1}{2}\log\left(B(t_1,t_2)\right) \geq \log\left(B\left(\frac{(t_1 + 2e')+t_1}{2}, \frac{(t_2 + 2(1-e'))+t_2}{2}\right)\right).$$
Now, applying Lemma \ref{lemma:conv_log_beta} with $\lambda = \frac{1}{2}$, $a_1 = t_1 + 2e'$, $a_2 = t_1$, $b_1 = t_2 + 2(1-e')$, and $b_2 = t_2$, we obtain:
$$\frac{1}{2}\log\left(B(t_1+2e',t_2+2(1-e'))\right) + \frac{1}{2}\log\left(B(t_1,t_2)\right) \geq \log\left(B\left(\frac{(t_1 + 2e')+t_1}{2}, \frac{(t_2 + 2(1-e'))+t_2}{2}\right)\right),$$
which implies the inequality \eqref{eq:nondec}, and thus confirms that ${\cal{B}}$ is non-decreasing in $n$ for fixed $x$.

Moreover, we can show that $\cB$ is strictly increasing in $n$ for fixed $x$. To see this, let $g(n)$ denotes a general function with $n$ being a variable of interest. We first show the following Lemma, which states the condition for which
$$g(n)^2 = g(n-1)g(n+1).$$

\begin{lemma}\label{lemma:eq_cond}
    The equation 
    $$g(n)^2 = g(n-1)g(n+1)$$
    holds if and only if $g(n)$ is a geometric progression, i.e., there exist constants $A \neq 0$ and $r \neq 0$ such that
    $$g(n) = Ar^n \text{ for all } n.$$
\end{lemma}

\noindent \textbf{Proof of Lemma \ref{lemma:eq_cond}:}
For the if part, suppose $g(n) = Ar^n$. Then, $g(n-1) = Ar^{n-1}$ and $g(n+1) = Ar^{n+1}$. Thus, we have
$$f(n-1)f(n+1) = A^2r^{n-1+n+1} = (Ar^n)^2 = g(n)^2.$$

For the only if part, suppose $g(n)$ satisfies the equation and assume $f(n) \neq n$ for all $n$. Take logarithms on both side, we have
$$2\log(g(n)) = \log(g(n-1)) + \log(g(n+1)).$$
Let $lg(n) = \log(g(n)),$ we have
$$lg(n+1) - lg(n) = lg(n) - lg(n-1).$$
Therefore, $lg(n)$ must be an arithmetic progression:
$$lg(n) = a+bn.$$
Therefore, we have
$$g(n) = \exp(a+bn) = \exp{(a)}\times (\exp{(b)})^n = Ar^n,$$
where $A = \exp{(a)}$ and $r = \exp{(b)}$. $\hfill \square$

\noindent It is straightforward to see that $B(a+ne',b+n(1-e')$ is not in the form of $Ar^n$. As the condition of $g(n) = Ar^n$ is necessary for equality to hold, therefore, we must have
$$B(t_1+2e',t_2+2(1-e'))B(t_1,t_2) > [B(t_1+e',t_2+(1-e'))]^2.$$
Consequently, $\cB$ is strictly increasing with respect to $n$.

%In summary, we have shown that ${\cal B}(n,x)$ is:
%\begin{itemize}
%    \item convex in $x$ for fixed $n$, with minimum at $x = e'$, and
%    \item non-decreasing in $n$ for fixed $x$.
%\end{itemize}
%These properties imply that for any $\theta^* < e'$, 

%back to comparing $B(t_1+2e,t_2+2(1-e))B(t_1,t_2)$ and $[B(t_1+e,t_2+(1-e))]^2$. Notice that
%\yy According to \eqref{eq:logc} the beta function \jj is log-convex, 

 %Exponentiating both sides yields
 %$$B(t_1,t_2)B(t_1+2e,t_2+2(1-e)) \geq [B(t_1+e,t_2+(1-e))]^2,$$
% \yy which is our objective \eqref{eq:nondec}. \jj 
%This completes the proof that ${\cal{B}}(n,x)$ is non-decreasing in $n$. Therefore, as $n$ increases, the value of $\cal{B}$ \yy is \jj  non-decreasing for any fixed value of $x$, which explains the upward shift of the convex curve observed in Figure \ref{fig:diff_plot}. 

Lastly, we use the two key properties of $\cal B$--its convexity in $x$ and monotonicity in $n$--to establish the condition under which ${\cal B} > 0$.
%\yy to explore the conditions that ensure \jj 
%$$F_{\text{Beta}}(\theta'; a+ne',b+n(1-e')) - F_{\text{Beta}}(\theta'; a+(n+1)e',b+(n+1)(1-e')) \geq 0.$$ 
Specifically, we show that {\bf for any $\theta^* < e$ (hence $\theta' < e'$), there exists an integer $n^\prime_{\min}$ such that for all $n > n^\prime_{\min}$, $\cB(n, x) > 0$ for all $x \in (0, \theta').$ }
We proceed in two parts:

\textbf{First}, suppose there exists some $n^\prime_{\min}$ such that $\cB(n^\prime_{\min}, x=e') \ge 0$. Since %by definition $\cB(n, x=0) > 0 $,  and 
%we have found the $n^\prime_{\min}$ since 
$\cB(n, x)$ is non-decreasing in $n$ and decreasing in $x$ over the interval $x \in [0, e'],$ it follows that for all $n \geq n_{\text{min}}'$, $\cB(n,x) \geq \cB(n_{\min}',x) > \cB(n_{\min}',e') \geq 0$ for $x \in (0, \theta')$. Therefore, the integrand in \eqref{eq:A_and_B_b1} is positive, and the integral itself is positive as well.

\textbf{Second}, %suppose 
alternatively suppose $\cB(n, x=e') < 0$ for all $n$. Since $\cB(n, x=0) >
0$ and $\cB(n, x)$ is continuous and strictly decreasing on $x \in [0,e']$,
the Intermediate value theorem guarantees the existence of a value
$\theta' = \theta^* + \theta_0 \in (0, e' = e + \theta_0)$ such that $\cB(n, x= \theta') = 0$ and $\cB(n,x) > 0$ for all $x < \theta'$. %when $\theta_0$ is known and given. 
Furthermore, since $\cB(n, x)$ is non-decreasing in $n$ for fixed $x$, we have 
$$\lim_{n \rightarrow \infty} {\cal B}(n,x = e') = \lim_{n \rightarrow \infty} \frac{B(a+(n+1)e',b+(n+1)(1-e')}{B(a+ne',b+n(1-e'))} - {e'}^{e'}(1-e')^{1-e'} $$
$$= {e'}^{e'}(1-e')^{1-e'} - {e'}^{e'}(1-e')^{1-e'} = 0,$$
i.e., it follows that as $n \to
\infty$, the corresponding root $\theta'$ approaches $e'$. 

Next, we show that there exists a monotonically increasing sequence of $\theta'$, denoted as $\bm{\Theta}' = \{\theta_1',\theta_2',\ldots,e'\}$, that is one-to-one corresponding to the sample sizes $\{1, 2, \ldots, \infty\}$ such that $\cB(n,x = \theta_n') = 0$. As shown previously, there exists a corresponding $\theta'$ for any sample size $n = 1, 2, \ldots$, we aim to prove that for any $\theta_n'$, we have $\theta_n' \leq \theta_{n+1}'$. We prove this by contradiction.

Suppose that there exists a $\theta_n'$ such that $\theta_n' > \theta_{n+1}'$. Because we showed previously that $\cB(n,x)$ decreases in $x$ for $x \in (0,e')$ for fixed $n$ and increases monotonically in $n$ for fixed $x$, for $\theta_n' > \theta_{n+1}'$, we have
$$\cB(n+1,x = \theta_{n+1}') = 0 > \cB(n+1, x = \theta_n') \geq \cB(n, x = \theta_n') = 0$$
which is a contradiction. Consequently, for all $\theta_n' \in \bm{\Theta}'$, we must have $\theta_1' \leq \theta_2' \leq \ldots, \leq e'$.  

Therefore, for any $\theta^* < e$, we have $\theta' \in (0, e')$, and there exists
$n^\prime_{\min}$ such that $\theta_{n^\prime_{\min}} \leq \theta' \leq \theta_{n^\prime_{\min}+1}$, and
$$\cB(n^\prime_{\min}, \theta') \geq 0 \; \text{ and } \; \cB(n^\prime_{\min}, x) > 0 \text{ for all } x \in (0, \theta').$$ This value of $n^\prime_{\min}$ ensures that for all
$n>n^\prime_{\min}$, $\cB(n, x) \ge \cB(n^\prime_{\min}, x) > 0$ over $(0, \theta')$. Consequently,
%Therefore, we have shown that for $\theta^* < e',$ there exists an $n^\prime_{\min} > 0$ such that when $n > n^\prime_{\min}$ equation 
$$
\eqref{eq:A_and_B_b1} = \int_0^{\theta'} {\cal A} \cdot \cB \; dx > \int_0^{\theta'} {\cal A} \cdot 0 \; dx = 0.
$$

Finally, although we have established the existence of such $n^\prime_{\min}$, it is possible that a smaller sample size %that ensures $\cB(n^\prime_{\min}, x) >0$ when $x \in (0, \theta^*)$. We can find a  sample size 
$n_{\min} \le n'_{\min}$ could still result in \eqref{eq:A_and_B_b1} being positive, even if $B(n_{\min}, x)$ is not strictly positive for all $x \in (0, \theta').$ $\hfill \square$

\bigskip
\newpage

%Figure \ref{fig:diff_e_n} illustrates the 
%value of $n_{\min}$ for different $e$ assuming $\theta^* = e - 0.04$.  

%In the BESS calculation, we find $n_{\min}$ numerically for a given $\theta^*$ and $e$ using the following Algorithm.  \jj  The required inputs  are: evidence $e$, $\theta^*$, $a, b$, and a maximum sample size $n_{\text{max}}$.

\subsection{Proof of Proposition 2}\label{subsec:prop_prop2}

Assume %that $e$ and 
$\bar{y}_0$ is prespecified, and $0 < \bar{y}_0+e < 1$. %fixed and known. %Further assume $a_1 = a_2 = \alpha$ and $b_1 = b_2 = \beta$. 
Define
$$\xi(e,n) = \int_{\bm{\theta} \in H_1} f(\bm{\theta}|\bar{y}_0,e,n)d\bm{\theta}= 1 - \int_{\bm{\theta} \in H_0} f(\bm{\theta}|\bar{y}_0,e,n)d\bm{\theta} = 1 - $$
$$\iint_{\theta_1-\theta_0 \leq \theta^*}f(\theta_1;a_1+n(\bar{y}_0+e),b_1+n(1-\bar{y}_0-e))f(\theta_0;a_0+n\bar{y}_0,b_0+n(1-\bar{y}_0)d\theta_1d\theta_0.$$
We seek conditions under which, for fixed $e$, %and $\bar{y}_0$,
\begin{equation}\label{eq:diff_func}
\xi(e,n+1) - \xi(e,n) = \int_{\bm{\theta} \in H_0} f(\bm{\theta}|\bar{y}_0, e,n)d\bm{\theta} - \int_{\bm{\theta} \in H_0} f(\bm{\theta}|\bar{y}_0, e,n+1)d\bm{\theta} \geq 0.
\end{equation}

Since $1 - \xi(e,n)$ involves the p.d.fs of two beta distributions, we rewrite \eqref{eq:diff_func} as
$$\xi(e,n+1) - \xi(e,n) =$$
\begin{equation}\label{eq:2arm_A_B_binary}
\begin{split}
    \iint_{\theta_1 - \theta_0 \leq \theta^*} \underbrace{\frac{\theta_1^{a_1+n(\bar{y}_0+e)-1}(1-\theta_1)^{b_1+n(1-\bar{y}_0-e)-1}\theta_0^{a_0+n\bar{y}_0-1}(1-\theta_0)^{b_0+n(1-\bar{y}_0)-1}}{C_3C_4}}_{{\cal{A}}_2}\times \\
    \left[\underbrace{\frac{C_3C_4}{C_1C_2} - \theta_1^{\bar{y}_0+e}(1-\theta_1)^{1-\bar{y}_0-e}\theta_0^{\bar{y}_0}(1-\theta_0)^{1-\bar{y}_0}}_{{\cal{B}}_2}\right]d\theta_1d\theta_0,
\end{split}
\end{equation}
where 
$$C_1 = B(a_1+n(\bar{y}_0+e),b_1+n(1-\bar{y}_0-e)), \;\; C_2 = B(a_0+n\bar{y}_0,b_0+n(1-\bar{y}_0)),$$
$$C_3 = B(a_1+(n+1)(\bar{y}_0+e),b_1+(n+1)(1-\bar{y}_0-e)), \;\; C_4 = B(a_0+(n+1)\bar{y}_0,b_0+(n+1)(1-\bar{y}_0)).$$
Similar to Section \ref{subsec:proof_prop1}, because ${\cal{A}}_2$ is positive for all $\theta_1$, $\theta_0$, and $n$, inequality \eqref{eq:diff_func} holds if ${\cal{B}}_2$ is non-negative. That is, if $\cB_2 \geq 0$,
$$ \iint_{\theta_1 - \theta_0 \leq \theta^*} {\cal A}_2\cdot \cB_2 \; d\theta_1d\theta_0 \geq \int_{\theta_1 - \theta_0 \leq \theta^*} {\cal A}_2 \cdot 0 \; d\theta_1d\theta_0 = 0.$$
Hence, we begin by analyzing $\cB_2$ with respect to $\theta_1$, $\theta_0$, and $n$. Note that for fixed $\bar{y}_0$ and $e$, $\cB_2$ in Equation \ref{eq:2arm_A_B_binary} is a function of $n$, $\theta_1$, and $\theta_0$, and is denoted by $\cB_2(n,\theta_1,\theta_0)$.

We begin by examining the dependence of $\cB_2(n,\theta_1,\theta_0)$ on $(\theta_1,\theta_0)$ for fixed $n$. As the term $\frac{C_3C_4}{C_1C_2}$ is finite, positive, and does not depend on $(\theta_1,\theta_0)$, it is equivalent to analyze
$$h(\theta_1,\theta_0) = \theta_1^{\bar{y}_0+e}(1-\theta_1)^{1-\bar{y}_0-e}\theta_0^{\bar{y}_0}(1-\theta_0)^{1-\bar{y}_0}.$$
It is straightforward to see that $h(\theta_1,\theta_0)$ is separable. Therefore, we analyze this function with respect to $\theta_1$ and $\theta_0$.
The first and second derivatives of %$\cB_2(n,\theta_1,\theta_0)$ 
$\log h(\theta_1,\theta_0)$ with respect to $\theta_1$ and $\theta_0$ are given by:
\begin{align*}
\frac{\partial \log h}{\partial \theta_1} = \frac{\bar{y}_0+e}{\theta_1} - \frac{1 - \bar{y}_0 - e}{1 - \theta_1},
%\theta_0^{\bar{y}_0}(1-\theta_0)^{1-\bar{y}_0}\theta_1^{\bar{y}_0+e-1}(1-\theta_1)^{-(\bar{y}_0+e)}(\theta_1 - (\bar{y}_0+e)), 
\\
\frac{\partial^2 \log h}{\partial \theta_1^2} = - \frac{\bar{y}_0+e}{\theta_1^2} - \frac{1 - \bar{y}_0 - e}{(1 - \theta_1)^2} < 0,
%\theta_0^{\bar{y}_0}(1-\theta_0)^{1-\bar{y}_0}(1-(\bar{y}_0+e))(\bar{y}_0+e)\theta_1^{\bar{y}_0+e-2}(1-\theta_1)^{-1-(\bar{y}_0+e)} > 0,
\\
\frac{\partial \log h}{\partial \theta_0} = \frac{\bar{y}_0}{\theta_0} - \frac{1 - \bar{y}_0}{1 - \theta_0},%\theta_1^{\bar{y}_0+e}(1-\theta_1)^{1-\bar{y}_0-e}\theta_0^{\bar{y}_0-1}(1-\theta_0)^{-\bar{y}_0}(\theta_0 - \bar{y}_0), 
\text{ and } \\
\frac{\partial^2 \log h}{\partial \theta_0^2} = - \frac{\bar{y}_0}{\theta_0^2} - \frac{1 - \bar{y}_0}{(1 - \theta_0)^2} < 0.
%\theta_1^{\bar{y}_0+e}(1-\theta_1)^{1-\bar{y}_0-e}(1-\bar{y}_0)\bar{y}_0\theta_0^{\bar{y}_0-2}(1-\theta_0)^{-1-\bar{y}_0} > 0.
\end{align*}
Moreover, it is straightforward to see the the Hessian is negative definite. Therefore, $h(\theta_1,\theta_0)$ is strictly log-concave with a global unique maximizer at $(\theta_1, \theta_0) = (\bar{y}_0+e,\bar{y}_0)$ \citep{boyd2004convex}. Consequently,
$\cB_2$ has the following properties:
\begin{itemize}
    \item 1) For any fixed $\theta_1$ and $n$, $\cB_2(n,\theta_1,\theta_0)$ is a strict convex function of $\theta_0$, attaining its global minimum at $\theta_0 = \bar{y}_0$; it decreases for $\theta_0 \in [0,\bar{y}_0)$ and increases for $\theta_0 \in (\bar{y}_0,1]$.
    \item 2) For any fixed $\theta_0$ and $n$, $\cB_2(n,\theta_1,\theta_0)$ is a strict convex function of $\theta_1$, attaining its global minimum at $\theta_1 = \bar{y}_0+e$; it decreases for $\theta_1 \in [0,\bar{y}_0+e)$ and increases for $\theta_1 \in (\bar{y}_0+e,1]$.
\end{itemize}
And $\cB_2(n,\theta_1,\theta_0)$ achieves a unique global minimum at $(\theta_1, \theta_0) = (\bar{y}_0 + e, \bar{y}_0)$. %and $\theta_1 = \bar{y}_0+e$.  

We then examine the dependence of $\cB_2(n,\theta_1,\theta_0)$ on $n$, holding $\theta_1$ and $\theta_0$ fixed. Since $h(\theta_1,\theta_0)$ %The term 
%$$-\theta_1^{\bar{y}_0+e}(1-\theta_1)^{1-\bar{y}_0-e}\theta_0^{\bar{y}_0}(1-\theta_0)^{1-\bar{y}_0}$$ 
%in $\cB_2$ 
does not depend on $n$. Thus, the $n$-dependence of $\cB_2$ arises entirely from the factor $\frac{C_3C_4}{C_1C_2}$, where
$$C_1 = B(a_1+n(\bar{y}_0+e),b_1+n(1-\bar{y}_0-e)), $$
$$C_2 = B(a_0+n\bar{y}_0,b_0+n(1-\bar{y}_0)),$$
$$C_3 = B(a_1+(n+1)(\bar{y}_0+e),b_1+(n+1)(1-\bar{y}_0-e)), $$
$$C_4 = B(a_0+(n+1)\bar{y}_0,b_0+(n+1)(1-\bar{y}_0)).$$
From Lemma~\ref{lemma:conv_log_beta} and Lemma~\ref{lemma:eq_cond}, both ratios $C_3/C_1$ and $C_4/C_2$ are positive and strictly increasing in $n$. Consequently, $\frac{C_3C_4}{C_1C_2}$ is strictly increasing in $n$ (for fixed $\bar{y}_0$ and $e$). Moreover, a direct calculation shows that 
$$\lim_{n \rightarrow \infty} \frac{C_3C_4}{C_1C_2} = {(\bar{y}_0+e)}^{\bar{y}_0+e}(1-\bar{y}_0-e)^{1-\bar{y}_0-e}{\bar{y}_0}^{\bar{y}_0}(1 - \bar{y}_0)^{1-\bar{y}_0},$$
which is a finite, positive constant. Hence, we have %Since the other term in $\cB_2$ does not depend on $n$, it follows that
$$\lim_{n \rightarrow \infty}\cB_2(n,\bar{y}_0+e,\bar{y}_0) = $$
$${(\bar{y}_0+e)}^{\bar{y}_0+e}(1-\bar{y}_0-e)^{1-\bar{y}_0-e}{\bar{y}_0}^{\bar{y}_0}(1 - \bar{y}_0)^{1-\bar{y}_0} - {(\bar{y}_0+e)}^{\bar{y}_0+e}(1-\bar{y}_0-e)^{1-\bar{y}_0-e}{\bar{y}_0}^{\bar{y}_0}(1 - \bar{y}_0)^{1-\bar{y}_0} = 0,$$
and
%Use the same argument as above, these results further imply that, as $n \rightarrow \infty$, 
$$\lim_{n \rightarrow \infty}\cB_2(n,\theta_1,\theta_0) 
= {(\bar{y}_0+e)}^{\bar{y}_0+e}(1-\bar{y}_0-e)^{1-\bar{y}_0-e}{\bar{y}_0}^{\bar{y}_0}(1 - \bar{y}_0)^{1-\bar{y}_0} - h(\theta_1, \theta_0)
> \lim_{n \rightarrow \infty}\cB_2(n,\bar{y}_0+e,\bar{y}_0)$$
for all $(\theta_1,\theta_0) \neq (\bar{y}_0+e,\bar{y}_0),$ where the inequality is due to $(\bar{y}_0+e,\bar{y}_0)$ being the unique global maximizer of $h(\theta_1,\theta_0)$. In other words, 
%This follows because 
%$$\lim_{n \rightarrow \infty} \frac{C_3C_4}{C_1C_2} =  {(\bar{y}_0+e)}^{\bar{y}_0+e}(1-\bar{y}_0-e)^{1-\bar{y}_0-e}{\bar{y}_0}^{\bar{y}_0}(1 - \bar{y}_0)^{1-\bar{y}_0}$$ 
%is a positive, finite constant. Using the same convexity and uniqueness arguments as before, we conclude that 
$$\lim_{n\rightarrow \infty}\cB_2(n,\theta_1,\theta_0)$$ 
%as a function of $(\theta_1,\theta_0)$, 
attains its unique global minimum at $(\theta_1, \theta_0) = (\bar{y}_0+e, \bar{y}_0)$, with a value of $0$, and for any $(\theta_1,\theta_0) \neq (\bar{y}_0+e,\bar{y}_0)$, the limit is strictly positive.

Next, we show that for any $\theta^* < e$, there exists an integer $n_{\text{min}}'$ such that for all $n \geq n_{\text{min}}'$, $\cB_2(n, \theta_1,\theta_0) > 0$ for every $(\theta_1,\theta_0)$ satisfying $\theta_1 - \theta_0 \leq \theta^*$.

First, %we fix $\theta_0 = \bar{y}_0$. %By definition, $\cB_2(n,\theta_1 = 0,\theta_0 = \bar{y}_0) = \frac{C_3C_4}{C_1C_2} > 0$. %Furthermore, as $\cB$ is a convex function, $\cB(n,\theta_1, \theta_0 = \bar{y}_0)$ is minimized at $\theta_1 = \bar{y}_0+e$, where $\cB(n,\theta_1,\theta_0 = \bar{y}_0)$ decreases monotonically for $\theta_1 \in [0,\bar{y}_0+e]$, and increases monotonically for $(\bar{y}_0+e,1]$. 
suppose there exists an integer $n_{\text{min}}'$ such that $$\cB_2(n_{\text{min}}',\bar{y}_0+e,\bar{y}_0) \geq 0.$$ 
Since %we have found the 
%$n_{\text{min}}'$ meets our requirement. Indeed, because 
$\cB_2$ is strictly convex in $(\theta_1, \theta_0)$ with its minimum at $(\bar{y}_0+e, \bar{y}_0)$, it follows that for any
%This is because for any 
%$n \geq n_{\text{min}}'$ and any 
$(\theta_1', \theta_0') \neq (\bar{y}_0+e,\bar{y}_0)$, 
%$$\cB_2(n,\theta_1 = \theta_1',\theta_0 = \bar{y}_0) \geq 
$$\cB_2(n_{\text{min}}', \theta_1', \theta_0') > \cB_2(n_{\text{min}}', \bar{y}_0+e, \bar{y}_0) \geq 0,$$ 
%Similarly, because $\cB_2$ is convex in $\theta_0$ with its minimum at $\theta_0 = \bar{y}_0$, we have for any $n \geq n_{\text{min}}'$ and any $\theta_0' \neq \bar{y}_0$,
%$$\cB_2(n, \theta_1', \theta_0') \geq \cB_2(n_{\text{min}}', \theta_1', \theta_0') \geq \cB_2(n_{\text{min}}', \theta_1', \bar{y}_0) \geq 0.$$
%Combining these inequalities, \yy we conclude that \bb %shows that 
%for any $n \geq n_{\text{min}}'$ and all $\theta_1,\theta_0 \in [0,1]$, 
%$$\cB_2(n,\theta_1,\theta_0) \geq 0,$$ 
and therefore \eqref{eq:diff_func} holds in this case.

%since $\cB(n,\theta_1,\theta_0)$ is increasing in $n$ and is a convex function with global minimum at $\theta_1 = \bar{y}_0+e$ and $\theta_0 = \bar{y}_0$. Thus, for all $n \geq n_{\text{min}}'$, 
%$$\eqref{eq:2arm_A_B_binary} = \iint_{\theta_1-\theta_0 \leq \theta^*} {\cal A} \cdot \cB(n,\theta_1,\theta_0) \; d\theta_1d\theta_0 \geq \iint_{\theta_1-\theta_0 \leq \theta^*} {\cal A} \cdot \cB(n,\bar{y}_0+e,\bar{y}_0) \; d\theta_1d\theta_0 $$
%$$\geq
%\iint_{\theta_1-\theta_0 \leq \theta^*} {\cal A} \cdot 0 \; d\theta_1d\theta_0 = 0.$$

Second, if for all finite $n$, 
$${\cal{B}}_2(n,\bar{y}_0+e,\bar{y}_0) < 0,$$ 
we invoke the following lemma.

\bigskip

\begin{lemma}\label{lemma:2arm_bin_step}
    Let $\cB_2$ be as defined in \eqref{eq:2arm_A_B_binary}, viewed as a function of $(n,\theta_1,\theta_0)$. %i.e., $\cB_2(n,\theta_1,\theta_0)$.
    If $\cB_2(n,\bar{y}_0+e,\bar{y}_0) < 0$ for all finite sample sizes $n$, then for each $n$ there exists a constant $\theta_n^* < e$ such that
    $${\cal{B}}_2(n,\theta_1,\theta_0) \geq 0 \text{ for all }(\theta_1,\theta_0) \text{ satisfying } \theta_1 - \theta_0 \leq \theta_n^*, \, \theta_1 \in [0,1], \text{ and } \theta_0 \in [0,1].$$
     Moreover, the sequence $(\theta_n^*)_{n \geq 1}$ is monotonically increasing in $n$, and %monotonically with respect to $n$, and %when $n \rightarrow \infty$, 
    $$\lim_{n \rightarrow \infty}\theta_n^* \rightarrow e.$$%, consequently,
    %$$\lim_{n \rightarrow \infty}\cB_2(n,\theta_1,\theta_0) \geq 0 \text{ for all } (\theta_1,\theta_0) \text{ satisfying } \theta_1 - \theta_0 \leq e, \, \theta_1 \in [0,1], \text{ and } \theta_0 \in [0,1].$$
    %\in \{(\theta_1,\theta_0);\theta_1 - \theta_0 \leq \theta^*, \theta_1 \in [0,1], \text{ and } \theta_0 \in [0,1]\}.$$
\end{lemma}

\paragraph{Proof of Lemma \ref{lemma:2arm_bin_step}:}
%To establish the existence of a $\theta^* < e$ for any fixed $n$, 
We 
begin by showing that for any finite $n$, if ${\cal{B}}_2(n,\bar{y}_0+e,\bar{y}_0) < 0$, then there exists a closed contour in the $(\theta_1,\theta_0)$-plane defined by ${\cal{B}}_2(n,\theta_1,\theta_0) = 0$. 
%, or equivalently, there exists values of $(\theta_1,\theta_0)$ such that
%$$\frac{C_3C_4}{C_1C_2} = \theta_1^{\bar{y}_0+e}(1-\theta_1)^{1-\bar{y}_0-e}\theta_0^{\bar{y}_0}(1-\theta_0)^{1-\bar{y}_0}.$$
%Using the 2-dimensional Intermediate value theorem (2d-IVT), %Not fixing $\theta_0 = \bar{y}_0$, from \eqref{eq:2arm_A_B_binary}, 

First, note that if $\theta_1 \in \{0,1\}$ or %regardless of $\theta_0$, and similarly when 
$\theta_0 \in \{0, 1\}$, we have
$${\cal{B}}_2(n,\theta_1,\theta_0) = \frac{C_3C_4}{C_1C_2} > 0.$$  %, regardless of $\theta_1$. 
Moreover, for any fixed $n$, the following properties hold: %$\cB_2(n,\theta_1,\theta_0) = \frac{C_3C_4}{C_1C_2} > 0$.
%Moreover, for any fixed $n$, we have 
\begin{enumerate}
    \item ${\cal{B}}_2(n,\theta_1,\theta_0)$ is continuous on $%(\theta_1,\theta_0) \in 
    [0,1]\times[0,1]$,
    \item On the boundary of the unit square -- i.e., when $\theta_1 = 0$ or $1$ (for any $\theta_0 \in (0,1)$) or when $\theta_0 = 0$ or $1$ (for any $\theta_1 \in (0,1)$) -- we have $\cB_2(n,\theta_1,\theta_0) > 0$. %for any values of $\theta_0 \in (0,1)$, and when $\theta_0 = 0$ or $1$, we also have $\cB_2(n,\theta_1,\theta_0) = \frac{C_3C_4}{C_1C_2} > 0$ for any values of $\theta_1 \in (0,1)$.%For any fixed $\theta_1$, $\cB_2(n, \theta_1,\theta_0)$ is monotonically decreasing in $\theta_0 \in [0,\bar{y}_0)$ and increasing in $\theta_0 \in [\bar{y}_0,1]$,
    \item By assumption, $\cB_2(n,\bar{y}_0+e,\bar{y}_0) < 0.$ %for any finite $n$.%For any fixed $\theta_0$, $\cB_2(n, \theta_1,\theta_0)$ is monotonically decreasing in $\theta_1 \in [0,\bar{y}_0+e)$ and increasing in $\theta_1 \in [\bar{y}_0+e,1]$.
\end{enumerate}
Let 
%By the two-dimensional Intermediate Value Theorem (2D-IVT), there must exist one point $(\theta_1',\theta_0')$ between each of the boundary points in %pair of points with the first point being one point from the set
$$\mathcal{BD} = \{(0,\theta_0);\theta_0 \in [0,1]\} \cup \{(1,\theta_0);\theta_0 \in [0,1]\} \cup \{(\theta_1,0);\theta_1 \in [0,1]\} \cup \{(\theta_1,1);\theta_1 \in [0,1]\}$$ 
denote the boundary of $[0,1]^2$.
By the two-dimensional Intermediate Value Theorem, for each boundary point $(\theta_{1},\theta_{0}) \in {\mathcal{BD}}$, there exists a continuous path to the interior point $(\bar{y}_0+e, \bar{y}_0)$ along which  ${\cal{B}}_2$ changes sign from positive (at the boundary) to negative (at the interior point). Consequently, there must exist at least one point $(\theta_1',\theta_0')$, called ``zero point", along this path such that $\cB_2(n,\theta_1',\theta_0') = 0$.

%\dd To see this, as $\cB_2$ is continuous in $\theta_1$ and $\theta_0$, there is always a pathway connecting any points in $\mathcal{BD}$, say a point named $(\theta_{1,1},\theta_{0,1})$, to the point $(\bar{y}_0+e,\bar{y}_0)$. As the first point is above zero while the last point is assumed to be less than zero, there has to have at least one intersection point on this pathway from $(\theta_{1,1},\theta_{0,1})$ to $(\bar{y}_0+e,\bar{y}_0)$, denoted as $(\theta_1',\theta_0')$, such that $\cB_2(n,\theta_1',\theta_0') = 0$. \bb
Because the boundary $\mathcal{BD}$ %that the set where the first point is picking from composed of infinitely many points, and 
encloses the unit square, there are infinitely many such zero points, forming a closed contour (level set) in the % $[0,1]\times[0,1]$, thus, there are infinitely many such zero points $(\theta_1',\theta_0')$, which forms a closed \yy contour \jj  (level set) in the 
$(\theta_1,\theta_0)$-plane. We denote this set by % points on this closed contour as
\begin{equation}\label{eq:set_co}
 \mathcal{CO}(n) = \{(\theta_1',\theta_0') \in (0,1)^2; \,\, \cB_2(n,\theta_1',\theta_0') = 0\}.    
\end{equation}

We now show that for any finite $n$, the contour $\mathcal{CO}(n)$ is a closed strictly convex curve. 
To see this, consider a straight line in the $(\theta_1,\theta_0)$-plane given by 
$$\theta_1 = {c_{s}}\cdot \theta_0 + {c_{i}},$$
where ${c_{s}} \in \mathbb{R}$ is the slope and ${c_{i}} \in \mathbb{R}$ %\in [-1,1]$ 
is the intercept. %Note that the intercept in $[-1,1]$ is because the line must pass through points inside the region of $[0,1]^2$. 
We examine the possible number of intersection points between this line and the contour of ${\mathcal{CO}}(n)$. %$$\cB_2(n,\theta_1,\theta_0) = 0$ (with $n$ fixed). 

Substituting $\theta_1' = {c_s}\cdot \theta_0' + {c_i}$ into $\cB_2(n, \theta_1',\theta_0') = 0$ and taking logarithms on both sides yields: %of the equation. This results in the expression
$$\log\left(\frac{C_3C_4}{C_1C_2}\right) = (\bar{y}_0+e)\log({c_s}\cdot\theta_0'+{c_i})+(1-\bar{y}_0-e)\log(1-({c_s}\cdot\theta_0'+{c_i})) + \bar{y}_0\log(\theta_0')+(1-\bar{y}_0)\log(1-\theta_0').$$
Differentiating the right-hand side twice with respect to $\theta_0'$ gives:
$$-\frac{(\bar{y}_0+e){c_s}^2}{({c_s}\cdot\theta_0'+{c_i})^2}-\frac{(1-\bar{y}_0-e){c_s}^2}{(1-({c_s}\cdot\theta_0'+c_i))^2}-\frac{\bar{y}_0}{\theta_0'^2} - \frac{1-\bar{y}_0}{(1-\theta_0')^2} < 0$$
for all $\theta_0' \in (0,1)$. 
Hence, the right-hand-side is strictly concave, which implies that there are at most two solutions for $\theta_0'$ to the equation $\cB_2(n,c_s\cdot\theta_0'+c_i,\theta_0') = 0$. In other words, any straight line intersects the contour at most twice. This implies that the %contour is a closed strictly convex curve in the %points between the line $\theta_1 = ce\cdot\theta_0+it$ and the contour $\cB_2(n,\theta_1,\theta_0) = 0$, or equivalently, the contour is closed and every intersection of the curve with a line consists of at most two points. This means that %Moreover, let $it = e$, in which case the line $\theta_1 = ce\cdot\theta_0+e$ passing through the point $(\bar{y}_0+e,\bar{y}_0)$ that is inside the contour, using the 2D-IVT again, it is straightforward to see that there are exactly two intersections between $\cB_2(n,\theta_1,\theta_0) = 0$ and the line $\theta_1 = ce\cdot\theta_0+e$. In fact, it is straightforward to see that for any line represented by $\theta_1 = ce\cdot\theta_0+it$, it the line passes through the contour, there are exactly two intersections between $\cB_2(n,\theta_1,\theta_0) = 0$ and the line $\theta_1 = ce\cdot\theta_0+it$. Consequently, geometrically, this 
contour is a closed strictly convex curve in the $(\theta_1,\theta_0)$-plane \citep{schneider2013convex}.

A closed strictly convex curve has the following properties \citep{eggleston1966convexity, schneider2013convex}:
\begin{itemize}
    \item[(1)] contains no straight-line segments, 
    \item[(2)] every tangent line touches the curve at exactly one point,
    \item[(3)] has strictly positive curvature everywhere,
    \item[(4)] has a unique supporting line at every point, and
    \item[(5)] is simple and divides the plane into interior and an exterior regions.
\end{itemize}
Denote ${\mathcal{CO}}^{-}(n)$ and ${\mathcal{CO}}^{+}(n)$ the sets for all points in the interior and the exterior regions of the contour ${\mathcal{CO}}(n)$.
From (5), all points $(\theta_1,\theta_0) \in {\mathcal{CO}}^{-}(n)$ satisfy %${\mathcal{CO}}^{-}(n)$ denotes the set of points that are inside the contour defined by $\cB_2(n,\theta_1,\theta_0) = 0$, 
${\cal{B}}_2(n,\theta_1,\theta_0) < 0$. %Note that 
By our assumption that %for any finite $n$, 
$\cB_2(n,\bar{y}_0+e,\bar{y}_0) < 0$ for all finite $n$, the point $(\bar{y}_0+e,\bar{y}_0)$ lies inside the contour for all %is always inside the contour for any 
finite $n$. Similarly, 
%inside this \yy contour, \jj  we have ${\cal{B}}_2(n,\theta_1,\theta_0) < 0$; \bb and 
all points $(\theta_1,\theta_0) \in {\mathcal{CO}}^{+}(n)$ satisfy %${\mathcal{CO}}^{+}(n)$ denotes the set of points that are outside the contour defined by $\cB_2(n,\theta_1,\theta_0) = 0$, 
${\cal{B}}_2(n,\theta_1,\theta_0) > 0$. 

\begin{figure}
    \centering
    \includegraphics[width=\linewidth]{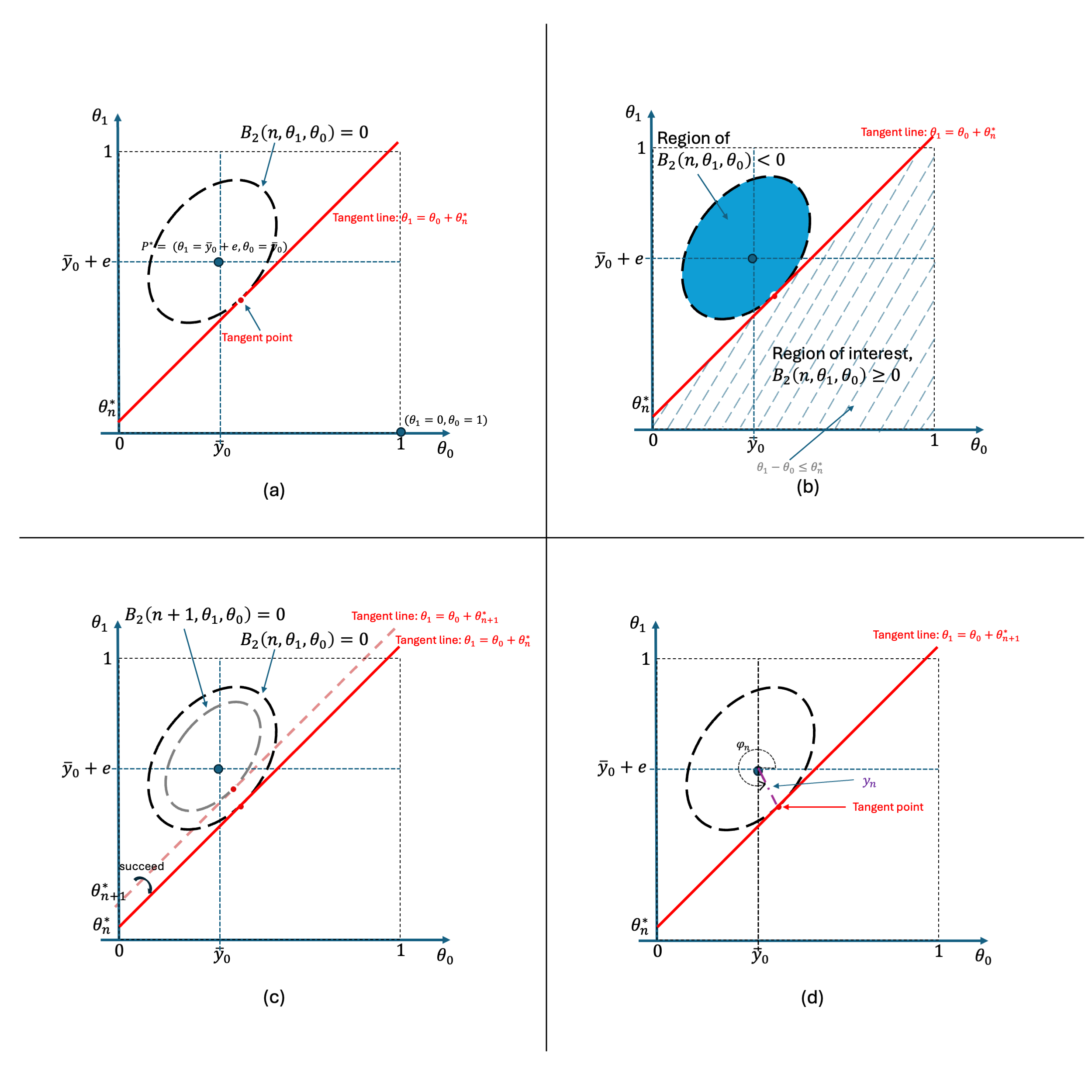}
    \caption{Illustration of the contour defined by $\cB_2(n,\theta_1,\theta_0) = 0$: (a) Tangent line $\theta_1 = \theta_0 + \theta_n^*$ touching the contour at the tangent point. (b) Region of interest $\{(\theta_1,\theta_0);\theta_1 - \theta_0 \leq \theta_n^*, \theta_1 \in (0,1), \theta_0 \in (0,1)\}$, shown as the blue dashed area; the blue shaded region denotes $\cB_2(n,\theta_1,\theta_0) < 0$. 
    (c) Shrinkage of the contour as $n$ increases, with the tangent line $\theta_1 = \theta_0 + \theta_{n+1}^*$ (dashed light red line) ``succeeding" the previous tangent line $\theta_1 = \theta_0 + \theta_{n}^*$ (solid red line), where $\theta_{n+1}^* > \theta_n^*$. (d) Radius $y_n$ (purple dashed line) with angle $\varphi_n$. In all panels, the black contour is the contour and red lines are tangent lines. }
    %The black dashed eclipse-like shape is the contour and the red line is the tangent line to the contour. The blue dashed region represents the region of interest, the blue shaded region represents a region where all $(\theta_1,\theta_0)$ values inside the region results in $\cB_2(n,\theta_1,\theta_0) < 0$. The gray dashed eclipse-like shape and the light red dashed line are the contour of $\cB_2(n+1,\theta_1,\theta_0) = 0$ and the tangent line $\theta_1 = \theta_0 + \theta_{n+1}^*$, the orange ball is the Euclidean ball, the purple dashed line is the Euclidean distance from the center $P^*$ to the tangent point, and the black dashed line is the shortest distance from $P^*$ to the tangent line that is perpendicular to the tangent line, respectively. We call the red dashed line being ``succeeded" by the light red dashed line in (c), where $\theta_{n+1}^* > \theta_n^*$. }
    \label{fig:contour_illu}
\end{figure}

Moreover, by properties (1) - (4), there exists a tangent line of the form 
$$\theta_1=\theta_0+c_i^{(n)},$$ 
where $c_i^{(n)}$ is the intercept value at a given $n$, that touches the contour $\cB(n,\theta_1,\theta_0) = 0$ at a point whose $\theta_0$-coordinate is greater than $\bar{y}_0$ %cross the line defined by $\theta_1 = -\frac{\bar{y}_0+e}{1-\bar{y}_0}\theta_0 + \frac{\bar{y}_0+e}{1-\bar{y}_0}$ 
(see Figure \ref{fig:contour_illu}a).
For each $n$, denote this intercept value by $\theta_n^*$, i.e., $\theta_n^* = c_i^{(n)}$. All points $(\theta_1,\theta_0)$ satisfying
$$\theta_1 - \theta_0 \leq \theta_n^*, \; \theta_1 \in (0,1), \, \theta_0 \in (0,1)$$
are in the exterior region of the contour, 
as illustrated in Figure \ref{fig:contour_illu}b,  hence, 
$\cB_2(n,\theta_1,\theta_0) \geq 0$. 
This establishes the first claim of the Lemma.

Next, we show that the contour shrinks as $n$ increases. By ``shrinking," we mean that increasing the sample size from $n$ to $n+1$ results in the new contour being strictly inside the previous one (see Figure \ref{fig:contour_illu}c). Mathematically, this results in
$$\cB_2(n+1,\theta_1,\theta_0) > 0$$ 
for all $(\theta_1,\theta_0) \in {\mathcal{CO}}^+(n) \cup {\mathcal{CO}}(n)$, recalling that $\mathcal{CO}(n)$ and $\mathcal{CO}^{+}(n)$ are the sets of points that are in the exterior and on the contour defined by $\cB_2(n,\theta_1,\theta_0) = 0$. 

We prove this by contradiction. Suppose there exists a point $(\theta_1',\theta_0') \in {\mathcal{CO}}^+(n) \cup {\mathcal{CO}}(n)$ such that 
$$\cB_2(n,\theta_1',\theta_0') \geq 0 \text{ but } \cB_2(n+1,\theta_1',\theta_0') \leq 0.$$
Since $\cB_2(n,\theta_1,\theta_0)$ is strictly increasing in $n$ for any fixed $(\theta_1,\theta_0)$, we have: %there exists one such point at sample size $n + 1$, where 
\begin{itemize}
    \item if $\cB_2(n,\theta_1',\theta_0') = 0$, then  $\cB_2(n+1,\theta_1',\theta_0') > 0$;
    \item if $\cB_2(n,\theta_1',\theta_0') > 0$, then $\cB_2(n+1,\theta_1',\theta_0') > 0$ as well.
\end{itemize}
In both cases,
%Recall that $\cB_2(n,\theta_1,\theta_0)$ strictly increases with $n$ for any fixed $\theta_1$ and $\theta_0$. Hence, if $\cB_2(n,\theta_1',\theta_0') = 0$, we have $\cB_2(n+1,\theta_1',\theta_0') > 0$. Furthermore, if $\cB_2(n,\theta_1',\theta_0') > 0$, we must also have $\cB_2(n+1,\theta_1',\theta_0') > 0$. Consequently, when 
$\cB_2(n,\theta_1',\theta_0') > 0$, %we must have  $\cB_2(n+1,\theta_1',\theta_0') > 0$. This 
contradicting the claim of $\cB_2(n+1,\theta_1',\theta_0') \leq 0$. 
%that $\cB_2(n+1,\theta_1',\theta_0') \leq 0$. Consequently, 
Therefore, when $n$ increases to $n+1$, every point on or outside the contour of $\mathcal{CO}(n)$ has a strictly positive $\cB_2$ value at $n+1$. This implies that % $\cB_2(n+1,\theta_1,\theta_0) > 0$. Hence, the points in the set 
$\mathcal{CO}(n+1)$ consists of points that previously lay inside ${\mathcal{CO}}(n)$, meaning %the contour. Consequently, 
the contour $\cB_2(n,\theta_1,\theta_0) = 0$ strictly encloses %the contour of 
$\cB_2(n+1,\theta_1,\theta_0) = 0$. %And it is straightforward to verify that an increase in $n$ results in a changed (increased) value of $\frac{C_3C_4}{C_1C_2}$ for any finite $n$, i.e., 
In other words, the contour shrinks as $n$ increases. 

We then show that the contour shrinks to a single point as $n \rightarrow \infty$. Recall that ${\mathcal{CO}}(n)$ denotes the set of all points on the contour $\cB_2(n,\theta_1,\theta_0) = 0$. We place the origin of a polar coordinate system at 
$$P^* = (\bar{y}_0+e,\bar{y}_0).$$ 
For each angle $\varphi \in [0,2\pi)$, the intersection of ${\mathcal{CO}}(n)$ with the ray from $P^*$ at angle $\varphi$ defines a radius $r_n(\varphi)$ such that 
$$P_n=P^* + r_n(\varphi)\nu(\varphi),$$ 
where $\nu(\varphi)$ is the unit vector in direction $\varphi$. 
From the result above, since ${\mathcal{CO}}(n+1)$ lies strictly inside ${\mathcal{CO}}(n)$, we have
$$r_{n+1}(\varphi) < r_n(\varphi) \,\, \forall \, \varphi, \, \forall \, n.$$
Thus, for each fixed $\varphi$, $r_n(\varphi)$ is strictly decreasing, nonnegative sequence, and therefore
$$\lim_{n \rightarrow \infty} r_n(\varphi) = r_{\infty}(\varphi) \geq 0.$$
Previously, we showed that 
$$\lim_{n \rightarrow \infty}\cB_2(n,\bar{y}_0+e,\bar{y}_0) = 0 \text{ and } \lim_{n \rightarrow \infty}\cB_2(n,\theta_1,\theta_0) > 0$$ 
for all $(\theta_1,\theta_0) \neq (\bar{y}_0+e,\bar{y}_0)$. This implies that, in the limit, the contour ${\mathcal{CO}}(n)$ collapses to $P^*$. 
Equivalently,
for any arbitrary radius $\epsilon_{\varphi} > 0$, there exists $N$ such that for all $n > N$, $$\sup_{\varphi}r_n(\varphi) < \epsilon_{\varphi}.$$ 
As $\epsilon_{\varphi}$ is arbitrary, we have $r_{\infty}(\varphi) = 0$ for all $\varphi$, and the limiting set is precisely the singleton $\{P^*\}$. 

A direct consequence of the shrinking of the contour with respect to $n$ is the sequence $(\theta_n^*)$ increases with $n$. In other words, recall that $\theta_1 = \theta_0 + \theta_n^*$ is the tangent line to the contour $\mathcal{CO}(n)$. When $n$ increases to $n+1$, there exists a parallel tangent line $\theta_1 = \theta_0 + \theta_{n+1}^*$ that ``succeeds" the previous tangent line $\theta_1 = \theta_0 + \theta_n^*$ (see Figure \ref{fig:contour_illu}c), with
$$\theta_n^* < \theta_{n+1}^*.$$

To see this, we first note that $\theta_n^* \neq \theta_{n+1}^*.$ If so, this means the tangent line $\theta_1 = \theta_0 + \theta_n^*$ coincides with the tangent line $\theta_1 = \theta_0 + \theta_{n+1}^*$. Let $(\theta_1',\theta_0')$ be the tangent point to the contour ${\mathcal{CO}}(n)$ and on the line $\theta_1 = \theta_0 + \theta_n^*$. By definition, $\cB_2(n, \theta_1', \theta_0') = 0$, and all other points $(\theta_1,\theta_0)$ on $\theta_1 = \theta_0 + \theta_n^*$ results in $\cB_2(n, \theta_1, \theta_0) > 0$. As $\cB_2$ strictly increases with $n$ for any fixed $(\theta_1,\theta_0)$, the point $(\theta_1',\theta_0')$ cannot be the tangent point on the contour $\cB_2(n+1, \theta_1, \theta_0) = 0$ because we must have $\cB_2(n+1, \theta_1', \theta_0') > 0$. There are also no other points on the line $\theta_1 = \theta_0 + \theta_n^*$ that can be the tangent point on the contour $\cB_2(n+1, \theta_1, \theta_0) = 0$, as it violates the property where $\cB_2$ strictly increases with $n$. Hence, $\theta_n^* \neq \theta_{n+1}^*$. Next, we show that it is also impossible to have $\theta_n^* > \theta_{n+1}^*$ using contradiction.

Suppose there exists a $n$ such that $\theta_n^* > \theta_{n+1}^*$, i.e., $\theta_{n+1}^* = \theta_{n}^* - \delta$, where $\delta > 0$. Let $(\theta_0' + \theta_{n+1}^*, \theta_0')$ be the tangent point to the contour $\cB_2(n+1, \theta_1, \theta_0) = 0$. In this case, we must have have $\cB_2(n+1, \theta_0’ + \theta_{n+1}^*, \theta_0’) = 0$. As $\cB_2$ strictly increase with $n$, this means that $\cB_2(n, \theta_0’ + \theta_{n+1}^*, \theta_0’) < 0$. However, the point $(\theta_0’ + \theta_{n+1}^*, \theta_0’)$ can be re-written as  $(\theta_0’ + \theta_{n}^* - \delta, \theta_0’)$, i.e., it lies in the region of $\theta_1 \leq \theta_0 + \theta_n^*$. Recall that previously, we established that all points in the region $\theta_1 \leq \theta_0 + \theta_n^*$ a positive, thus, we must have $\cB_2(n, \theta_0’ + \theta_{n}^* - \delta, \theta_0’) = \cB_2(n, \theta_0’ + \theta_{n+1}^*, \theta_0’) \geq 0$, a contradiction. Hence, we must have $\theta_n^* < \theta_{n+1}^*$ for all $n$.   

Finally, a direct consequence of the contour monotonically shrinking to $P^*$ as $n \rightarrow \infty$ is 
$$\lim_{n \rightarrow \infty}\theta_n^* = e.$$
To see this, recall the polar coordinate. For any fixed $n$, let $P_n^*$ be the polar coordinate of the tangent point for sample size $n$, $\varphi_n$ be the angle, and $y_n$ be the radius (the distance from the tangent point $P_n^*$ to $P^*$, see Figure \ref{fig:contour_illu}d). Since previously, we have shown that $r_{\infty}(\varphi) = 0$ for all $\varphi$, this implies that
$$y_{\infty} = r_\infty(\varphi_{\infty}) = 0.$$
Consequently, the tangent line $\theta_1 = \theta_0 + \theta_{\infty}^*$ must pass through the point $P^* = (\bar{y}_0+e,\bar{y}_0).$ Further recall that the set $\mathcal{CO}(\infty) = \{P^*\}$, $P^*$ is also the tangent point. In this case, we have 
$\bar{y}_0 + e = \bar{y}_0 + \theta_{\infty}^*,$
or equivalently, 
$$\lim_{n \rightarrow \infty} \theta_n^* = \theta_{\infty}^* = e.$$
\hfill $\square$

\bigskip

Using Lemma~\ref{lemma:2arm_bin_step}, we conclude that for any fixed $n$, there exists a $\theta_n^* < e$ such that $\cB_2(n,\theta_1,\theta_0) \geq 0$ for all $(\theta_1,\theta_0)$ satisfying $\theta_1 - \theta_0 \leq \theta_n^*$. In other words, for the countable set of sample sizes $n \in \{1, 2, \ldots, \infty\}$, there exists a corresponding countable set of thresholds $\Theta^* = \{\theta^*_1, \theta^*_2, \ldots, e\}$, with each $\theta_n^* < e$ denoting the largest value such that $\cB_2(n,\theta_1,\theta_0) \geq 0$ whenever $\theta_1 - \theta_0 \leq \theta_n^*$.

Now, for any given threshold $\theta^*$, we can identify the smallest $\theta_n^* \in \Theta^*$, denoted as $\theta_{\min}^*$, such that $\theta_{\min}^* \geq \theta^*$. %if $(\theta_1, \theta_0)$ such that $\theta_1 - \theta_0 \leq \theta^*$, this pair of values must also satisfy $\theta_1 - \theta_0 \leq \theta_{\min}^*$. %Denote this value by ${\theta^*}' = \theta_n^*$, and 
Let $n_{\text{min}}'$ be the corresponding sample size of $\theta_{\min}^*$.
Since when $n \geq n_{\min}'$, $\cB_2(n,\theta_1,\theta_0) \geq 0$ for all values in the set of $\Theta_{s1} = \{(\theta_1,\theta_0) \in [0,1]^2; \theta_1 - \theta_0 \leq \theta_{\min}^*\}$, we also have
$\cB_2(n,\theta_1,\theta_0) \geq 0$ for all values in the set of $\Theta_{s2} = \{(\theta_1,\theta_0) \in [0,1]^2; \theta_1 - \theta_0 \leq \theta^*\}$ when $n \geq n_{\min}'$. This is because $\Theta_{s2}$ is a subset of $\Theta_{s1}$ and $\cB_2(n,\theta_1,\theta_0)$ is an increasing function in $n$.
Hence, for all $n \geq n_{\text{min}}'$, we have
$$\int_{\bm{\theta} \in H_0}f(\bm{\theta}|e,n)d\bm{\theta} - \int_{\bm{\theta} \in H_0}f(\bm{\theta}|e,n+1)d\bm{\theta} \geq 0.$$
%where $H_0' = \{\theta_1 - \theta_0 \leq {\theta^*}'\}$. 

Finally, similar to Section \ref{subsec:proof_prop1}, it is possible that a smaller sample size $n_{\min} \leq n_{\min}'$ could still result in 
$$\int_{\bm{\theta} \in H_0}f(\bm{\theta}|e,n)d\bm{\theta} - \int_{\bm{\theta} \in H_0}f(\bm{\theta}|e,n+1)d\bm{\theta} \geq 0$$
for all $n \geq n_{\min}$, even if $\cB_2(n_{\min},\theta_1,\theta_0)$ is not strictly positive for all $(\theta_1,\theta_0)$ in the region of $\theta_1 - \theta_0 \leq \theta^*$.

$\hfill \square$

\bigskip

\bigskip
\newpage

\newpage

\subsection{Proof of Theorem 2:}

Let $n_{\text{max}}$ be the maximum sample size, and let %By Lemma~\ref{lemma:opt_y_bar_0}, for each sample size $n \in \{1, 2, \ldots, n_{\text{max}}\}$, the possible values of the optimizer $\bar{y}_0^*(n)$ lie in the set
%$${\cal S} = \left\{\min\left[\max\left(0,\frac{1-e}{2}+\frac{\beta-\alpha}{2n}\right),1-e\right];n = \{1, 2, \ldots, n_{\text{max}}\}\right\} \cup \{0,1-e\}.$$
$${\cal S} = \{\bar{y}_{0n}^*;n = 1, \ldots, n_{\max}\}$$
be the set that contains the minimizer of $\bar{y}_0$, i.e., 
$$\bar{y}_{0n}^* = \arg\min_{\bar{y}_0} \{\text{Pr}(H = H_1\mid \bar{y}_1,\bar{y}_0,n);\bar{y}_1 - \bar{y}_0 = e\},$$
for each $n$ from $n = 1$ to $n = n_{\max}$. 

Assume fixed $e$. Recall the notation 
$\xi = 1 - \int_{\bm{\theta} \in H_0'}f(\bm{\theta}\mid \bar{y}_0,e,n)d\bm{\theta}$, and denote $\xi(n,\bar{y}_0)$ as a function of both $n$ and $\bar{y}_0$.
By proposition 2, for a given $\theta^* < e$ and for each $\bar{y}_0 \in {\cal S}$, there exists a $n_{\text{min}}$ such that for all $n \geq n_{\text{min}}$,
$$\xi(n+1,\bar{y}_0) - \xi(n,\bar{y}_0) = \int_{\bm{\theta} \in H_0}f(\bm{\theta}\mid \bar{y}_0,e,n)d\bm{\theta} - \int_{\bm{\theta} \in H_0}f(\bm{\theta}\mid \bar{y}_0,e,n+1)d\bm{\theta} \geq 0.$$
%where $H_0' = \{\theta_1 - \theta_0 \leq {\theta^*}'\}$.
Define 
$$n_{\text{min}}^* = \max \{n_{\text{min}}; n_{\text{min}} \text{ is associated with each } \bar{y}_0 \in {\cal S}\}.$$
%and let ${\theta^*}'_{n_{\text{min}}^*}$ be the corresponding threshold at $n_{\min}^*$. %${\theta^*}'$.
Then, for all $n \in \{n_{\text{min}}^*, \ldots, n_{\text{max}}\}$, Proposition 2 implies:
$$\xi(n+1,\bar{y}_0^*(n+1)) \geq \xi(n,\bar{y}_0^*(n+1)).$$
Additionally, since $\bar{y}_0^*(n)$ is the optimizer at sample size $n$, we have:
$$\xi(n,\bar{y}_0^*(n+1)) \geq \xi(n,\bar{y}_0^*(n)).$$
Combining the two inequalities, we obtain:
$$\xi(n+1,\bar{y}_0^*(n+1)) \geq \xi(n,\bar{y}_0^*(n)).$$
Therefore, the posterior probability satisfies: 
$$\text{Pr}(H = H_1|e,n+1) \geq \text{Pr}(H = H_1|e,n)$$
for all $n \geq n_{\text{min}}^*$. 

%\yy
%Note that Theorem 2 may not hold if $\bar{y}_{0n}^*$ is not the minimizer. We again demonstrate this with a counter example. Let $\theta^* = 0.05$, $q = 0.5$, $e$ be fixed at 0.1, and $a_1 = b_1 = a_0 = b_0 = 0.5$. Suppose $\bar{y}_0 = 0.1$ and $n = 10$ ($n_{\min} = 1$), we have $\Pr(H = H_1|\bar{y}_0,e,n) = 0.66$. Now, when $n^\star = 15 > n$ but $\bar{y}_0^* = 0.5$ ($n_{\min} = 1$), we have $\Pr(H = H_1|\bar{y}_0^*,e,n^\star) = 0.65 < \Pr(H = H_1|\bar{y}_0,e,n) = 0.66$. Thus, although $n_{\min} = 1$, an increase in $n$ does not correspond to an increase in confidence in this case.

$\hfill \square$

\bigskip
\newpage

\subsection{Proof of Theorem 3}

%To mathematically prove that $\text{Pr}(H=H_1|e,n)$ increases monotonically with $n$ under a fixed evidence $e$ supporting $H_1$, it suffices to show that the integral term $\int_{\bm{\theta} \in H_1} f(\bm{\theta}|e,n)d\bm{\theta}$ in Equation (3) increases monotonically with $n$, given a fixed $e$ under certain conditions.
Based on Appendix Sections~\ref{subsec:pH1_onearm} and~\ref{subsec:pH1_twoarm_cont}, for trials with continuous outcomes in either one- or two-arm settings, Equation (9),
$$\xi = \int_{\bm{\theta}\in H_1} f(\bm{\theta}\mid e',n)d\bm{\theta}$$ %the integral in Equation (3) 
can be expressed as
%$$\int_{\theta^*}^{\infty} \frac{1}{\sqrt{\frac{2\pi}{\frac{1}{b}+\frac{n}{\sigma^2}}}}\exp{\left\{-\frac{1}{2}\left[\frac{1}{b}+\frac{n}{\sigma^2}\right]\left(\theta - \frac{1}{\frac{1}{b}+\frac{n}{\sigma^2}}\left[\frac{a}{b}+\frac{ne}{\sigma^2}\right]\right)^2\right\}} d\theta $$
$$\xi = 1 - \Phi(\theta';\mu_n,\sigma_n^2),$$
where $\Phi(\cdot|\mu_n,\sigma_n^2)$ is the CDF of a normal distribution with mean $\mu_n$ and variance $\sigma_n^2$, and
$$\mu_n = \frac{1}{\frac{1}{b}+\frac{n}{\sigma^2}}\left[\frac{a}{b}+\frac{ne'}{\sigma^2}\right],\;\; \sigma_n^2 =  \left[\frac{1}{b}+\frac{n}{\sigma^2}\right]^{-1}.$$
% is the CDF of a normal distribution with mean $\mu$ and variance $\Sigma$. 
An increase in $\xi$ with respect to $n$ corresponds to a decrease in the Normal CDF value, or equivalently, a decrease in the associated z-score: 
$$z(n) = \frac{\theta' - \mu_n}{\sqrt{\sigma_n^2}}.$$
Substituting the expressions for $\mu_n$ and $\sigma_n^2$, we get:
$$z(n) = 
%\frac{\theta^* - \frac{ne' + a\sigma^2/b}{\frac{\sigma^2}{b}+n}}{\frac{1}{\frac{1}{\sigma}\sqrt{\frac{\sigma^2}{b}+n}}} = 
\theta'\frac{1}{\sigma}\sqrt{A+n} - ne'\frac{1}{\sigma}\frac{1}{\sqrt{A + n}} - \frac{a\sigma}{b}\frac{1}{\sqrt{A+n}},$$
where $A = \frac{\sigma^2}{b}$. 

To determine when $z(n)$ decreases % conditions such that the z-score $z(n)$ decreases 
monotonically with $n$, we compute the derivatives:% $z(n)$ with respect to $n$. Specifically, we have
$$h(n) = \frac{d}{dn} z(n) = \frac{1}{\sigma\sqrt{A+n}}\left(\frac{\theta'}{2} - e' + \frac{ne'}{2(A+n)}+\frac{a\sigma^2}{2b(A+n)}\right).$$
To ensure $z(n)$ is strictly decreasing in $n$, we require $h(n) < 0$. Since the leading factor $\frac{1}{\sigma\sqrt{A+n}}$ is positive, this condition reduces to:
$$\frac{\theta'}{2} - e' + \frac{ne'}{2(A+n)} +\frac{a\sigma^2}{2b(A+n)}  < 0.$$
Now, note that %Because $A > 0$, we bound $\frac{ne'}{2(A+n)}$ as 
$$\frac{ne'}{2(A+n)} < \frac{e'}{2},$$ 
since $\frac{n}{A+n} < 1$. Therefore,
%Thus, we have
$$\frac{\theta'}{2} - e' + \frac{ne'}{2(A+n)} + \frac{a\sigma^2}{2b(A+n)} < \frac{\theta' - e'}{2} + \frac{a\sigma^2}{2b(A+n)} = \frac{\theta^* - e}{2} + \frac{a\sigma^2}{2b(A+n)},$$
recall that $\theta' = \theta^* + \theta_0$ and $e' = e + \theta_0$ in one-arm trial and $\theta' = \theta^*$ and $e' = e$ in two-arm trial.
Thus, a sufficient condition for $h(n) < 0$ is: 
$$e > \theta^* + \frac{a}{b}\left(\frac{1}{b}+\frac{n}{\sigma^2}\right)^{-1}.$$

For any given $e$, $\theta^*$, and hyperparameters $a$ and $b$, we can identify the smallest sample size $n_{\text{min}}$ by equating $e$ and $\theta^* + \frac{a}{b}\left(\frac{1}{b}+\frac{n}{\sigma^2}\right)^{-1}$ and solving for $n$. Since $n_{\min}$ has to be a positive integer, this results in the following equation:
$$n_{\min} = \max\left(\left\lfloor\left(\frac{a}{(e - \theta^*)b} - \frac{1}{b}\right)\sigma^2\right\rfloor,1\right) = \max\left(\left\lfloor\left(\frac{[a - (e-\theta^*)]\sigma^2}{(e - \theta^*)b}\right)\right\rfloor,1\right). $$

%$$e > \theta^* + \frac{a}{b}\left(\frac{1}{b}+\frac{n_{\text{min}}}{\sigma^2}\right)^{-1}$$
%holds. 
Since the function
%Let $n_{\text{min}}$ be a given minimum sample size for the trial. Since 
$\left(\frac{1}{b}+\frac{n}{\sigma^2}\right)^{-1}$ is monotonically decreasing in $n$, the condition $e > \theta^* + \frac{a}{b}\left(\frac{1}{b}+\frac{n_{\min}}{\sigma^2}\right)^{-1}$ holds for all
$n \geq n_{\min}$. %Therefore, if
%$$\frac{a}{b}\left(\frac{1}{b}+\frac{n_{\text{min}}}{\sigma^2}\right)^{-1} \geq \frac{a}{b}\left(\frac{1}{b}+\frac{n}{\sigma^2}\right)^{-1}$$
%for all $n \geq n_{\text{min}}$. Hence, if
%$$e > \theta^* + \frac{a}{b}\left(\frac{1}{b}+\frac{n_{\text{min}}}{\sigma^2}\right)^{-1},$$
%then the inequality 
%$$e > \theta^* + \frac{a}{b}\left(\frac{1}{b}+\frac{n}{\sigma^2}\right)^{-1}$$ 
%also holds for all $n \geq n_{\text{min}}$. As a result, the z-score $z(n)$ decreases monotonically, and consequently, $\xi(n)$ increases monotonically in $n$ for all $n \geq n_{\text{min}}$.

Finally, in the special case where $a = 0$, the condition simplifies to $e > \theta^*$. Hence, as long as $e > \theta^*$, $n_{\min} = 0$ in this case.
%Thus, for continuous outcomes, the monotonicity property requires a condition on $e$, which is $e > \theta^* + \frac{a}{b}\left(\frac{1}{b}+\frac{n}{\sigma^2}\right)^{-1}$. When $a = 0$, this condition reduces to the general assumption of $e > \theta^*$. 
$\hfill \square$

\newpage

\subsection{Proof of Proposition 3}\label{subsec:prop_3}

Using the definition of the gamma CDF, we write:
$$F_{\text{Gamma}}(\theta';a+ne',b+n) - F_{\text{Gamma}}(\theta';a+(n+1)e',b+n+1) = $$
\begin{equation}\label{eq:gamma_A_B}
    \int_0^{\theta'} \underbrace{\frac{(b+n+1)^{a+(n+1)e'}x^{a+ne'-1}\exp{(-(b+n)x)}}{\Gamma(a+(n+1)e')}}_{{\cal{A}}_g}\left[\underbrace{\frac{\Gamma(a+(n+1)e')(b+n)^{a+ne'}}{\Gamma(a+ne')(b+n+1)^{a+(n+1)e'}} - x^{e'}\exp{(-x)}}_{{\cal{B}}_g}\right]dx.
\end{equation}

The term ${\cal A}_g$ is strictly positive for all $x \geq 0$, $a, b > 0$, $n \geq 0$, and $e' > 0$. As in Section \ref{subsec:proof_prop1}, the integrand $\cB_g$ represents the difference between % in equation \eqref{eq:gamma_A_B} is also a difference between 
a constant (in $x$) and the function $x^{e'}\exp{(-x)}$. It follows that the overall difference in the gamma CDF % is straightforward to see that the difference of the two gamma CDFs 
is positive if $\cB_g > 0$, that is, 
$$\int_0^{\theta'} {\cal A}_g \cdot {\cal B}_g \; dx \geq \int_0^{\theta'} {\cal A}_g \cdot 0 \; dx = 0.$$

We now analyze the conditions under which $\cB_g > 0$. First, fix $e'$ and $n$, and consider $\cB_g$ as a function of $x$. Taking the %the first and second 
derivative with respect to $x$, we obtain:
$$\frac{d{{\cal{B}}_g}}{dx} = \exp{(-x)}x^{e'-1}(x-e').$$%, \,\, \frac{d^2{\cal{B}}}{dx^2} = \exp{(-x)}x^{e'-2}(e'-(x-e')^2).$$
This derivative shows that $\cB_g(x)$ is strictly decreasing for $x \leq e'$, strictly increasing for $x \geq e'$, and attains its minimum at $x = e'$. %being the only critical point. 
Since $\cB_g(x)$ is positive both at $x = 0$ and as $x \rightarrow \infty$ (since 
$\lim_{x \rightarrow \infty} x^{e'}\exp{(-x)} = 0$), the condition $\cB_g > 0$ holds if:% We conclude that
$$\frac{\Gamma(a+(n+1)e')(b+n)^{a+ne'}}{\Gamma(a+ne')(b+n+1)^{a+(n+1)e'}} \geq {e'}^{e'}\exp{(-e')}.$$
Moreover, it can be shown that as $n \rightarrow \infty$, the left-hand side converges to the right-hand side:
$$\lim_{n \rightarrow \infty}\frac{\Gamma(a+(n+1)e')(b+n)^{a+ne'}}{\Gamma(a+ne')(b+n+1)^{a+(n+1)e'}} = {e'}^{e'}\exp{(-e')}.$$ 
That is, $\lim_{n \rightarrow \infty}\cB_g(n,x = e') = 0$.

Next, we consider $\cB_g$ as a function of $n$, fixing $e'$ and $x$. Since $x^{e'}\exp{(-x)}$ does not depend on $n$, the monotonicity of $\cB_g$ in $n$ depends on:% then analyze the relationship between $n$ and  
$$R_g(n) = \frac{\Gamma(a+(n+1)e')(b+n)^{a+ne'}}{\Gamma(a+ne')(b+n+1)^{a+(n+1)e'}}.$$
To analyze whether $R_g(n)$ is increasing in $n$, consider the ratio
\begin{equation}\label{eq:R_g_ratio}
    \frac{R_g(n+1)}{R_g(n)} = \frac{\Gamma(a+(n+2)e')\Gamma(a+ne')}{\Gamma(a+(n+1)e')^2}\frac{(b+n+1)^{2(a+(n+1)e')}}{(b+n)^{a+ne'}(b+n+2)^{a+(n+2)e'}}.
%\underbrace{\left[\frac{(b+n+1)^2}{(b+n)(b+n+2)}\right]^{a+ne'}}_{(2)}\underbrace{\left[\frac{b+n+1}{b+n+2}\right]^{2e'}}_{(3)} \geq 1.
\end{equation}
If $\frac{R_g(n+1)}{R_g(n)} > 1$, $R_g(n)$ is strictly increasing with $n$. Taking the logarithm of the ratio, we have
$$\log\left(\frac{R_g(n+1)}{R_g(n)}\right) = \log\Gamma(a + (n+2)e') + \log\Gamma(a+ne') - 2\log\Gamma(a+(n+1)e') +$$
$$2(a+(n+1)e')\log(b+n+1) - (a+ne')\log(b+n) - (a+(n+2)e')\log(b+n+2).$$
Let $x$ be a generic variable, and define the function 
$$h(x) := \log\Gamma(a+xe') - (a+xe')\log(b+x)$$
on $[0,\infty)$. It is straightforward to see that
$$\log\left(\frac{R_g(n+1)}{R_g(n)}\right) = h(n+2) + h(n) - 2h(n+1).$$
It is suffice to show that $h(x)$ is convex, because if so, then by convexity of $h(x)$, we have
$$h(n+2) + h(n) - 2h(n+1) \geq 0,$$
with strict positive if $h(x)$ is strict convex. This then would imply that $R_g(n)$ increases with $n$. 
%three components: 1) the first term is a ratio of gamma functions; 2) the second term compares growth rates of polynomial powers; and 3) the third term decays slowly with $n$.
%To establish monotonicity of $R_g(n)$, we show that the product of the three terms in \eqref{eq:R_g_ratio} is greater than or equal to 1 under suitable conditions. 
The following lemma provides the proof for $h(x)$ being a strict convex function of $x$.  %conditions under which this holds.
%The following Lemma is used to show that under certain conditions, the above claim is true.

\begin{lemma}\label{lemma:count_one_arm_step}
    Assume the hyperparameters and the evidence $a,b,e' > 0$, and define the function on $[0, \infty)$
    %$h(x) := $
    %, and $e'$ be given hyperparameters and evidence. If
    \begin{equation}\label{eq:one_arm_gamma_cond}
        h(x) := \log\Gamma(a+xe') - (a+xe')\log(b+x).%\frac{\Gamma(a+2e')\Gamma(a)}{\Gamma(a+e')^2}\left[\frac{(b+1)^2}{b(b+2)}\right]^a\left[\frac{b+1}{b+2}\right]^{2e'} \geq 1,
    \end{equation}
    %then $\frac{R_g(n+1)}{R_g(n)} \geq 1$ for all $n > 0$.
    This function is strictly convex for $x > 0$.
\end{lemma}

\paragraph{Proof of Lemma \ref{lemma:count_one_arm_step}:}

Taking the first and second derivatives of $h(x)$ with respect to $x$, we have
$$h'(x) = e'\psi(a+xe') - e'\log(b+x) - \frac{a+xe'}{b+x}, \text{ and}$$
$$h''(x) = (e')^2\psi'(a+xe') - \frac{2e'}{b+x}+\frac{a+xe'}{(b+x)^2},$$
where $\psi(\cdot)$ and $\psi'(\cdot)$ are the digmma and the trigamma functions, respectively. Let $z = a+xe' > 0$, $t = b+x > 0$, and use the property $\psi'(u) > \frac{1}{u}$ for any $u > 0$, we have
$$h''(x) > \frac{(e')^2}{z} - \frac{2e'}{t} + \frac{z}{t^2} = \frac{(e')^2t^2 - 2e'tz + z^2}{zt^2} = \frac{(e't - z)^2}{zt^2} = \frac{(e'b - a)^2}{(a+xe')(b+x)^2} \geq 0.$$
Hence, $h(x)$ is strictly convex on $[0, \infty)$.

$\hfill \square$

By Lemma~\eqref{eq:one_arm_gamma_cond}, we conclude that $\frac{R_g(n+1)}{R_g(n)} > 1$, and $R_g(n)$ strictly increases with $n$. Recall that the term $x^{e'}\exp(-x)$ in $\cB_g$ does not depend on $n$, this result further implies that the function $\cB_g(n, x)$ strictly increases with $n$ for any fixed $x$. 

Next, %assume that condition~\eqref{eq:one_arm_gamma_cond} is satisfied. 
We now show that for any $\theta^* < e$, there exists an integer $n_{\text{min}}' > 0$ such that for all $n > n_{\text{min}}'$, $\cB_g(n,x) > 0$ for all $x \in (0,\theta')$ (recall that $\theta' = \theta^* + \theta_0$).

First, observe that $\cB_g(n,x = 0) > 0$ by definition. If there exists an $n_{\text{min}}'$ such that ${\cal B}_g(n_{\text{min}}', x = e') \geq 0$, then this $n_{\text{min}}'$ satisfies our requirement, since $\cB_g(n,x)$ is non-decreasing in $n$ for each fixed $x$, and thus remains non-negative for all $n \geq n_{\text{min}}'$ and $x \in (0, \theta')$. 

Now consider that ${\cal B}_g(n, x = e') < 0$ for all finite $n$. Fix any $n > 0$. Since ${\cal B}_g(n, x = 0) > 0$ and ${\cal B}_g(n, x)$ is continuous and strictly decreasing in $x$ over $[0,e']$, by the Intermediate Value Theorem, there exists some $\theta' \in (0,e')$ such that 
$${\cal B}_g(n, x = \theta') = 0, \; \text{ and } \; \cB_g(n, x) > 0 \text{ for all } x \in (0, \theta').$$ 
Moreover, because $\cB_g(n, x)$ strictly increases in $n$ for fixed $x$, the corresponding $\theta^*$ for which $\cB_g(n,x = \theta') = 0$ increases with $n$. In particular, as $n' \rightarrow \infty$, we have $\theta' \rightarrow e'$, i.e., with a fixed $\theta_0$, $\theta^* \rightarrow e$ as $n \rightarrow \infty$.

Lastly, we show that there exists a monotonically increasing sequence of $\theta'$, denoted as $\bm{\Theta}' = \{\theta_1',\theta_2',\ldots,e'\}$, that is one-to-one corresponding to the sample sizes $\{1, 2, \ldots,  \infty\}$ such that $\cB_g(n,x = \theta_n') = 0$. As previously, we have shown that there exits a corresponding $\theta'$ for any sample size $n = 1, 2, \ldots$, we aim to proof that for any $\theta_n'$, we have $\theta_n' \leq \theta_{n+1}'$. Using the same argument as in Proposition 1, suppose that there exists a $\theta_n'$ such that $\theta_n' > \theta_{n+1}'$, we have
$$\cB_g(n+1,x=\theta_{n+1}') = 0 > \cB_g(n+1,x=\theta_{n}') \geq \cB_g(n,x=\theta_{n}') = 0,$$
which is a contradiction. Consequently, for all $\theta_n' \in \bm{\Theta}'$, we must have $\theta_1' \leq \theta_2' \leq \ldots \leq e'.$ Therefore, for any $\theta^* < e$, we have $\theta' \in (0,e')$, and there exists $n_{\min}'$ such that $\theta_{n_{\min}}' \leq \theta' \leq \theta_{n_{\min}+1}'$, and
%$$\cB_g(n_{\min}',\theta') \geq 0 \text{ and } \cB_g(n_{\min}',x) > 0 \text{ for all } x \in \{0, \theta'\}.$$
%In other words, for any $\theta^* < e$, we have $\theta' \in (0,e')$, and there exists an $n_{\text{min}}'$ such that 
$$\cB_g(n_{\text{min}}',\theta') = 0, \; \text{ and } \; \cB_g(n_{\text{min}}',x) > 0 \text{ for all } x \in (0,\theta').$$  
This value of $n_{\min}'$ ensures that for all $n > n_{\min}'$, $\cB_g(n,x) \geq \cB_g(n_{\min}',x) > 0$ over $(0, \theta')$. Consequently,
%Since $\cB_g(n,x)$ is non-decreasing in $n$, it follows that for all $n > n_{\text{min}}'$, we have $\cB_g(n,x) \geq \cB_g(n_{\text{min}}',x) > 0$ for all $x \in (0, \theta')$.

%Therefore, we conclude that for any $\theta^* < e$ (with a fixed $\theta_0$), there exists an integer $n_{\text{min}}' > 0$ such that for all $n > n_{\text{min}}'$, 
$$\eqref{eq:gamma_A_B} = \int_0^{\theta'} {\cal A}_g\cdot\cB_g \; dx > \int_0^{\theta'} {\cal A}_g\cdot 0 \; dx = 0.$$

Finally, note $n_{\text{min}}'$ guarantees $\cB_g(n_{\text{min}}',x) > 0$ for all $x \in (0,\theta')$. However, it is possible to find a smaller sample size $n_{\text{min}} \leq n_{\text{min}}'$ such that \eqref{eq:gamma_A_B} remains positive, even if $\cB_g(n_{\text{min}},x)$ is not strictly positive over the entire interval $x \in (0,\theta')$. $\hfill \square$

\bigskip
\newpage

\subsection{Proof of Theorem 4}

Assume $e$ and $\bar{y}_0$ are fixed and known. Define 
$$H(e,n) = \int_{\bm{\theta}\in H_0} f(\bm{\theta}|\bar{y}_0,e,n)d\bm{\theta} = $$
$$\iint_{\theta_1 - \theta_0 \leq \theta^*} f(\theta_1;a_1+n(\bar{y}_0+e),b_1+n)f(\theta_0;a_0+n\bar{y}_0, b_0+n)d\theta_1d\theta_0.$$
We aim to show that under certain conditions, 
$$H(e,n) - H(e,n+1) \geq 0.$$
Since $H(e,n)$ is composed of two gamma densities, we can express the difference $H(e,n) - H(e,n+1)$ as:
%\begin{align}\label{eq:two_gamma_A_B}
%    \iint_{\theta_1 - \theta_0 \leq \theta^*} \underbrace{\frac{(b_1+n+1)^{a_1+(n+1)(\bar{y}_0+e)}{\theta_1}^{a_1+n(\bar{y}_0+e)}\exp{(-(b_1+n)\theta_1)} (b_2+n+1)^{a_2+(n+1)\bar{y}_0}{\theta_0}^{a_2+n\bar{y}_0}\exp{(-(b_0+n)\theta_0)} }{\Gamma(a_1+(n+1)(\bar{y}_0+e))\Gamma(a_2+(n+1)\bar{y}_0)}}_{\cal{A}} \times \\
%    \left[\underbrace{\frac{\Gamma(a_1+(n+1)(\bar{y}_0+e))(b_1+n)^{a_1+n(\bar{y}_0+e)} \Gamma(a_2+(n+1)\bar{y}_0)(b_2+n)^{a_2+n\bar{y}_0} }{\Gamma(a_1+n(\bar{y}_0+e))(b_1+n+1)^{a_1+(n+1)(\bar{y}_0+e)} \Gamma(a_2+n\bar{y}_0)(b_2+n+1)^{a_2+(n+1)\bar{y}_0} } - \theta_1^{(\bar{y}_0+e)}\theta_0^{\bar{y}_0}\exp{(-\theta_1)}\exp{(-\theta_0)}}_{\cal{B}}\right]dx.
%\end{align}
$$
    \iint_{\theta_1 - \theta_0 \leq \theta^*} \underbrace{\frac{{\theta_1}^{a_1+n(\bar{y}_0+e) \color{black} -1  }\exp{(-(b_1+n)\theta_1)}{\theta_0}^{a_0+n\bar{y}_0 \color{black} -1 }\exp{(-(b_0+n)\theta_0)} }{C_{g1}C_{g2}}}_{{\cal{A}}_{g2}} \times $$
\begin{equation}\label{eq:two_gamma_A_B}
    \left[\underbrace{\frac{C_{g1}C_{g2}}{C_{g3}C_{g4}} - \theta_1^{(\bar{y}_0+e)}\theta_0^{\bar{y}_0}\exp{(-\theta_1)}\exp{(-\theta_0)}}_{{\cal{B}}_{g2}}\right]dx.
\end{equation}
where 
$$C_{g1} = \frac{\Gamma(a_1+(n+1)(\bar{y}_0+e))}{(b_1+n+1)^{a_1+(n+1)(\bar{y}_0+e)}}, \; C_{g2} = \frac{\Gamma(a_0+(n+1)\bar{y}_0)}{(b_0+n+1)^{a_0+(n+1)\bar{y}_0}},$$
$$C_{g3} = \frac{\Gamma(a_1+n(\bar{y}_0+e))}{(b_1+n)^{a_1+n(\bar{y}_0+e)}}, \; C_{g4} = \frac{\Gamma(a_0+n\bar{y}_0)}{(b_0+n)^{a_0+n\bar{y}_0}}.$$
The integrand ${\cal A}_{g2}$ is strictly positive. To determine when the full expression is non-negative, we analyze the behavior of ${\cal B}_{g2}$.

We begin by examining the dependence of $\cB_{g2}(n, \theta_1, \theta_0)$ on $(\theta_1,\theta_0)$ for fixed $n$. Similar to Proposition \ref{subsec:prop_prop2}, the first term in this function is finite, positive, and does not dependent on $(\theta_1,\theta_0)$, it is equivalent to analyze
$$h_{g}(\theta_1,\theta_0) = \theta_1^{(\bar{y}_0+e)}\theta_0^{\bar{y}_0}\exp{(-\theta_1)}\exp{(-\theta_0)},$$
which is a separable function. Through straightforward computation, $h_g(\theta_1,\theta_0)$ is strictly log-concave with a global unique maximizer at $(\theta_1,\theta_0) = (\bar{y}_0+e,\bar{y}_0)$ \cite{boyd2004convex}. Equivalently, the function $\cB_{g2}$ is strictly convex with respect to $(\theta_1,\theta_0)$ for any fixed $n$, with a unique global minimizer at $P^* = (\bar{y}_0+e,\bar{y}_0)$. Moreover,    
%By taking partial derivatives with respect to $\theta_1$ and $\theta_0$, it is straightforward to verify:
\begin{itemize}
    \item For any fixed $\theta_1$ and $n$, 
    $\cB_{g2}(n,\theta_1,\theta_0)$ is a strict convex function of $\theta_0$; it decreases for $\theta_0 \in [0, \bar{y}_0)$ increases for $\theta_1 \in (\bar{y}_0, \infty)$.
    \item For any fixed $\theta_0$ and $n$, $\cB_{g2}(n,\theta_1,\theta_0)$ is a strict convex function of $\theta_1$; it decreases for $\theta_1 \in [0, \bar{y}_0+e)$ and increases for $\theta_1 \in (\bar{y}_0+e, \infty)$.
\end{itemize}

Next, we examine the dependence of $\cB_{g2}(n,\theta_1,\theta_0)$ on $n$, holding $\theta_1$ and $\theta_0$ fixed. Since $h_g(\theta_1,\theta_0)$ does not dependent on $n$, the $n$-dependence of $\cB_{g2}$ arises entirely from the factor $\frac{C_3C_4}{C_1C_2}$. From Proposition \ref{subsec:prop_3} and Lemma \ref{lemma:count_one_arm_step}, since both ratios $C_3/C_1$ and $C_4/C_2$ are positive and strictly increasing in $n$, consequently, the the factor $\frac{C_3C_4}{C_1C_2}$ is strictly increasing in $n$ (for fixed $\bar{y}_0$ and $e$). Moreover, a direct calculation shows that %as $n \rightarrow \infty$, we have:
$$\lim_{n \rightarrow \infty} \frac{C_{g1}C_{g2}}{C_{g3}C_{g4}} = {(\bar{y}_0+e)}^{(\bar{y}_0+e)}\bar{y}_0^{\bar{y}_0}\exp{(-\bar{y}_0 \color{black} -  e)}\exp{(-\bar{y}_0)},$$
which is a finite, positive constant. Hence, we have %and thus 
$$\lim_{n \rightarrow \infty}\cB_{g2}(n, \bar{y}_0+e, \bar{y}_0) = $$
$${(\bar{y}_0+e)}^{(\bar{y}_0+e)}\bar{y}_0^{\bar{y}_0}\exp{(-\bar{y}_0 \color{black} - e)}\exp{(-\bar{y}_0)} - {(\bar{y}_0+e)}^{(\bar{y}_0+e)}\bar{y}_0^{\bar{y}_0}\exp{(-\bar{y}_0 \color{black} - e)}\exp{(-\bar{y}_0)} = 0,$$
and 
$$\lim_{n \rightarrow \infty}\cB_{g2}(n,\theta_1,\theta_0) = {(\bar{y}_0+e)}^{(\bar{y}_0+e)}\bar{y}_0^{\bar{y}_0}\exp{(-\bar{y}_0 \color{black} - e)}\exp{(-\bar{y}_0)} - h_g(\theta_1,\theta_0) > \lim_{n \rightarrow \infty} \cB_{g2}(n,\bar{y}_0+e,\bar{y}_0)$$
for all $(\theta_1,\theta_0) \neq (\bar{y}_0+e,\bar{y}_0)$ (due to $(\bar{y}_0+e,\bar{y}_0)$ being the unique global maximizer of $h_g(\theta_1,\theta_0)$). Hence, $\lim_{n \rightarrow \infty} \cB_{g2}(n, \theta_1,\theta_0)$ attains its unique global minimum as $(\bar{y}_0+e,\bar{y}_0)$, with a value of $0$, and for any $(\theta_1,\theta_0) \neq (\bar{y}_0+e,\bar{y}_0)$, the limit is strictly positive.

%Now, consider the monotonicity of ${\cal B}_{g2}$ with respect to $n$. The only term in $\cB_{g2}$ that depends on $n$ is the ratio $\frac{C_{g1}C_{g2}}{C_{g3}C_{g4}}$. By Lemma \ref{lemma:count_one_arm_step},
%if the following conditions are satisfied: 
%$$\frac{\Gamma(a_1+2(\bar{y}_0+e))\Gamma(a_1)}{\Gamma(a_1+\bar{y}_0+e)^2}\left[\frac{(b_1+1)^2}{b_1(b_1+2)}\right]^{a_1}\left[\frac{b_1+1}{b_1+2}\right]^{\bar{y}_0+e} \geq 1,$$
%$$\frac{\Gamma(a_0+2\bar{y}_0)\Gamma(a_0)}{\Gamma(a_0+\bar{y}_0)^2}\left[\frac{(b_0+1)^2}{b_0(b_0+2)}\right]^{a_0}\left[\frac{b_0+1}{b_0+2}\right]^{\bar{y}_0} \geq 1,$$
%then both $\frac{C_{g1}}{C_{g3}}$ and $\frac{C_{g2}}{C_{g4}}$ are positive, non-decreasing functions of $n$, and hence so is $\frac{C_{g1}C_{g2}}{C_{g3}C_{g4}}$. As a result, $\cB_{g2}$ is  \yy  a \bb non-decreasing function in $n$ for fixed $(\theta_1,\theta_0)$.

Next, we show that for any $\theta^* < e$, there exists an $n_{\text{min}}'$ such that for all $n \geq n_{\text{min}}'$, $\cB_{g2}$ is positive in  the region $\{\theta_1 - \theta_0 \leq \theta^*\}$.

Similar to Section \ref{subsec:prop_prop2}, suppose there exists an integer $n_{\text{min}}'$ such that 
$$\cB_{g2}(n_{\text{min}}', \bar{y}_0+e, \bar{y}_0) \geq 0.$$ 
Since $\cB_{g2}$ is strictly convex in $(\theta_1,\theta_0)$ with its minimum at $(\bar{y}_0+e,\bar{y}_0)$, this implies that $n_{\text{min}}'$ satisfies the required condition. %Indeed, for any $\theta_1' \neq \bar{y}_0+e$, we have
%$$\cB_{g2}(n_{\text{min}}', \theta_1 = \theta_1',\theta_0 = \bar{y}_0) \geq \cB_{g2}(n_{\text{min}}', \theta_1 = \bar{y}_0+e,\theta_0 = \bar{y}_0) \geq 0.$$
%Moreover, for fixed $\theta_1'$, any $n \geq n_{\text{min}}'$, and any $\theta_0' \neq \bar{y}_0$, we have
%$$\cB_{g2}(n,\theta_1 = \theta_1',\theta_0 = \theta_0') \geq \cB_{g2}(n_{\text{min}}', \theta_1 = \theta_1',\theta_0 = \theta_0') \geq \cB_{g2}(n_{\text{min}}', \theta_1 = \theta_1',\theta_0 = \bar{y}_0).$$
%Combining these inequalities, it follows that for all $n \geq n_{\text{min}}'$, we have $\cB_{g2}(n,\theta_1,\theta_0) \geq 0$, and thus 
%$$H(e,n) - H(e,n+1) \geq 0.$$

Second, if for all finite $n$, 
$$\cB_{g2}(n,\bar{y}_0+e,\bar{y}_0) < 0,$$ 
we invoke the following lemma.

\begin{lemma}\label{lemma:count_2arm_n}
    Let $\cB_{g2}$ be as defined in \eqref{eq:two_gamma_A_B}, viewed as a function of $(n, \theta_1, \theta_0)$. If $\cB_{g2}(n,\bar{y}_0+e,\bar{y}_0) < 0$ for all finite sample sizes $n$, then for each $n$ there exists a constant $\theta_n^* < e$ such that
    $$\cB_{g2}(n,\theta_1,\theta_0) \geq 0 \text{ for all } (\theta_1,\theta_0) \text{ satisfying } \theta_1 - \theta_0 \leq \theta_n^*, \theta_1 > 0, \text{ and } \theta_0 \geq 0.$$
    Moreover, the sequence $(\theta_n^*)_{n \geq 1}$ is monotonically increasing in $n$, and 
    $$\lim_{n \rightarrow \infty} \theta_n^* \rightarrow e.$$
\end{lemma}

\noindent The proof of Lemma \ref{lemma:count_2arm_n} is similar to the proof of Lemma \ref{lemma:2arm_bin_step} in Section \ref{subsec:prop_prop2}. Specifically, we can show that $\cB_{g2}(n,\theta_1,\theta_0) = 0$ forms a closed contour following the steps in the proof of Lemma \ref{lemma:2arm_bin_step}. Moreover, for any finite $n$, the contour 
$${\mathcal{CO}}_g(n) = \{(\theta_1',\theta_0') \in (0,\infty)^2; \cB_{g2}(n,\theta_1',\theta_0') = 0\}$$
is also a closed strictly convex curve, because there are at most two intersection points between the contour and the line $\theta_1 = c_{sg}\theta_0 + c_{ig}$, i.e., the second derivative of $\log(h_g(c_{sg}\cdot\theta_0+c_{ig},\theta_0))$ with respect to $\theta_0$ is
$$- \frac{(\bar{y}_0+e)c_{sg}^2}{(c_{sg}\cdot\theta_0+c_{ig})^2} - \frac{\bar{y}_0}{\theta_0^2} < 0.$$
Then, the rest of the proof then follows the proof of Lemma \ref{lemma:2arm_bin_step}.

Using Lemma \ref{lemma:count_2arm_n}, we conclude that for any fixed $n$, there exists a $\theta_n^* < e$ such that $\cB_{g2}(n,\theta_1,\theta_0) \geq 0$ for all $(\theta_1, \theta_0)$ satisfying $\theta_1 - \theta_0 \leq \theta_n^*$. Follow the last part of Section \ref{subsec:prop_prop2}, for any given $\theta^*$, we can identify the smallest $\theta_n^* \geq \theta^*$ (with a corresponding sample size $n_{\min}'$). When $n \geq n_{\min}'$, $\cB_{g2}(n,\theta_1,\theta_0) \geq 0$ for all values of $(\theta_1,\theta_0)$ satisfying $\theta_1 - \theta_0 \leq \theta^*$. And it is also possible to find a smaller sample size $n_{\min} \leq n_{\min}'$ could still result in $H(e,n) - H(e,n+1) \geq 0$ for all $n \geq n_{\min}$, even if $\cB_{g2}$ is not strictly positive for all $(\theta_1,\theta_0)$ in the region of $\theta_1 - \theta_0 \leq \theta^*$.

$\hfill \square$

\bigskip
\newpage

\subsection{Additional Simulation Setup and Results in Section 5} 

\paragraph{Additional simulation details and results}

\begin{figure}[h]
    \centering
    \includegraphics[width=\textwidth]{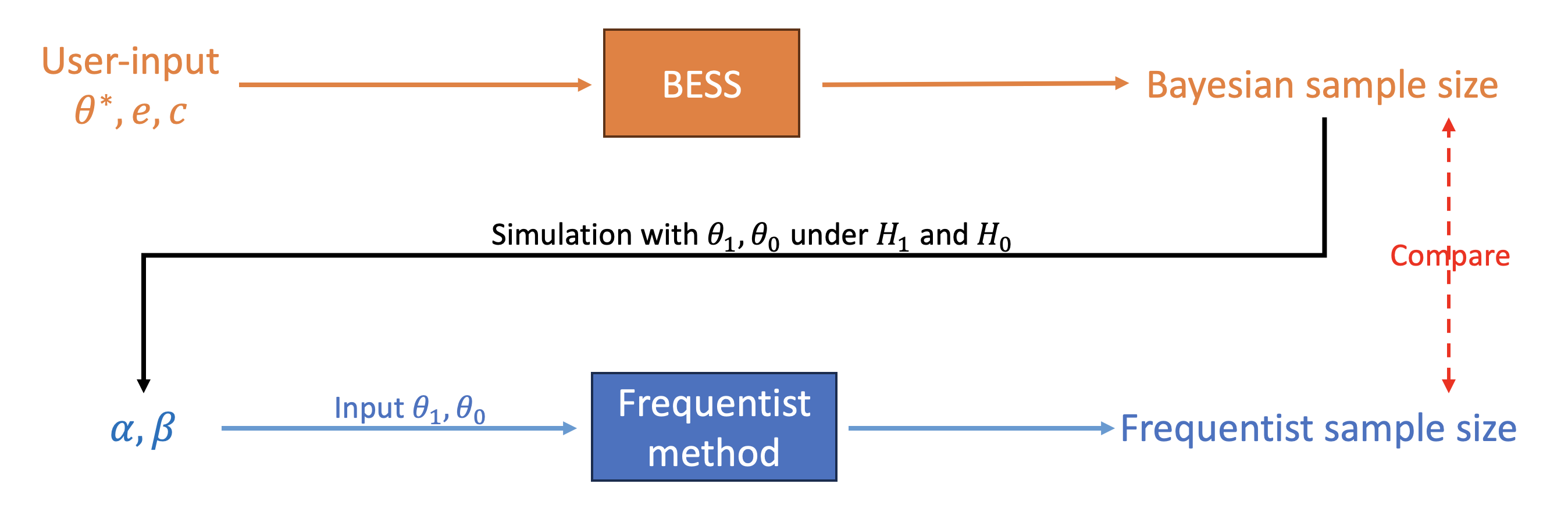}
    \caption{Flowchart of simulation process to compare sample sizes estimated under BESS and that of the frequentist method.}
    \label{fig:flow_comp}
\end{figure}

Figure \ref{fig:flow_comp} shows a flowchart for the simulation process in Section 5.1.

\noindent Table \ref{tab:verify_tab} shows the estimated sample sizes via BESS and the oracle sample size with the standard SSE method for other outcome and trial types.

\begin{table}[h]
    \caption{Simulation results compare BESS with Standard SSE when the type I  error rate and power are matched between both methods in one- and two-arm trial with all three outcomes. %For continuous outcomes in one- and two-arm trials, the monotonic conditions are checked in both cases. 
    For all three outcomes %binary outcome 
    in one- and two-arm trials, $n_{\text{min}}$ are one. %For count-data outcomes in one-and two-arm trials, $n_{\text{min}}$s are both zero. 
    The results show the estimated sample sizes of the two methods are similar for various levels of evidence $e$ and confidence $c$ in each trial and outcome type.  }
    \label{tab:verify_tab}
    \centering
    \resizebox{\linewidth}{!}{
    \begin{tabular}{cc|cc|cc|cc}
    \hline
       Trail & Outcome & Evidence & Confidence & type I & power & BESS & Standard SSE \\ 
        type & type & $e$ & $c$ & error rate $\alpha$ & $1 - \beta$ & $n$ & $n$ \\
    \hline
        \multirow{3}{*}{One-arm} & Binary & 0.35 & 0.8 & 0.14 & 0.52 & 120 & 118 \\
        & Continuous & 0.4 & 0.8 & 0.21 & 0.50 & 67 & 67 \\
        & Count & 0.375 & 0.8 & 0.18 & 0.52 & 56 & 55 \\
    \hline
        \multirow{2}{*}{Two-arm} 
        & Continuous & 0.15 & 0.8 & 0.21 & 0.51 & 71 & 72 \\
        & Count & 0.3 & 0.8 & 0.26 & 0.97 & 72 & 72 \\
    \hline
    \end{tabular}}
\end{table}

\noindent Table \ref{tab:my_label} show the results when $\theta_0$ and $\theta_1$ are mis-specified in the standard SSE.

\begin{table}[h]
    \caption{Standard SSE sample size and simulated type I error rate, power, false positive rate, and false negative rate when $\theta_1 - \theta_0$ is mis-specified by the frequentist method to match $e$ in BESS for two-arm trial with binary outcome. The oracle results corresponding to $\theta_1 = 0.4$ and $\theta_0 = 0.25$ -- shown in Table 2 -- for the frequentist sample size, FDR, and FOR are reproduced in brackets. }
    \label{tab:my_label}
    \centering
    \resizebox{\linewidth}{!}{
    \begin{tabular}{cc|cccc|c|cccc}
        \hline
        Evidence & Confidence & \multicolumn{4}{c|}{Planned} & Standard & \multicolumn{4}{c}{Simulation} \\
        $e$ & $c$ & $\theta_1$ & $\theta_0$ & $\alpha$ & $1-\beta$ & SSE $n$ %sample size $n$ 
        & $\alpha$ & $1-\beta$ & FDR & FOR \\
        \hline
        \multirow{3}{*}{0.1} & 0.7 & 0.35 & 0.25 & 0.36 & 0.73 & 161 (42) & 0.30 & 0.93 & 0.24 (0.33) & 0.10 (0.29) \\
        & 0.8 & 0.35 & 0.25 & 0.23 & 0.82 & 455 (117) & 0.18 & 0.99 & 0.15 (0.22) & 0.01 (0.19) \\
        & 0.9 & 0.35 & 0.25 & 0.12 & 0.92 & 1102 (284) & 0.08 & 1.00 & 0.07 (0.12) & 0.00 (0.08) \\
        \hline
        \multirow{3}{*}{0.2} & 0.7 & 0.45 & 0.25 & 0.45 & 0.59 & 3 (6) & 0.37 & 0.49 & 0.43 (0.43) & 0.45 (0.43) \\
        & 0.8 & 0.45 & 0.25 & 0.28 & 0.51 & 8 (16) & 0.25 & 0.41 & 0.38 (0.35) & 0.43 (0.41) \\
        & 0.9 & 0.45 & 0.25 & 0.19 & 0.48 & 18 (39) & 0.12 & 0.30 & 0.29  (0.28) & 0.44 (0.39) \\
        \hline
    \end{tabular}}
\end{table}

\paragraph{Additional simulation details in sensitivity of prior}

In   the  previous simulation,   we assume a flat prior $\text{Beta}(0,0)$ for BESS. In particular, this prior is used to find the sample size as well as to compute  $\text{Pr}(H = H_1\mid \bm{y}_n)$   
for each simulated trial. In \cite{morita2008determining}, the authors show that the prior effective sample size of $\text{Beta}(a,b)$ for a binomial likelihood is quantified as $(a+b).$ Therefore, $\text{Beta}(0,0)$ is considered noninformative in that it includes no prior information for the posterior inference.  

However, BESS can accommodate prior information, if available, as part of sample size estimation.    
Assuming there exist $n_0$ patients per arm as prior data,  we demonstrate the sensitivity of incorporating these prior information through simulation.
First consider the simulation process for a single trial with the following two steps: 1) generating the prior data, 2) assume evidence $e = 0.15$ and $c = 0.8$, estimate the sample size via BESS using the informative priors constructed from the prior data. For step 1, assuming there are $n_0 = 10$ external patients data per arm that is available to be incorporated, we generate these prior data similar to the simulation process in Section 5.1 
with $\theta_1 = 0.4$ and $\theta_0 = 0.25$. Denote the generated data $\bm{y}_j^0 = \{y_{ij}^0; i = 1, \ldots, n_0\}, \, j = 1, 0$ as the   binary    outcomes of these 10 patients' external data for the treatment and the control arms, respectively. We consider an informative prior for BESS as 
$$(\theta_1,\theta_0)|H = H_j \sim   \prod_{j=1}^2 \text{Beta}\left(\sum_{i=1}^{n_0} y_{ij}^0, n_0 - \sum_{i=1}^{n_0} y_{ij}^0\right) I\{(\theta_1,\theta_0)\in H_j\}.   $$ 
We find sample size via BESS's algorithm 2 with the informative prior, $e$, and $c$.
Following this process, we simulate 1,000 trials and compute the average sample size to compare to BESS with vague prior.

\bigskip
\newpage

\subsection{Metrics Used in Section 6.2}
\label{subsec:metrics}

The metrics include: 1) type I error rate, 2) type II error rate, 3) false positive rate, and 4) false negative rate. These rates are defined below:

\medskip

\noindent {\bf type I error rate:} the proportion of simulated trials in which the null is true but falsely rejected: 
$$\text{type I error rate} = \frac{\# \text{ rejected simulated trials from null}}{\# \text{ simulated trials from null}}.$$
    
\noindent{\bf type II error rate:} the proportion of simulated trials in which the alternative is true but falsely rejected: 
$$\text{type II error rate} = \frac{\# \text{ accepted simulated trials from alternative}}{\# \text{ simulated trials from alternative}}.$$
    
\noindent{\bf false discovery rate (FDR):} the proportion of simulated trials in which the null is rejected when null is true:
$$\text{FPR} = \frac{\# \text{ rejected simulated trials from null}}{\# \text{ rejected simulated trials}}.$$
    
\noindent {\bf false omission rate (FOR):} the proportion of simulated trials in which the null is accepted when null is not true:
$$\text{FNR} = \frac{\# \text{ accepted simulated trials from alternative}}{\# \text{ accepted simulated trials}}.$$

\bigskip
\newpage

\end{document}